# Dynamic Nanodomains Dictate Macroscopic Properties in Lead Halide Perovskites

Milos Dubajic,[1,2,*] James R. Neilson,[3,4,5,*] Johan Klarbring,[6,7] Xia Liang,[6] Stephanie A. Boer,[8] Kirrily C. Rule,[9] Josie E. Auckett,[8] Leilei Gu,[10] Xuguang Jia,[11,12] Andreas Pusch,[2] Ganbaatar Tumen-Ulzii,[1] Qiyuan Wu,[2] Thomas A. Selby,[1] Yang Lu,[1] Julia C. Trowbridge,[3] Eve M. Mozur,[3] Arianna Minelli,[5] Nikolaj Roth,[5] Kieran W. P. Orr,[1,13] Arman Mahboubi Soufiani,[2,14] Simon Kahmann,[1] Irina Kabakova,[15] Jianning Ding,[11,16] Tom Wu,[17,18] Gavin J. Conibeer,[2] Stephen P. Bremner,[2,*] Aron Walsh,[6,*] Michael P. Nielsen,[2,*] and Samuel D. Stranks[1,13,*]

[1]Department of Chemical Engineering and Biotechnology,
University of Cambridge, Philippa Fawcett Drive, Cambridge, CB3 0AS, UK
[2]School of Photovoltaic & Renewable Engineering, UNSW Sydney, Kensington 2052, Australia
[3]Department of Chemistry, Colorado State University,
Fort Collins, Colorado 80523-1872, United States of America
[4]School of Materials Science & Engineering, Colorado State University,
Fort Collins, Colorado 80523-1872, United States of America
[5]Inorganic Chemistry Laboratory, University of Oxford, Oxford, UK
[6]Department of Materials, Imperial College London, London SW7 2AZ, United Kingdom
[7]Department of Physics, Chemistry and Biology (IFM),
Linköping University, SE-581 83, Linköping, Sweden
[8]Australian Synchrotron, ANSTO, 800 Blackburn Road, Clayton, VIC 3168, Australia
[9]Australian Nuclear Science and Technology Organisation,
Locked Bag 2001, Kirrawee, DC NSW 2232, Australia
[10]Taizhou Institute of Science and Technology, Nanjing University of
Science and Technology, Taizhou 225300, Jiangsu Province, China
[11]School of Microelectronics and Control Engineering,
Jiangsu Province Cultivation Base for State Key Laboratory of Photovoltaic Science and Technology,
Jiangsu Collaborative Innovation Center of Photovoltaic Science and Engineering,
Changzhou University, Changzhou, 213164, Jiangsu, China
[12]Faculty of Engineering and IT, University Technology Sydney, 2007 Sydney, Australia
[13]Department of Physics, Cavendish Laboratory, University of Cambridge,
JJ Thomson Avenue, Cambridge, CB3 0HE, UK
[14]Helmholtz-Zentrum Berlin für Materialien und Energie GmbH, Division Solar Energy, 12489 Berlin, Germany
[15]School of Mathematical and Physical Sciences, University Technology Sydney, 2007 Sydney, Australia
[16]School of Mechanical Engineering, Jiangsu University, Zhenjiang, 212013, Jiangsu, China
[17]School of Materials Science and Engineering, Faculty of Science,
University of New South Wales, UNSW Sydney, Kensington, 2052, Australia
[18]Department of Applied Physics, The Hong Kong Polytechnic University, Kowloon, Hong Kong, China

Empirical A-site cation substitution has advanced the stability and efficiency of hybrid organic-inorganic lead halide perovskites solar cells and the functionality of X-ray detectors. Yet, the fundamental mechanisms underpinning their unique performance remain elusive. This multi-modal study unveils the link between nanoscale structural dynamics and macroscopic optoelectronic properties in these materials by utilising X-ray diffuse scattering, inelastic neutron spectroscopy and optical microscopy complemented by state-of-the-art machine learning-assisted molecular dynamics simulations. Our approach uncovers the presence of dynamic, lower-symmetry local nanodomains embedded within the higher-symmetry average phase in various perovskite compositions. The properties of these nanodomains are tunable via the A-site cation selection: methylammonium induces a high density of anisotropic, planar nanodomains of out-of-phase octahedral tilts, while formamidinium favours sparsely distributed isotropic, spherical nanodomains with in-phase tilting, even when crystallography reveals cubic symmetry on average. The observed variations in the properties of dynamic nanodomains are in agreement with our simulations and are directly linked to the differing macroscopic optoelectronic and ferroelastic behaviours of these compositions. By demonstrating the influence of A-site cation on local nanodomains and consequently, on macroscopic properties, we propose leveraging this relationship to engineer the optoelectronic response of these materials, propelling further advancements in perovskite-based photovoltaics, optoelectronics, and X-ray imaging.

_________

* milos.dubajic@hotmail.com
* james.neilson@colostate.edu
* stephen.bremner@unsw.edu.au
* a.walsh@imperial.ac.uk
* michael.nielsen@unsw.edu.au
* sds65@cam.ac.uk

Hybrid halide perovskites, with their extraordinary optoelectronic properties, have emerged as a promising class of materials for light emission, detection and energy conversion [1–3]. This is particularly evident in their application as absorber material in photovoltaics, rivaling established commercial technologies [4]. The combination of hybrid halide perovskite top cells with silicon bottom cells in a tandem arrangement is widely seen as the most promising route to improved efficiencies in large-scale photovoltaics [5]. The remarkable progress of perovskite X-ray detectors is highlighted by their ability to offer single-photon-sensitive, low-dose and energy-resolved X-ray imaging, due to the use of thick and uniform perovskite single-crystals, making them competitive with traditional semiconductor materials like silicon and CdZnTe [2].

Despite these extraordinary advancements, uncovering the fundamental microscopic mechanisms responsible for these remarkable properties remains an open scientific challenge, particularly when contrasted with conventional inorganic semiconductors. In hybrid halide perovskites, short-range correlations of atomic species can give rise to local crystallographic structure distinct from the average (global) structure and may play a critical role in defining their macroscopic properties. Fundamentally, the presence of hidden local order modifies the electronic and phononic band structure of the material [6]. A range of hitherto unexplained macroscopic properties in various material classes, including certain battery materials [7, 8], relaxor ferroelectrics [9] and superconductors [10], have emerged from complex local structure landscapes.

In the cubic perovskite aristotype, some of the possible local atomic correlations are: B-site off-centering, octahedral tilt correlations involving B-X bond correlations and A-site correlations, all graphically represented in Supplementary Fig. S2 a. The precise description of the local structure in these materials has not yet been established and currently ranges from polymorphous networks [11], B-site off-centering correlations [12], to local orthorhombic phases embedded in average tetragonal and cubic global structures [13, 14], static in-phase and anti-phase octahedral correlations [15], noncentrosymmetric dynamic local nanodomains [16] and two-dimensional dynamic octahedral sheets [17, 18] with no consensus on whether these local nanodomains are dynamic or static. In halide perovskites, the local structure may be intimately connected to their superior optoelectronic properties [19–21]. It has been suggested that the presence of local dynamic nanodomains stimulates the formation of large ferroelectric polarons [18, 22–24], potentially protecting charge carriers against non-radiative recombination and possibly explaining in part their remarkable photovoltaic performance [25–27].

The highest performing optoelectronic devices are now employing formamidinium (FA) cations, with a movement away from methylammonium (MA). Recent advancements in perovskite solar cells have led to record-breaking efficiencies primarily by employing FA cations in perovskite polycrystalline films. Similarly, $FAPbBr_3$ single crystals have been empirically demonstrated to outperform $MAPbBr_3$ in X-ray detection applications [28, 29]. $FAPbBr_3$ single crystals exhibit a suite of advantageous properties in comparison to $MAPbBr_3$, including lower dark currents and enhanced thermal and phase stability [28, 29]. Furthermore, $FAPbBr_3$ exhibits a fivefold enhancement in carrier lifetime, a fourfold increase in diffusion lengths, and a tenfold reduction in dark carrier concentration, compared to $MAPbBr_3$ [30]. This is somewhat perplexing, as prevailing understanding posits that the electronic structure, and consequently device performance, should not be significantly influenced by the choice of A-site cation. The lack of a comprehensive explanation for these differences underscores the need to investigate whether local structural variations could be the missing link.

Here, we employ single-crystal diffuse scattering to directly probe the local structure of $MAPbBr_3$ and $FAPbBr_3$ in reciprocal space. We both develop a classical model and run large scale machine learning molecular dynamics (MD) simulations [31, 32] that both reproduce the experimental diffuse scattering patterns, allowing us to determine the precise real-space arrangement of the local structure. We uncover the presence of lower-symmetry local phases consisting of pockets of dynamically tilted octahedra within higher-symmetry average structures. The properties of these local octahedral tilts are directly dictated by the nature of the A-site cation: methylammonium cations foster the formation of densely packed planar nanodomains characterised by out-of-phase octahedral tilting, whereas formamidinium results in more sparse spherical nanodomains where octahedra exhibit in-phase tilting along the corresponding crystallographic axis. We propose that these local octahedral correlations give rise to the superior performance of FA hybrid halide perovskites over MA analogues, and we show they impact macroscopic properties including ferroelasticity. Enhanced control over local structure in hybrid halide perovskites could thus be leveraged to engineer additional desirable macroscopic properties, potentially driving further advancements in perovskite-based photovoltaics, optoelectronics and X-ray imaging and even other more exotic device types.

**A-site cations dictate local octahedral correlations in the cubic phase**

To explore the impact of the organic A-site cation on the local structure in lead halide perovskites, we performed X-ray diffuse scattering on $MAPbBr_3$ and $FAPbBr_3$ single crystals. Diffuse scattering arises when X-rays scatter from atoms that dynamically or statically deviate from perfect periodicity, while still maintaining some degree of spatial correlation within the crystal. Although we measure diffuse scattering across a large portion of reciprocal space, for our analysis in the main text, we present the X-ray scattering function $S(\mathbf{q})$ across HK1.5 reciprocal planes of $MAPbBr_3$ and $FAPbBr_3$ in Fig. 1 **a** and **c** (top left quadrants), respectively, for nominally cubic ($Pm3m$) phases at room temperature. The observed diffuse scattering indicates the presence of local spatial correlations in these materials.

We conducted molecular dynamics (MD) simulations using machine learning potentials on $20^3$ pseudo-cubic unit cells, with further details provided in the Methods Section and in Ref. [31]. Using the resulting MD trajectories, we calculated the X-ray scattering function $S(\mathbf{q}, E)$. The top right quadrants of Fig. 1 **a** and **c** illustrate $S(\mathbf{q})$ across the HK1.5 plane for $MAPbBr_3$ and $FAPbBr_3$, respectively, obtained by fully energy (E) integrating $S(\mathbf{q}, E)$ derived from the MD trajectories. These results align closely with the experimentally observed $S(\mathbf{q})$, which is inherently fully energy integrated, as inelastic and elastic X-ray scattering events are experimentally indistinguishable. By selectively integrating MD simulated $S(\mathbf{q}, E)$ over a certain energy range, we further divided it into two distinct components. The first component, dominated by quasi-elastic diffuse scattering (QEDS), is represented in the bottom-right quadrants of Fig. 1 **a** and **c**. QEDS primarily arises from scattering associated with very low energy transfers ($< 1$ meV). The second component, which is influenced by contributions from zone edge acoustic and optical phonons, effectively represents thermal diffuse scattering (TDS) and is depicted in the bottom-left quadrants of Fig. 1 **a** and **c**. This component is characterized by broad diffuse peaks at the $X$ high symmetry points in the Brillouin zone (details about Brillouin zone symmetry points can be found in Supplementary Fig. S14). The QEDS features broad diffuse peaks at the $R$ points in $MAPbBr_3$ and at the $X$ and $M$ points in $FAPbBr_3$.

We separated the total experimental diffuse scattering into QEDS (top quadrants in Fig. 1 **b** and **d**) and TDS components following the procedure detailed in Supplementary Section VI. Upon comparing the bottom right quadrants of Fig. 1 **a** and **c** with the top quadrants in Fig. 1 **b** and **d**, we note a resemblance between the experimental QEDS and the MD obtained QEDS pattern in $MAPbBr_3$. While in $FAPbBr_3$, these two patterns mostly agree, there is also additional intensity in the MD QEDS pattern at the X point, which is not observed in the experimentally derived QEDS. We find that this MD QEDS X point scattering originates from zone centre acoustic phonons (the detailed explanation is given in Supplementary Section V B) which we underestimate through our experimental derivation of QEDS. Throughout this manuscript we will provide evidence for our statement that this QEDS at R points in $MAPbBr_3$ and M points in $FAPbBr_3$ originates from the anharmonic lattice contributions [17, 33, 34], which give rise to low energy excitations in the form of dynamic and locally tilted octahedral nanodomains, spatially and temporally correlated, resulting in quasi-elastic scattering with X-rays.

To determine the size, shape and symmetry of these local nanodomains from the experimental QEDS data we develop a classical local octahedra tilting model. We first calculate the X-ray structure factors for a lattice comprising of octahedrally tilted nanodomains within a cubic matrix. To accurately represent short-range correlations in real space, we broaden each generated 3D Bragg peak, corresponding to the local structure, in inverse proportion to the three-dimensional correlation length values, which are used as fitting parameters in aligning our model with experimental data (a detailed description and a validation of the model can be found in Supplementary Section IV A). Our model successfully describes the experimentally observed QEDS (Fig. 1 **b** and **d**) and also replicates QEDS patterns across other reciprocal space planes (Supplementary Section IV B).

The QEDS patterns of $MAPbBr_3$ and $FAPbBr_3$ exhibit distinct characteristics. In $MAPbBr_3$, the 3D QEDS forms rod-like structures in reciprocal space, indicative of planar (anisotropic) local structure nanodomains in real space (Fig. 1 **e**). Conversely, in $FAPbBr_3$ the 3D QEDS consists of ellipsoids, reflecting a more isotropic local structure in real space (Fig. 1 **f**).

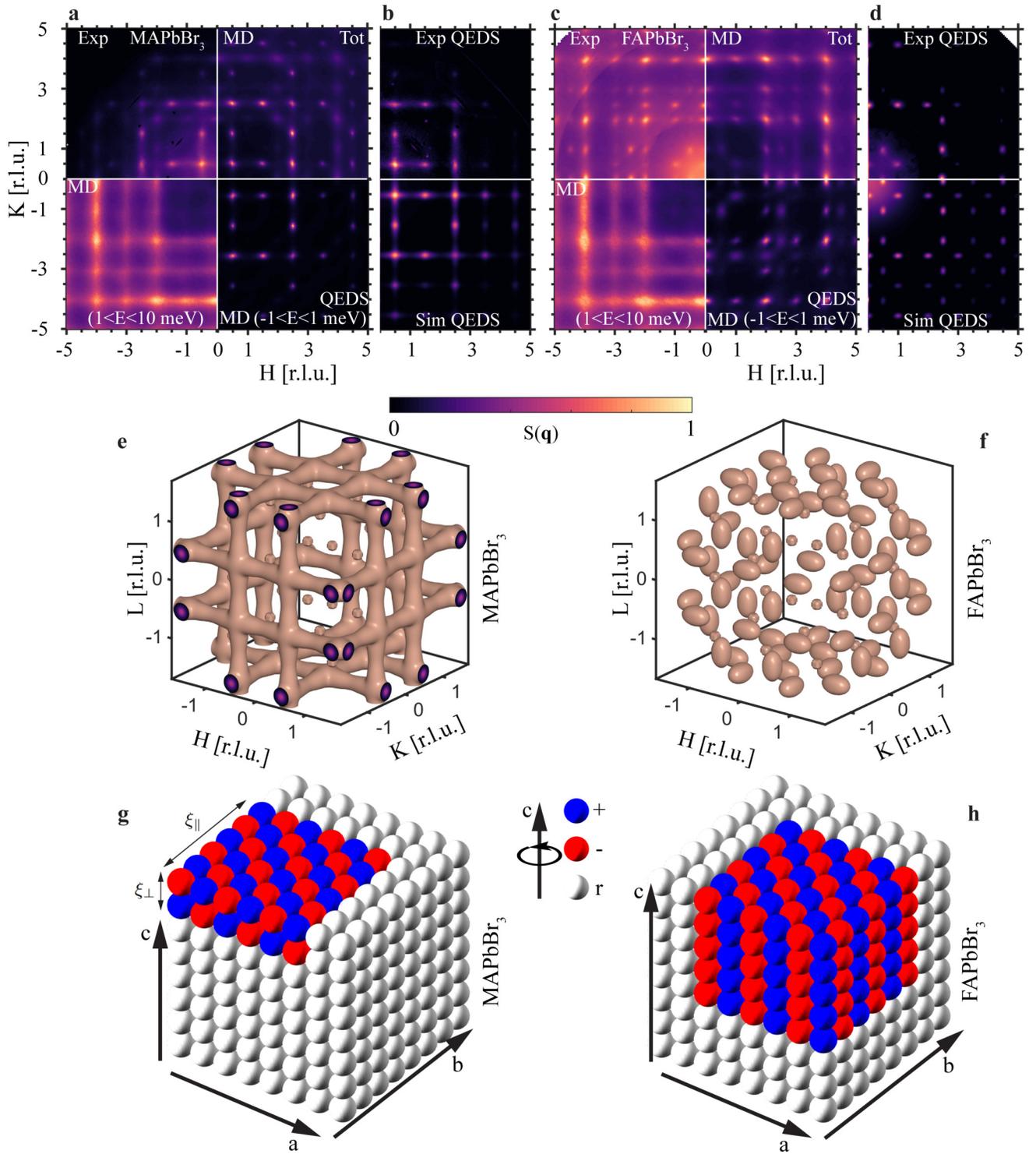

FIG. 1. **Experimental, MD and classically simulated X-ray diffuse scattering reveals the striking contrast in shape and distribution of nanodomains of locally tilted octahedra due to the A-site cation in lead halide perovskites.** (a)-(d) Experimental and simulated X-ray scattering function $S(\mathbf{q})$ at HK1.5 reciprocal space plane at 300 K, in average cubic $Pm\bar{3}m$ phase. (a) MAPbBr$_3$ and (c) FAPbBr$_3$ quadrants: top left/right- total $S(\mathbf{q})$ (energy transfer integrated) experimental/MD simulated diffuse scattering; bottom left- MD simulated $S(\mathbf{q}, E)$ integrated over $1 < E < 10$ meV; bottom right-MD simulated $S(\mathbf{q}, E)$ integrated over $-1 < E < 1$ meV. (b) MAPbBr$_3$ and (d) FAPbBr$_3$ two-quadrant panels: top-Experimentally derived QEDS, bottom-resultant QEDS obtained by fitting experimental data to classical local structure model. $S(\mathbf{q})$ intensities in each quadrant are normalised to 1. Each pixel in 2D images is coloured after applying bilinear interpolation on raw $S(\mathbf{q})$ using surf function in Matlab. Isosurface representation of 3D QEDS 300 K diffuse scattering in (e) MAPbBr$_3$, (f) FAPbBr$_3$, shows the clear difference between the two compositions. Local octahedra correlations at a single snapshot in time inferred from experimental diffuse scattering in MAPbBr$_3$ (g) and FAPbBr$_3$ (h). Each octahedron is represented by a sphere, where the colour represents the respective tilt angle along the $c$ crystallographic direction, with blue corresponding to positive, red to negative and white to a random close to zero tilt angle. Local octahedra tilts in MAPbBr$_3$ (g) form planar/pancake-like correlations, with octahedra correlated out-of-phase in $c$ direction, while in FAPbBr$_3$, correlations are more isotropic, forming spherical nanodomains of octahedra that are in-phase correlated along $c$ direction. For simplicity, the nanodomains are shown as cuboids instead of cylinders. The size of the nanodomains is defined with correlation lengths $\xi_\perp$ and $\xi_\parallel$ as depicted in (g).

Specifically, in MAPbBr$_3$ local nanodomains, characterised by out-of-phase tilted octahedra akin to tetragonal $I4/mcm$ symmetry and in FAPbBr$_3$ nanodomains with in-phase tilted octahedra akin to tetragonal $P4/mbm$ symmetry, exhibit short-range spatial correlations while in the long-range limit, the crystal symmetry remains cubic. Importantly, our analysis reveals a variation in the nanodomain shapes between these two lead bromides. By fitting experimental QEDS to our classical model we find that in MAPbBr$_3$, nanodomains resemble planar or pancake-like structures with measured in-plane diameters of $\xi_\parallel \approx 21$ Å and out-of-plane diameters of $\xi_\perp \approx 6$ Å (Fig. 1 **g**), in agreement with recent findings reported in [18]. $\xi_\perp \approx 6$ Å corresponds to roughly one unit cell meaning that on average two sheets of $c$-axis octahedral tilts are correlated along the $c$ direction. In contrast, nanodomains in FAPbBr$_3$ more closely resemble spheres, with in-plane and out-of-plane diameters of $\xi_\perp \approx 20$ Å and $\xi_\parallel \approx 30$ Å, respectively (Fig. 1 **h**). Further, our real-space examination of MD trajectories at 300 K for both compounds utilises spatial correlation functions to determine the size, shape, and local symmetry of the nanodomains, all of which significantly align with the conclusions drawn from fitting the experimental data to our classical model. The results from the real-space MD trajectory analysis are further detailed in the Supplementary Section vii. Additionally, our analysis of MD trajectories in real space enables us to estimate the volumetric density of dynamic local nanodomains (Supplementary Fig. S20), revealing a significantly higher density in MAPbBr$_3$, approximately double that observed in FAPbBr$_3$. This observation aligns with both experimental and MD simulated $S(\mathbf{q})$ data, which show that the intensity of TDS relative to QEDS signals in FAPbBr$_3$ is markedly higher than in MAPbBr$_3$. Further details on how TDS and QEDS relate to the density of dynamic nanodomains are given in Supplementary Section vi. Overall, these findings underscore the rich information contained in the diffuse X-ray scattering and simulation data, revealing the critical role of the organic A-site cation in influencing the local structure of halide perovskites.

### Local dynamic nanodomains are twinned

Utilising our classical model we show that local nanodomains exhibit twinning. The QEDS patterns in the cubic phases are modelled as the sum of three non-merohedrally twinned components of the local $I4/mcm$ (MAPbBr$_3$) and $P4/mbm$ (FAPbBr$_3$) structure nanodomains. These components utilise the same rotation operators that define the global twins emerging during the ferroelastic cubic-to-tetragonal phase transition in MAPbBr$_3$. We will demonstrate this in the example of MAPbBr$_3$. The local twinning rules are analogous in FAPbBr$_3$ and are shown in Supplementary Fig. S5. In Fig. 2 **a** we present the kernel twin-component $D_1$ of the local tetragonal $I4/mcm$ phase in MAPbBr$_3$ already depicted in Fig. 1 **g**. The resulting $S(\mathbf{q})$ isosurface of local structure component $D_1$ is shown in Fig. 2 **e**. The diffuse scattering rods align with the $L$ reciprocal space direction ($c$ in real space), due to short-range $c$-axis tilt correlations along the same direction. Similarly, after applying the relevant symmetry operations, we can derive $S(\mathbf{q})$ isosurfaces of $D_2$ and $D_3$ components, as shown in Fig. 2 **f** and **g**, respectively. The experimental QEDS is a sum of the three components (Fig. 2 **h**). Analogously, real space local structure is a sum of the real space components (Fig. 2 **d**). Thus, the local structure consists of mutually orthogonal (twinned) nanodomains of locally tilted octahedra embedded in the nearly cubic matrix as depicted in Fig. 2 **d** for MAPbBr$_3$ and in Supplementary Fig. S5 for FAPbBr$_3$. This configuration results in a diffuse scattering pattern that is effectively twinned, which is likely common across various halide perovskite compositions.

### Probing spatio-temporal local structure dynamics

We further utilise inelastic neutron spectroscopy, employing the cold and thermal triple axis spectrometers with distinct energy resolutions, Sika and Taipan, respectively, to probe dynamics of the local structure. Upon selecting only elastic scattering to probe QEDS, we detected broadened superstructure peaks at $R$ Brillouin zone points in MAPbBr$_3$ over a range of temperatures within the average cubic phase (Fig. 3 **a**). This corroborates our X-ray diffuse scattering results, confirming the existence of local structure discs featuring out-of-phase octahedral tilting with $I4/mcm$ symmetry. The correlation lengths are obtained as $1/HWHM$ of the fitted Lorentzian peaks (fitting procedure is outlined in Supplementary Section x), and correspond to the diameters of the in-plane octahedral correlations ($\xi_\parallel$) of the disc-like nanodomains (see Supplementary Fig. S7 **a**) with values presented in Fig. 3 **b**. While the two peaks in Fig. 3 **a** stem from the same local correlations, their corresponding correlation lengths differ. Thus, we employed a weighted average to derive a final value of $\xi_\parallel \approx 1.4$ nm at room temperature (in agreement with X-ray-derived correlation lengths presented in Supplementary Table III), with a tendency of $\xi_\parallel$ to increase as we approach the cubic-tetragonal phase transition at lower temperatures.

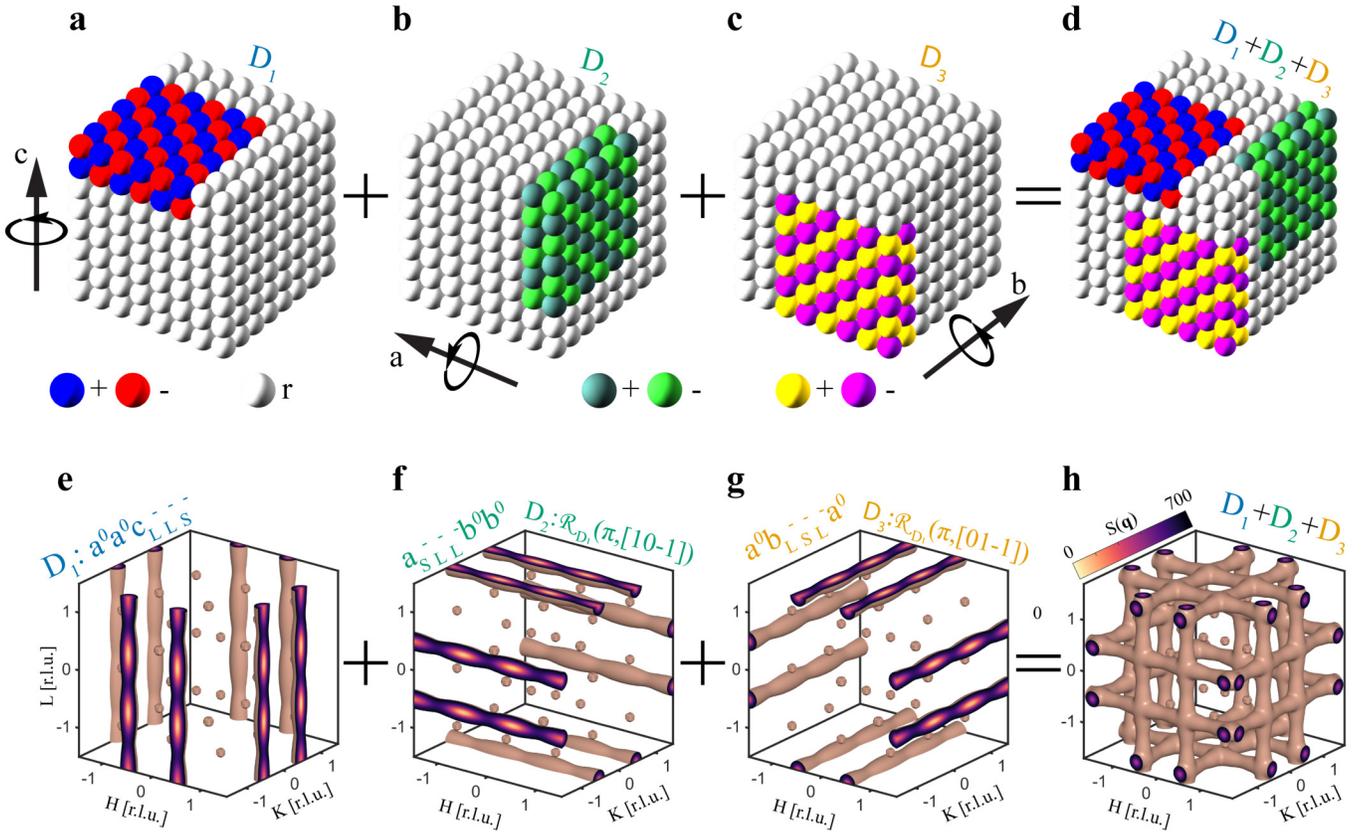

FIG. 2. **Classical X-ray diffuse scattering model reveals three types of locally tilted octahedral nanodomains in cubic MAPbBr₃.** (**a**) $D_1$ local nanodomain component in real space consists of correlated octahedral tilts, with short-range anti-phase ordering of $c$-axis tilts (one $I4/mcm$ $c$ axis unit) along $c$ and out-of-phase long-range ordering of $c$-axis tilts along $a$ and $b$, denoted as $a^0a^0c_{LLS}^{---}$ in modified Glazer notation (see Supplementary Section iii for details about the notation). Note that e.g. $c$-axis tilts along $a$ and $b$ are always out-of-phase correlated in perovskite lattice due to rigid octahedra connection rules. (**b**) $D_2$ nanodomain component in real space consists of octahedra which exhibit short-range anti-phase ordering of $a$-axis tilts along $a$ and long-range ordering of $a$-axis tilts along $b$ and $c$ denoted as $a_{SLL}^{---}b^0b^0$. (**b**) $D_3$ nanodomain component in real space consists of octahedra which exhibit short-range anti-phase ordering of $b$-axis tilts along $b$ and long-range ordering of $b$-axis tilts along $a$ and $c$ denoted as $a^0b_{LSL}^{---}a^0$. Each octahedron is represented by a sphere, where the colour represents the octahedron tilt angle along $c$ (**a**), $a$ (**b**) and $b$ (**c**) crystallographic directions. Colours represent positive/negative tilts while white corresponds to random tilt angle. (**e**-**g**) $S(\mathbf{q})$ QEDS isosurfaces computed from real space nanodomains $D_1$, $D_2$ and $D_3$, respectively. (**f**) The reciprocal space of the second local structure component ($D_2$) is derived by rotating $D_1$ in (**a**) 180° around the $[10\bar{1}]$ vector of the $D_1$ coordinate system, which is for this particular crystal symmetry, equivalent to 90° rotation around [010]. (**g**) The reciprocal space of the third local structure component ($D_3$) is achieved by rotating $D_1$ in (**a**) 180° around the $[01\bar{1}]$ vector of the $D_1$ coordinate system, which is for this particular crystal symmetry, equivalent to 90° rotation around [100]. (**h**) Experimentally observed $S(\mathbf{q})$ QEDS isosurfaces is the cumulative sum of all three components. (**d**) In real space, the local structure is characterised as the combination of three distinct local $I4/mcm$ planar nanodomains that are mutually orthogonal, i.e. twinned.

In FAPbBr₃ elastic scans were performed and a QEDS peak was observed at the M point as shown in Fig. 3 **c**. The appearance of these QEDS peaks at the M points corroborates the findings from our X-ray diffuse scattering investigations that identified dynamic, local nanodomains of in-phase octahedral tilting with $P4/mbm$ symmetry within globally cubic FAPbBr₃. Contrasting with MAPbBr₃, the local structure in FAPbBr₃ demonstrates longer correlation lengths, approximately $\xi_\parallel \approx 4$ nm at room temperature with a tendency to increase as phase transition is approached (Fig. 3 **d**). We repeated the elastic scans in MAPbBr₃ and FAPbBr₃ using Sika, with an order of magnitude better energy resolution which confirmed the presence of the same QEDS signals (Supplementary Fig. S30).

Moreover, to experimentally determine the lifetimes of the dynamic nanodomains, we also performed quasi-elastic energy-resolved scans at the same R and M points, which were resolution-limited in energy, deriving the conclusion that the dynamic nanodomain lifetimes are greater or equal to approximately 7 ps (Supplementary Fig. S32). Because of strong incoherent quasi-elastic scattering from hydrogen atoms present in the organic cations across both compositions, we utilised MD-derived $S(\mathbf{q},E)$ to estimate the lifetimes. In Fig. 3 **e** and **g** we show the $S(\mathbf{q},E)$ along the R-M directions in the reciprocal space, in MAPbBr$_3$ and FAPbBr$_3$, respectively. The diffuse scattering rods along the R–M direction in both compositions originate from low energy, quasi-elastic scattering, as we previously corroborated. By fitting vertical cross-sections of $S(\mathbf{q},E)$ in Fig. 3 **e** and **g** at every $\mathbf{q}$ to Lorentzian functions we calculate the lifetimes as $1/FWHM$, presented in Fig. 3 **f** and **h**. Across both compositions, lifetimes exhibit modulation along the R-M line, with MAPbBr$_3$ showing sharp peaks in $\mathbf{q}$ with a maximum of 4 ps at the R point, and FAPbBr$_3$ presenting broader peaks with a maximum of 1.5 ps at the M points. These are the lifetimes of pure out-of-phase (in-phase) spatially correlated octahedrally tilted nanodomains in MAPbBr$_3$ (FAPbBr$_3$) which we associate with diffuse scattering peaks in $\mathbf{q}$ space at R (M) points. These modes both exhibit the largest spatial and longest temporal correlations. While correlations that are not pure R or M modes are also possible, they are both weakly temporally (as seen in Fig. 3 **f** and **h**) and spatially correlated as we recently outlined in [32]. Thus, all other excitations along the R–M line can be intuitively understood as a thermal bath of background short-lived overdamped phonons. Overall, our results are consistent with those obtained from X-ray diffuse scattering, affirming the presence of dynamic (within the sub-10 ps range) out-of-phase and in-phase tilted local octahedral nanodomains in the average cubic phase of these materials. Investigating these temporal dynamics requires instruments with high energy resolution, as resolution-limited quasi-elastic lines could erroneously suggest static nanodomains.

**Local structure in the low-temperature phases and its impact on ferroelastic twins**

Upon the first low-temperature phase transition of MAPbBr$_3$ and FAPbBr$_3$, distinct superstructure Bragg peaks emerge at the $\mathbf{q}$ coordinates of the diffuse scattering peaks previously present in the room temperature phases, as shown in Fig. 3 **a** and **b** and Supplementary Fig. S7. Notable differences are observed between MAPbBr$_3$ and FAPbBr$_3$, as depicted in Fig. 4 **a** and **b**, respectively. In MAPbBr$_3$, QEDS indicates the presence of local structure in the lower symmetry phase at $T = 200$ K, whereas the absence of strong QEDS in FAPbBr$_3$ indicates that there are no additional local octahedral tilted nanodomains in the lower symmetry phase distinct from what is observed on average.

In MAPbBr$_3$ we simulate QEDS with classical model (Fig. 4 **a**) to reveal locally tilted octahedra that exhibit a tilting pattern like in the average MAPbBr$_3$ orthorhombic ($Pnma$) phase as visualised in Supplementary Fig. S6. Thus, local orthorhombic nanodomains of $Pnma$ symmetry are embedded within the globally tetragonal $I4/mcm$ phase.

The observed gradual increase in the correlation lengths (Fig. 3 **b**) and lifetimes (Supplementary Fig. S24) of locally tilted nanodomains within the cubic phase of MAPbBr$_3$, upon approaching the average tetragonal phase, indicates that the phase transition is mediated by the progressive growth and slowing down of dynamic nanodomains. These nanodomains ultimately evolve into macroscopic static twins at the phase transition point, implying that the ferroelastic twins in the tetragonal phase of MAPbBr$_3$ are seeded from these dynamic nanodomains. This is supported by the fact that twinning laws in the average tetragonal phase mirror those of the dynamic nanodomains in the average cubic phase.

The formation of macroscopic twins in MAPbBr$_3$ is observed in optical microscope images of the cleaved single-crystal surface. At 250 K, where MAPbBr$_3$ is in the globally cubic phase (Fig. 4 **b**), only dynamic local nanodomains are present, with no observable macroscopic twins. However, at 200 K in the average tetragonal phase (Fig. 4 **e**), the emergence of macroscopic ferroelastic twins is clearly imaged. The twin boundaries are graphically represented in Fig. 4 **c**, aligning with previous findings in halide perovskites [35, 36]. Moreover, Fig. 4 **f** illustrates that additional twin types form during the transition from the tetragonal to the orthorhombic phase. The crystallographic relationship between nanodomains of local orthorhombic structure in the average tetragonal phase, as detailed in Supplementary Table II, remains consistent with the global twinning laws observed during the tetragonal-orthorhombic phase transition, also evident in reciprocal space (Supplementary Fig. S3). These collective observations further reinforce the notion that in MAPbBr$_3$, the emergence of global twinning and ferroelastic phase transition is a direct consequence of the local structure, with the macroscopic twinning governed by the symmetry relationships among the local nanodomains.

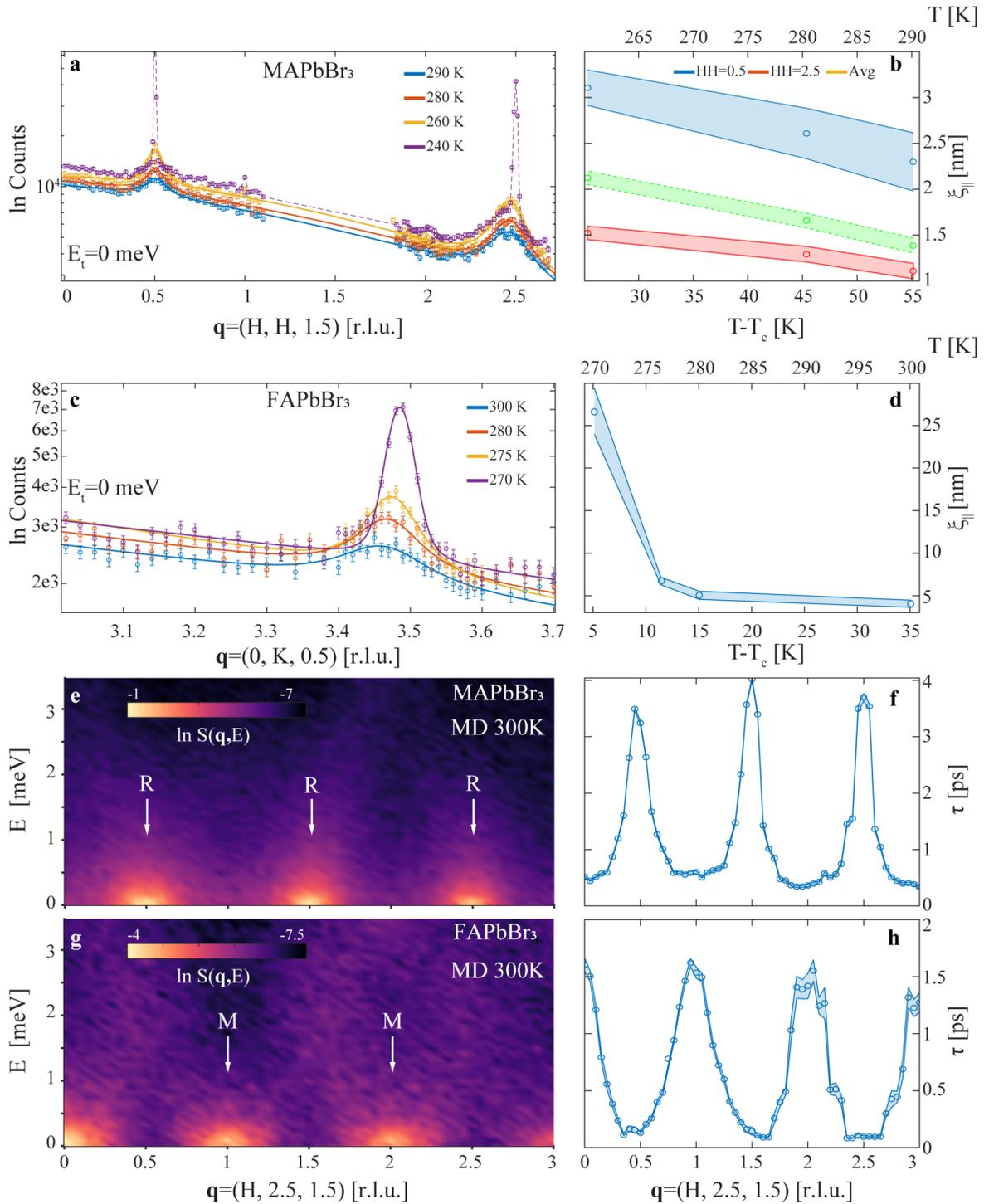

FIG. 3. **Elastic neutron scattering and MD simulations confirm the presence of quasi-elastic dynamic local octahedral nanodomains within cubic MAPbBr$_3$ and FAPbBr$_3$.** (**a**) Elastic scans (zero neutron energy transfer, $E_t = 0$) across the [H H 1.5] direction in the reciprocal space of MAPbBr$_3$ for various temperatures in the average cubic phase. Experimental data with measurement uncertainty represented with error bars and the corresponding fits given in full lines. (**b**) The correlation lengths obtained by fitting two diffuse scattering peaks at R points, **q** = (0.5, 0.5, 1.5) and **q** = (2.5, 2.5, 1.5), shown in (**a**). Their weighted average is depicted by the green curve. (**c**) Elastic scans across the [0 K 0.5] direction in the reciprocal space of FAPbBr$_3$ for various temperatures in the average cubic phase. Experimental data with measurement uncertainty represented with error bars and the corresponding fits given in full lines. (**d**) The derived correlation lengths for the observed diffuse scattering at M point, **q** = (0, 3.5, 0.5). In (**b**) and (**d**) the shaded area around the data points represents 95% confidence intervals of the performed fits while $T_c$ denotes the phase transition temperature. The data was collected with a thermal triple-axis spectrometer, Taipan. (**e**) MD computed $S(\mathbf{q}, E)$ for MAPbBr$_3$ at 300 K for **q** = (H, 2.5, 1.5) which corresponds to a horizontal cut of HK1.5 plane for K = 2.5. **q** with integer/non-integer $H$ values correspond to M/R points in the Brillouin zone. Each vertical section of (**e**) manifests as a quasi-elastic Lorentzian, centred at zero energy transfer. In (**f**), the lifetimes associated with each quasi-elastic Lorentzian line are derived along **q**. The shaded area around the data points represents 95% confidence intervals of the performed fits. (**g**)-(**h**) Analogous plots are shown for FAPbBr$_3$ MD computed $S(\mathbf{q}, E)$ at 300 K.

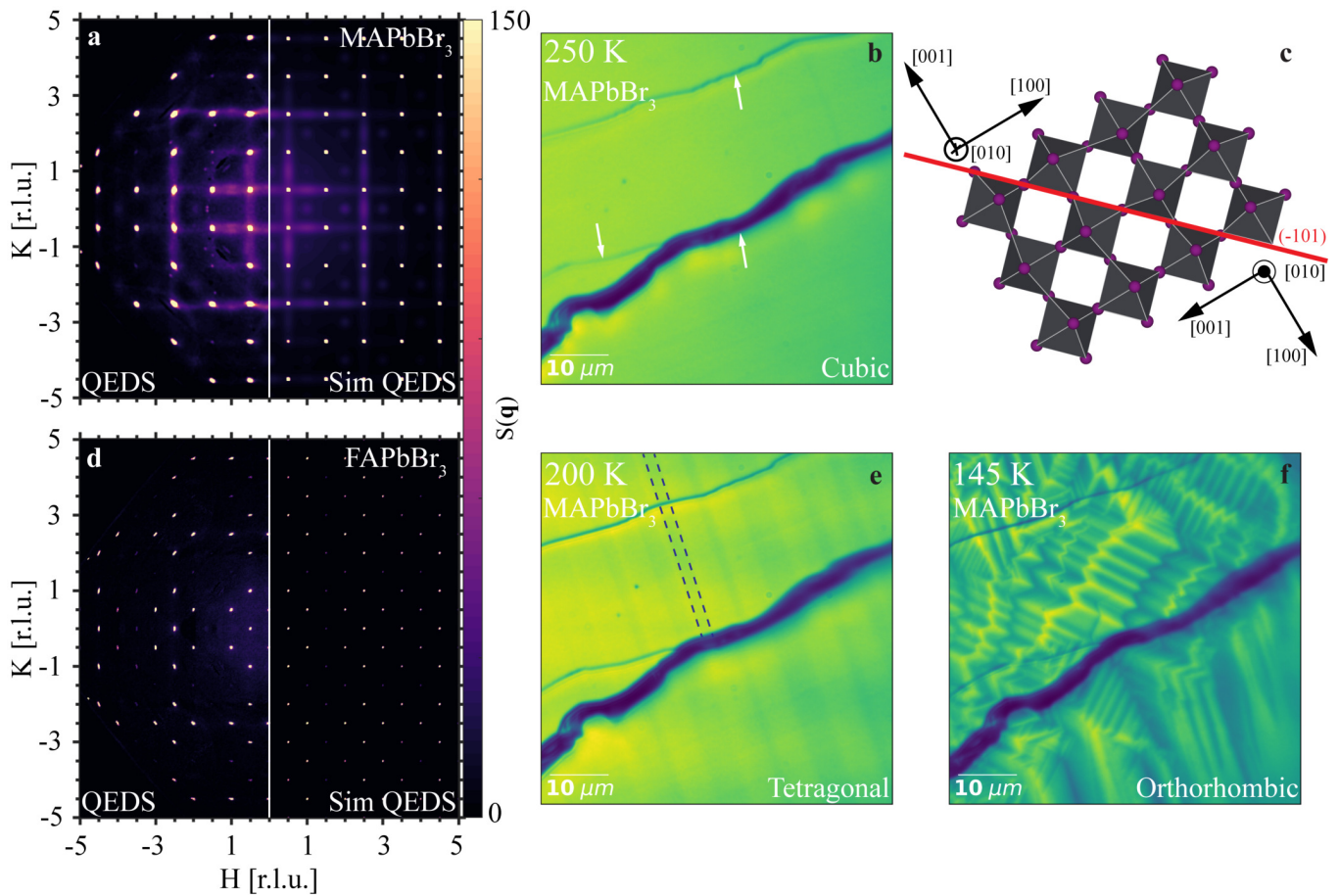

FIG. 4. **Local structure in the low-temperature phases seeds the macroscopic ferroelastic twins.** (**a**) 2D X-ray $S(\mathbf{q})$ intensity distribution across HK1.5 reciprocal space plane of MAPbBr$_3$ at $T = 200$ K. Experimental QEDS pattern (left panel) and simulated using the classical model (right panel). Optical microscope image of the cleaved surface of MAPbBr$_3$ in the average (**b**) cubic, (**e**) tetragonal, (**f**) orthorhombic phase. In (**b**) white arrows show crystal cracks and in (**e**), the boundaries of one twin domain are emphasised with dashed blue lines. (**c**) Graphical representation of the two twin components in the tetragonal MAPbBr$_3$ where axes of the two twin components are labelled in the pseudocubic notation, with a clear twin boundary, which we image in (**e**). (**d**) Experimental (left panel) and simulated (right panel) 2D $S(\mathbf{q})$ intensity distribution across HK1.5 reciprocal space plane of FAPbBr$_3$ at $T = 200$ K, confirming the absence of strong diffuse scattering and no local structure in an average cubic $Im\bar{3}$ symmetry.

We find that FAPbBr$_3$ transitions from cubic $Pm\bar{3}m$ to cubic $Im\bar{3}$ (see further discussion on the assignment of space group in Supplementary Section viii), which is distinct from the commonly suggested $P4/mbm$ space group found in the literature. Under optical microscopy (Supplementary Fig. S33) and through crystallographic analysis (Supplementary Fig. S26), macroscopic twinning is not detected. This observation is consistent with the lack of reported ferroelasticity in FAPbBr$_3$ [37]. Given that $Im\bar{3}$ forms when each of the three angular components of the octahedra have non-zero and identical tilting angles, it suggests that the suppression of diffuse scattering might result from the disruption of local octahedral correlations when all three components of octahedral angles are locked and exhibit strong infinite correlations in all three directions ($a^+a^+a^+$), as defined by $Im\bar{3}$ space group.

### Conclusions

We find that high-symmetry cubic halide perovskite structures are composed of spatially and temporally correlated, dynamically tilted octahedra, resembling local nanodomains of lower symmetry phases. In methylammonium-based perovskites, we observed densely packed plane-like tetragonal nanodomains of out-of-phase correlated octahedral tilts in the average cubic phase and plane-like orthorhombic nanodomains in the average tetragonal phase. By contrast, for formamidinium-based systems, we identified the presence of local tetragonal spherical nanodomains of in-phase tilted octahedra sparsely packed in the average cubic phase, while the local nanodomains were not present in the next lower symmetry average phase ($Im\bar{3}$, $a^+a^+a^+$ tilting pattern in Glazer notation). In both systems, the local structure possesses a dynamic nature with sub-10 ps lifetimes and no static components are present, which aligns with recent predictions [31, 32].

In MAPbBr$_3$, the dynamic nanodomains nucleate to grow into macroscopic ferroelastic twins, both at cubic-tetragonal and tetragonal-orthorhombic transition. Conversely, in FAPbBr$_3$, although these dynamic nanodomains grow as the phase transition nears, they do not nucleate into macroscopic twins but rather into a singular crystal of the $Im\bar{3}$ cubic phase, in line with the observed absence of ferroelastic phase transition. Thus, we establish a connection between emerging macroscopic properties in these materials, such as ferroelasticity, and the onset of dynamic local nanodomains, revealing the fundamental microscopic mechanism behind the ferroelastic phase transition.

We propose several reasons why the properties of the local structure are strongly dependent on the A-site cation. The tolerance factors of FAPbBr$_3$ are closer to unity when compared to those of MAPbBr$_3$ [38], indicating a tendency of FAPbBr$_3$ to form structures closer to ideal cubic symmetry. This aligns with our observations of less pronounced and more isotropic dynamic disorder in FAPbBr$_3$. Furthermore, the choice of A-site cation significantly influences both short-range physical (with FA being larger than MA) and chemical (FA exhibiting stronger hydrogen bonding with inorganic sublattice) interactions in these compounds. These factors collectively influence the local dynamic strains instigated by the A-site cation [39, 40], thereby affecting the shape, density and symmetry of the fluctuating local nanodomains.

We thus propose an explanation for the notably superior optoelectronic performance observed in FA based over MA based lead halide perovskites. Although both materials exhibit cubic symmetry on average during room temperature applications, our findings suggest that differences in optoelectronic performance arise from their contrasting local structural properties. The dynamic local octahedral tilting leads to electronic and local band gap fluctuations that directly impact charge carrier dynamics [21, 41]. The observed lower degree of dynamic disorder in FAPbBr$_3$, compared to MAPbBr$_3$, may be linked to recently noted significantly lower intrinsic defect densities in FAPbBr$_3$ [42]. This, in turn, results in longer-lived charge carriers and higher potential for efficient charge carrier collection in FAPbBr$_3$ devices, such as that observed in FAPbBr$_3$ based X-ray detectors compared to MAPbBr$_3$. Furthermore, these significant distinctions in local structure may also provide a convincing rationale for the superior efficiency of FA-based solar cell materials over their MA-based counterparts. Our findings open up exciting new research avenues and are likely to have significant implications for the future design and development of perovskite-based photovoltaics, optoelectronics and X-ray imaging, as well as new applications in which local order is controlled to dictate device properties.

## METHODS

### Synthesis of perovskite single crystals

MAPbBr$_3$: A mixture of 1M of PbBr$_2$ and 1M MABr was dissolved in 1 mL DMF. To ensure complete dissolution, the solution was stirred vigorously at 25 °C for 6 h and then filtered with a 0.45 μm filter head before use. After adding a small MAPbBr$_3$ crystal to the filtered solution, the solution was transferred to an oven. To make the crystals larger, the solution was further heated to 85 °C with a rate of 10 °C per 30 min, and the crystal reached its full size after 24 hours. The obtained crystals were separated and dried to obtain MAPbBr$_3$.

FAPbBr$_3$:A mixture of 1M of PbBr$_2$ and 1 M FABr was dissolved in 1 mL DMF/GBL (1:1). To ensure complete dissolution, the solution was stirred vigorously at 25 °C for 6 h and then filtered with a 0.45 μm filter head before use. After adding a small FAPbBr$_3$ crystal to the filtered solution, the solution was transferred to an oven. To make the crystals larger, the solution was further heated to 55° C with a rate of 10 °C per 30 min, and the crystal reached its full size after 24 hours. The obtained crystals were separated and dried to obtain FAPbBr$_3$.

FAPbBr$_3$ grown at Colorado State University: CH(NH$_2$)$_2$CH$_3$COO and HBr were obtained from Sigma Aldrich Corporation. PbBr$_2$ and other solvents were procured from VWR and used without further purification. In a typical preparation, approximately 0.4 g of CH(NH$_2$)$_2$CH$_3$COO were dissolved in 8 mL of hydrobromic acid (47% v/v) at 80 °C for 15 minutes. Subsequently, 1.1 g of PbBr$_2$ (1.25:1.0 mole ratio CH(NH$_2$)$_2$CH$_3$COO:PbBr$_2$) were added, and the solution was stirred until all the powder had dissolved. CH(NH$_2$)$_2$PbBr$_3$ was precipitated using ethanol as an antisolvent, and the powder was washed with ethanol. Single crystals were grown using an anti-solvent method. A small vial containing 0.5 mL of a filtered 1 M solution of $CH(NH_2)_2PbBr_3$ in a 1:1 mixture of dimethylformamide and $\gamma$-butyrolactone by volume was placed in a sealed, larger vial containing approximately 5 mL of ethanol. Crystal growth reactions were carried out over three days.

### Single crystal X-ray diffuse scattering measurements

Single crystal perovskite samples were selected under a polarizing microscope (Leica M165Z) and picked up on a MicroMount (MiTeGen, USA) consisting of a thin polymer tip with a wicking aperture. Single crystal diffraction measurements on MAPbBr$_3$ at 300 K, 200 K, and 100 K were carried out on a Bruker D8 Quest Single Crystal diffractometer at different temperatures using $I\mu S$ Incoatec Microfocus source with $Mo-K\alpha$ radiation ($\lambda = 0.710\,723$ Å). The single crystals, mounted on the goniometer using a cryo loop for intensity measurements, were coated with immersion oil type NVH and then quickly transferred to the nitrogen stream generated by an Oxford Cryostream 700 series. Symmetry-related absorption corrections using the program SADABS were applied and the data were corrected for Lorentz and polarisation effects using Bruker APEX3 software. Precession images were generated in CrysAlisPro software. The average structure was solved by program SHELXT (with intrinsic phasing) [43] and the full-matrix least-square refinements were carried out using SHELXL-2014 [44] through the Olex2 [45] software suite.

X-ray diffuse scattering measurements on MAPbBr$_3$ and FAPbBr$_3$ at 300 K and 200 K were carried out on MX1 beamline at Australian Synchrotron using X-rays of 12.9 keV with X-ray flux of $36 \times 10^{11}\,s^{-1}$ incident on an area of 120 µm x 120 µm. The crystal dimensions measured for MAPbBr$_3$ were 200 µm x 70 µm x 50 µm, while those for FAPbBr$_3$ were 100 µm x 70 µm x 70 µm. The single crystals, mounted on the goniometer using a cryo loop for intensity measurements, were coated with immersion oil type NVH and then quickly transferred to the nitrogen stream generated by an Oxford Cryostream 800 series. CrysAlisPro [46] was used for indexing, determination and refinement of the orientation matrix. In the process of data analysis, precession images were first unwrapped using CrysAlisPro. Detailed examination of the diffraction patterns revealed that the diffuse scattering adhered to Laue symmetry. Accordingly, Laue symmetry averaging was then applied to the data, also using CrysAlisPro.

X-ray diffuse scattering measurements on MAPbBr$_3$ and FAPbBr$_3$ at 300 K and 200 K were carried out at I19-1 beamline at Diamond Light Source using X-rays of 18 keV with X-ray flux of $1.542 \times 10^{13}\,s^{-1}$ incident on an area of 100 µm x 100 µm. The single crystals, mounted on the goniometer using a cryo loop for intensity measurements, were coated with immersion oil type NVH and then quickly transferred to the nitrogen stream generated by an Oxford Cryostream 800 series. In the process of data analysis, precession images were unwrapped using CrysAlisPro. To enhance the signal-to-noise ratio while minimizing the X-ray dose received by the samples, we repeated the measurement between 10 to 25 times. Subsequently, we summed up these frames in post-processing. This approach allowed us to extend the dynamic range of the measurement, enabling us to capture both Bragg peaks without saturating the detector and to detect weak diffuse scattering.

X-ray diffuse scattering measurements on FAPbBr$_3$ at 300 K were carried out on a Rigaku Synergy S diffractometer fitted with a Dectris EIGER2 R 1M detector. All data sets were collected under copper radiation ($\lambda$= 1.5406 Å). Crystals were mounted on a 0.2 mm diameter MiTeGen loop using Paratone-N oil as a cryoprotectant. An exposure time of 90 s ($\theta = 40°$) or 120 s ($\theta = 80°$) was needed to detect diffuse features at room temperature (300 K). Each measurement involved a full $\phi$-scan carried out in a single run. CrysAlisPro was used for indexing, determination and refinement of the orientation matrix. For the diffuse scattering analysis the scattering data were reconstructed on a three-dimensional grid defined by $-10 \le h, k, l \le +20$ with voxel sizes of $\Delta h = \Delta k = \Delta l = 0.05$ r.l.u., resulting in an array of 401 × 401 × 401 voxels. For this purpose, the crystal orientation as refined with CrysAlisPro was converted using a customized code to serve as input for the Meerkat program [47]. After careful inspection of the diffraction data it was observed that the diffuse scattering also follows Laue symmetry. The data were subsequently averaged for Laue symmetry using Meerkat.

### Molecular Dynamics Simulations

To perform large-scale molecular dynamics (MD) simulations MAPbBr$_3$ and FAPbBr$_3$ Allegro [48, 49] machine learning force fields (MLFFs) were trained. The training, validation and test sets were constructed based on density functional theory (DFT) using an on-the-fly structure selection process implemented in VASP [50, 51], where a Gaussian approximation potential (GAP)-style potential is fit on-the-fly, and training structures are picked from the MD simulations based on a Bayesian error prediction [52]. Structure selection runs were performed using NpT ensemble MD for each of the materials using $2 \times 2 \times 2$ pseudo-cubic supercells at 6 separate temperatures, 100, 160, 210, 270, 350 and 450 K. The r$^2$SCAN exchange-correlation functional [53], a plane wave basis set cut-off energy of 500 eV, and a $2 \times 2 \times 2$ $\Gamma$-centred $k$-point grid was adopted. The energy threshold for electronic convergence was set to $10^{-5}$ eV, and a Gaussian smearing with a width of 50 meV was applied for the smearing of electronic band occupancy. To avoid issues related to the incomplete basis set when large volume changes occur during the on-the-fly MD run, an additional single-point DFT calculation was performed on all structures, and these recalculated forces, energies and stresses made up the final training, validation and test set. The full sets consisted of 2985 and 2640 structures for MAPbBr$_3$ and FAPbBr$_3$, respectively.

Separate Allegro MLFFs were trained for MAPBBr$_3$ and FAPbBr$_3$, using radial cutoffs of 6.5 Å, 2 layers and 32 tensor features with full O(3) symmetry and $l_{max} = 2$, a two-body latent multi-layer perceptron (MLP) with dimensions [64, 128, 256, 512] and later latent MLP of dimension [512], both with SiLU non-linearities and a single-layer final edge-energy MLP with dimension 128 and no linearity. Atomic distances were embedded using trainable Bessel-functions. The models used the efficient mixed precision scheme described in Ref. [49].

The training and validation sets contained 2388 and 299, and 2112 and 264 structures for MAPbBr$_3$ and FAPbBr$_3$, respectively and were shuffled after each epoch. The training was performed with the Adam optimizer in pytorch [54] for 1858 and 2141 epochs for MAPbBr$_3$ and FAPbBr$_3$, respectively, using a batch size of 5 and a learning rate of 0.001. A loss function with a 1:1:1 weighing of the Allegro per atom mean squared energy, force and stress terms, respectively, was used. The trained models achieved root mean squared errors (RMSE) on energies, force components and stress tensor components of 0.2 meV/atom, 10 meV/Å, and 0.5 kbar and 0.2 meV/atom, 7 meV/Å and 0.3 kbar on hold out test sets containing 298 and 264 structures for MAPbBr$_3$ and FAPbBr$_3$ respectively. Parity and error distribution plots are provided in Supplementary Section xii.

The Allegro MD simulations were performed in the Large-scale Atomic/Molecular Massively Parallel Simulator (LAMMPS) package [55], using the pair_allegro patch [56]. Large simulation cells constructed as $20 \times 20 \times 20$ pseudo-cubic unit cells were used for both materials and a 0.5 fs timestep was used to integrate the classical equations of motion. For each material, two initial configurations were constructed, one with randomly oriented FA/MA molecules and one with perfectly aligned molecules. These initial configurations were equilibrated using fixed-shape NPT dynamics at 400 K for 200 ps, except for FAPbBr$_3$ with aligned molecules where 150 ps equilibration time was used. Then, for a specific temperature of interest, a further 50 ps of equilibration was performed before running 0.5 ns of NVE dynamics, resulting in 1 ns of production NVE trajectory data for each material at each temperature. Results were cross-checked between the runs with different initial configurations and no qualitative changes were observed, indicating that the structures had been sufficiently equilibrated with respect to the molecular orientations. Based on these MD trajectories, real-space structural dynamics analysis was performed with the PDynA package [31].

To connect to the single crystal X-ray diffuse scattering measurements, the dynamical structure factor $S(\mathbf{q}, E)$ was calculated from the MD trajectories using the pynamic structure factor (psf) package [57]. For each material and initial configuration, the 0.5 ns trajectories were divided into 10 blocks of 50 ps. $S(\mathbf{q}, E)$ was then averaged over these blocks and (in fractional reciprocal lattice units) over the two initial configurations. We extracted $S(\mathbf{q}, E)$ for $0 \leq$ H,K,L $\leq 5$ and expanded these values to the whole plane by mirroring in the coordinate axes. $q$-dependent atomic form factors, approximated as a sum of Gaussians, were used as described in the SI of [18] and implemented in psf [57].

### Inelastic Neutron scattering measurements

Constant energy, variable $\mathbf{q}$ (momentum) scans were performed with the thermal triple axis spectrometer, Taipan at Australian Nuclear Science and Technology Organisation (ANSTO), Sydney, Australia. Taipan was aligned with o-40'-40'-o collimation, in a configuration where incident neutron energy was varied with fixed final scattered neutron energy of 14.87 meV. A graphite filter was used on the scattered side to remove higher order scattering. The two large single crystals (about 1 cm$^3$ in volume), MAPbBr$_3$ and FAPbBr$_3$ were aligned in HHL and HK0 planes, respectively.

The cold triple axis spectrometer, SIKA at ANSTO, was aligned with o-60'-60'-60' collimation using a large double-focusing pyrolytic graphite (PG) monochromator and analyser which were oriented with a fixed final energy of 5 meV. This allowed an energy resolution of approximately FWHM of 0.09 meV. The cooled Be filter was used to remove unwanted higher order reflections. Samples were wrapped in thin Teflon tape to prevent perovskite crystal surface from reacting with Al, and then mounted on an Al plate before being inserted into the cryostat. Counting times were approximately 5 minutes per point as the incident flux of 1.5 M neutrons.

### Low temperature optical microscopy measurements

Samples were measured in an Oxford HiRes2 Microstat cooled with liquid helium and kept under vacuum during the measurement. The emergence of ferroelastic nanodomains was observed using the Photon Etc. IMA microscopy system. 20× (Nikon TU Plan Fluor, 0.45 NA) and 100× long working distance objective (Nikon TU Plan Fluor, 0.8 NA) with appropriate chromatic aberration corrections were used for all measurements. Single crystals were cleaved immediately prior to measurements to obtain a fresh and uniform surface and then mounted in open cycle liquid helium cryostat. The samples were cooled down/heated up with a maximum 1°/min rate to avoid any unexpected thermal stress effects. The images were taken with a high sensitivity CCD camera (Hamamatsu ORCA Flash 4.0 V3 sCMOS camera) with $2048 \times 2048$ $6.5 \times 6.5 \, \mu m^2$ pixels that is thermoelectrically cooled to $-10°C$.

### DSC measurements

Differential scanning calorimetry measurements were performed on a Netzsch DSC Proteus 204 F1 in an Al crucible under a $N_2$ atmosphere. The samples were cooled and heated in the temperature range of $115 - 330$ K with a scan rate of 10 K/min. The masses of the $MAPbBr_3$ and $FAPbBr_3$ samples were 19.9 mg and 20.3 mg, respectively.

### AUTHOR CONTRIBUTIONS

### CONFLICTS OF INTEREST

### DATA AVAILABILITY

The data that support the findings of this study is available to download at the University of Cambridge Apollo Repository [DOI to be added at acceptance].

### CODE AVAILABILITY

The Python and Matlab code that support the findings of this study are publicly available at GitHub [link to be added at acceptance].

### ACKNOWLEDGMENTS

We thank Diamond Light Source for providing beamtime at the I19-1 (proposal CY33123). The authors would like to acknowledge the ANSTO beam time received on Sika and Taipan through proposal numbers P14150 and DB9600. The authors thank the staff from the Mark Wainwright Analytical Centre at UNSW Sydney for the X-ray and DSC measurements. Via our membership of the UK's HEC Materials Chemistry Consortium, which is funded by EPSRC (EP/X035859/1), this work also used the ARCHER2 UK National Supercomputing Service (http://www.archer2.ac.uk). The training of the machine-learned force fields were enabled by the Berzelius resource provided by the Knut and Alice Wallenberg Foundation at the National Supercomputer Centre. We acknowledge the National Academic Infrastructure for Supercomputing in Sweden (NAISS) partially funded by the Swedish Research Council through grant agreement no. 2022-06725 for awarding this project access to the LUMI supercomputer, owned by the EuroHPC Joint Undertaking, hosted by CSC (Finland) and the LUMI consortium. This work was supported by the National Natural Science Foundation of China (Grant No. 51572037, 91648109, 51335002, 51702024), the Natural Science Foundation of Jiangsu Province (Grant No. BK20200981), Changzhou Sci Tech Program (Grant No. CJ20190050). This work was supported by the Australian Research Council Discovery Project (ARC DP) (DP190101973). M.D. acknowledges UKRI guarantee funding for Marie Skłodowska-Curie Actions Postdoctoral Fellowships 2022 (EP/Y024648/1) and support by AINSE Limited through a PGRA award. M.D. acknowledges helpful discussions with Ekhard Salje, Richard Mole and Tiarnan Doherty. M.P.N. recognises the support of the UNSW Scientia Program and the ARC Centre of Excellence in Exciton Science (Grant No. CE17100026). J.R.N, A.M., and N.R. acknowledge A.L. Goodwin and an ERC Grant (Advanced Grant 788144) for support. J.R.N. acknowledges the Leverhulme Trust for granting a Visiting Professorship at the Inorganic Chemistry Laboratory, University of Oxford. The work at Colorado State University was supported by grant DE-SC0023316 funded by the U.S. Department of Energy, Office of Science. T.A.S. acknowledges funding from EPSRC Cambridge NanoDTC, (EP/S022953/1). S.K. is grateful for an Early Career Fellowship supported by the Leverhulme Trust (ECF-2022-593) and the Isaac Newton Trust (22.08(i)). K.W.P.O. acknowledges an EPSRC studentship (project reference: 2275833). J. K. acknowledges support from the Swedish Research Council (VR) program 2021-00486. The authors thank the Leverhulme Trust (RPG-2021-191) for funding.

# Supplementary Information: Dynamic Nanodomains Dictate Macroscopic Properties in Lead Halide Perovskites

Milos Dubajic,[1, 2, *] James R. Neilson,[3, 4, 5, *] Johan Klarbring,[6, 7] Xia Liang,[6] Stephanie A. Boer,[8] Kirrily C. Rule,[9] Josie E. Auckett,[8] Leilei Gu,[10] Xuguang Jia,[11, 12] Andreas Pusch,[2] Ganbaatar Tumen-Ulzii,[1] Qiyuan Wu,[2] Thomas A. Selby,[1] Yang Lu,[1] Julia C. Trowbridge,[3] Eve M. Mozur,[3] Arianna Minelli,[5] Nikolaj Roth,[5] Kieran W. P. Orr,[1, 13] Arman Mahboubi Soufiani,[2, 14] Simon Kahmann,[1] Irina Kabakova,[15] Jianning Ding,[11, 16] Tom Wu,[17, 18] Gavin J. Conibeer,[2] Stephen P. Bremner,[2, *] Aron Walsh,[6, *] Michael P. Nielsen,[2, *] and Samuel D. Stranks[1, 13, *]

[1] Department of Chemical Engineering and Biotechnology,
University of Cambridge, Philippa Fawcett Drive, Cambridge, CB3 0AS, UK
[2] School of Photovoltaic & Renewable Engineering, UNSW Sydney, Kensington 2052, Australia
[3] Department of Chemistry, Colorado State University,
Fort Collins, Colorado 80523-1872, United States of America
[4] School of Materials Science & Engineering, Colorado State University,
Fort Collins, Colorado 80523-1872, United States of America
[5] Inorganic Chemistry Laboratory, University of Oxford, Oxford, UK
[6] Department of Materials, Imperial College London, London SW7 2AZ, United Kingdom
[7] Department of Physics, Chemistry and Biology (IFM),
Linköping University, SE-581 83, Linköping, Sweden
[8] Australian Synchrotron, ANSTO, 800 Blackburn Road, Clayton, VIC 3168, Australia
[9] Australian Nuclear Science and Technology Organisation,
Locked Bag 2001, Kirrawee, DC NSW 2232, Australia
[10] Taizhou Institute of Science and Technology, Nanjing University of
Science and Technology, Taizhou 225300, Jiangsu Province, China
[11] School of Microelectronics and Control Engineering,
Jiangsu Province Cultivation Base for State Key Laboratory of Photovoltaic Science and Technology,
Jiangsu Collaborative Innovation Center of Photovoltaic Science and Engineering,
Changzhou University, Changzhou, 213164, Jiangsu, China
[12] Faculty of Engineering and IT, University Technology Sydney, 2007 Sydney, Australia
[13] Department of Physics, Cavendish Laboratory, University of Cambridge,
JJ Thomson Avenue, Cambridge, CB3 0HE, UK
[14] Helmholtz-Zentrum Berlin für Materialien und Energie GmbH, Division Solar Energy, 12489 Berlin, Germany
[15] School of Mathematical and Physical Sciences, University Technology Sydney, 2007 Sydney, Australia
[16] School of Mechanical Engineering, Jiangsu University, Zhenjiang, 212013, Jiangsu, China
[17] School of Materials Science and Engineering, Faculty of Science,
University of New South Wales, UNSW Sydney, Kensington, 2052, Australia
[18] Department of Applied Physics, The Hong Kong Polytechnic University, Kowloon, Hong Kong, China

---

* milos.dubajic@hotmail.com
* james.neilson@colostate.edu
* stephen.bremner@unsw.edu.au
* a.walsh@imperial.ac.uk
* michael.nielsen@unsw.edu.au
* sds65@cam.ac.uk

## i. PHASE TRANSITIONS, AVERAGE AND LOCAL STRUCTURE IN HALIDE PEROVSKITES

We used differential scanning calorimetry (DSC) to determine phase transition temperatures in our materials. DSC is a thermal analysis technique employed to determine the heat flow in or out of a sample by simultaneously applying heat to both the sample and a reference material whilst recording their respective temperatures. The internal energy changes in the sample can subsequently be inferred from the measured temperature difference [1]. DSC is a highly effective tool for identifying the temperature points at which structural phase transitions occur [2–4]. As the lattice adopts a new configuration during the phase transition, heat is exchanged with the environment, resulting in a peak on the DSC curve. DSC measurements were carried out on MAPbBr$_3$ and FAPbBr$_3$ single crystals. The corresponding results are presented in Fig. S1 **a** and **b**, respectively. Temperature points indicative of potential phase transition events are depicted above the corresponding DSC curve peaks in Fig. S1.

DSC peaks in the heating cycle at $T = 235$ K, $T = 155$ K, and $T = 150$ K can be assigned to cubic-tetragonal (tetragonal $I4/mcm$ shown in Fig. S2 **b**) , tetragonal-incommensurate and incommensurate-orthorhombic [5–7] phase transitions in MAPbBr$_3$, based on prior DSC investigations [8–11]. Numerous efforts have been made to identify transition temperatures and space groups for each phase of FAPbBr$_3$ [9, 12–16]. While a consensus exists in the literature on space groups, temperature transition points remain ambiguous. FAPbBr$_3$ undergoes a series of structural phase transitions with decreasing temperature. Consequently, the $T = 264$ K transition (as seen in Fig. S1 **b**) has previously been assigned to the cubic-tetragonal transition (tetragonal $P4/mbm$ shown in Fig. S2 **b**), whereas it remains unclear which individual transitions correspond to the other observed DSC peaks (the orthorhombic $Pnma$ phase depicted in Fig. S2 **d** was most commonly identified at low temperatures). The measured DSC curves show a good agreement with previously reported DSC measurements in MAPbBr$_3$ [8–10] and FAPbBr$_3$ [9, 13, 17].

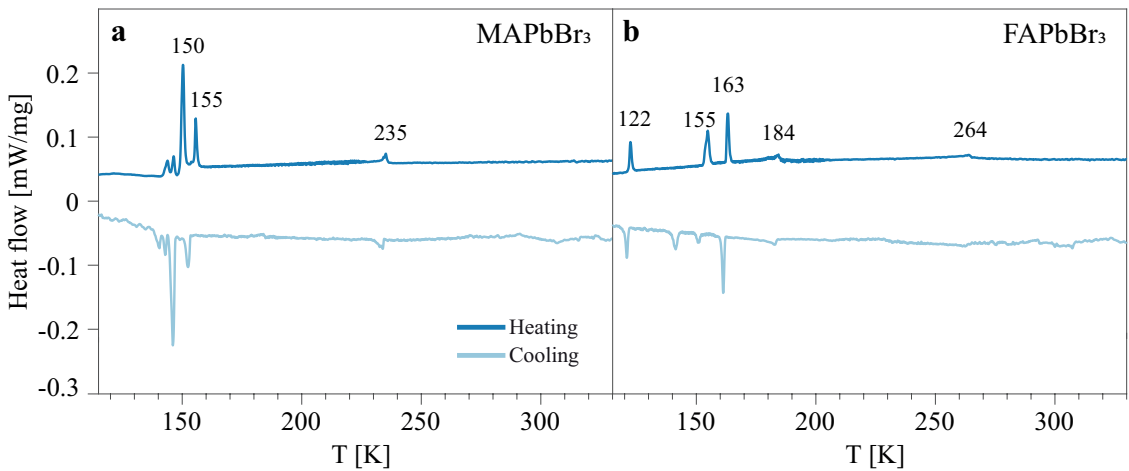

FIG. S1. **Differential scanning calorimetry (DSC) measurements.** DSC heating and cooling cycles of (**a**) MAPbBr$_3$ (**b**) and FAPbBr$_3$.

The data presented in Fig. S1 confirms that between $T = 300$ K and $T = 200$ K in these materials, only one phase transition occurs. As we refined the space groups at those two temperatures the DSC data supports our claims from the main text that MAPbBr$_3$ transitions from cubic $Pm\bar{3}m$ to twinned $I4/mcm$ at 235 K while FAPbBr$_3$ transitions from cubic $Pm\bar{3}m$ to cubic $Im\bar{3}$ at 264 K.

## ii. TWINNING OBSERVATION USING SINGLE CRYSTAL XRD

In our case, we observe non-merohedral twinning, a type of twinning where the twin domains are usually related to each other by a symmetry operation of the higher symmetry phase that is not a symmetry operation of the lower symmetry phase. This occurs during the ferroelastic cubic-to-tetragonal phase transition in MAPbBr$_3$ while this twinning is not present during the phase transition in FAPbBr$_3$. We have managed to capture this transition in MAPbBr$_3$ in both reciprocal space (Fig. S3 **a**, **b** and **c**) and in real space (Fig. S34 **a** and **b**). Based on X-ray scattering factors in reciprocal space we identified the symmetry relationships that define this non-merohedral twinning (given in Table I and depicted in Fig. S4 **a**). It can be observed that the peak intensities of the experimental and simulated data do not exactly match. This is a result of our assumption that all twin components occur with the same probability, which is in general not the case. However, our the simulated pattern does match the experimental X-ray Bragg pattern in all the planes presented in Fig. S3, which confirms our twinning laws correctly capture non-merohedral twinning which occurs during cubic-tetragonal and tetragonal-orthorhombic phase transitions in MAPbBr$_3$. As twinning laws are generally given through rotation operators that act on a reference twin component (in our case $D_1$), we visualise them in Fig. S4.

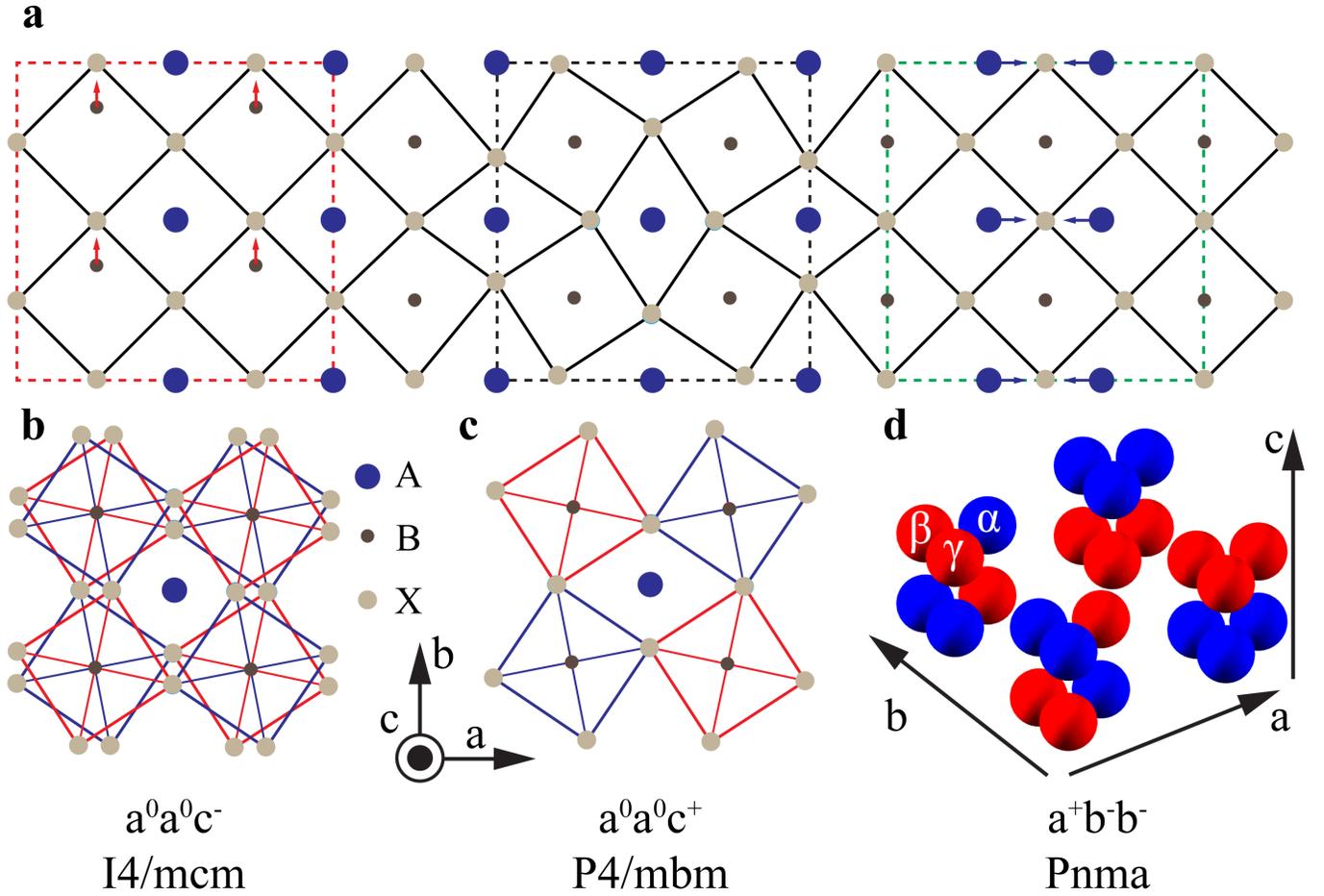

FIG. S2. **Graphical illustration of local and average structures in perovskite-type** $ABX_3$ **lattice.** (**a**) Depiction of various short-range atomic correlations. From left to right: B-site off-centering within the red rectangle, octahedral tilt correlations within the black rectangle, and A-site correlations within the green rectangle. Octahedral tilt patterns across various space groups with corresponding Glazer notation given in brackets (**b**) tetragonal $I4/mcm$ ($a^0a^0c^-$), (**c**), tetragonal $P4/mbm$ ($a^0a^0c^+$), (**d**) orthorhombic $Pnma$ ($a^+b^-b^-$). In (**b**) and (**c**) octahedron edges are depicted in blue and green and indicate positive and negative tilts along the c direction. In (**d**), each octahedron is depicted with three spheres in a triangular arrangement, with colours representing the tilt angle's direction: blue for positive and red for negative, along the $a$ ($\alpha$), $b$ ($\beta$), and $c$ ($\gamma$) axes.

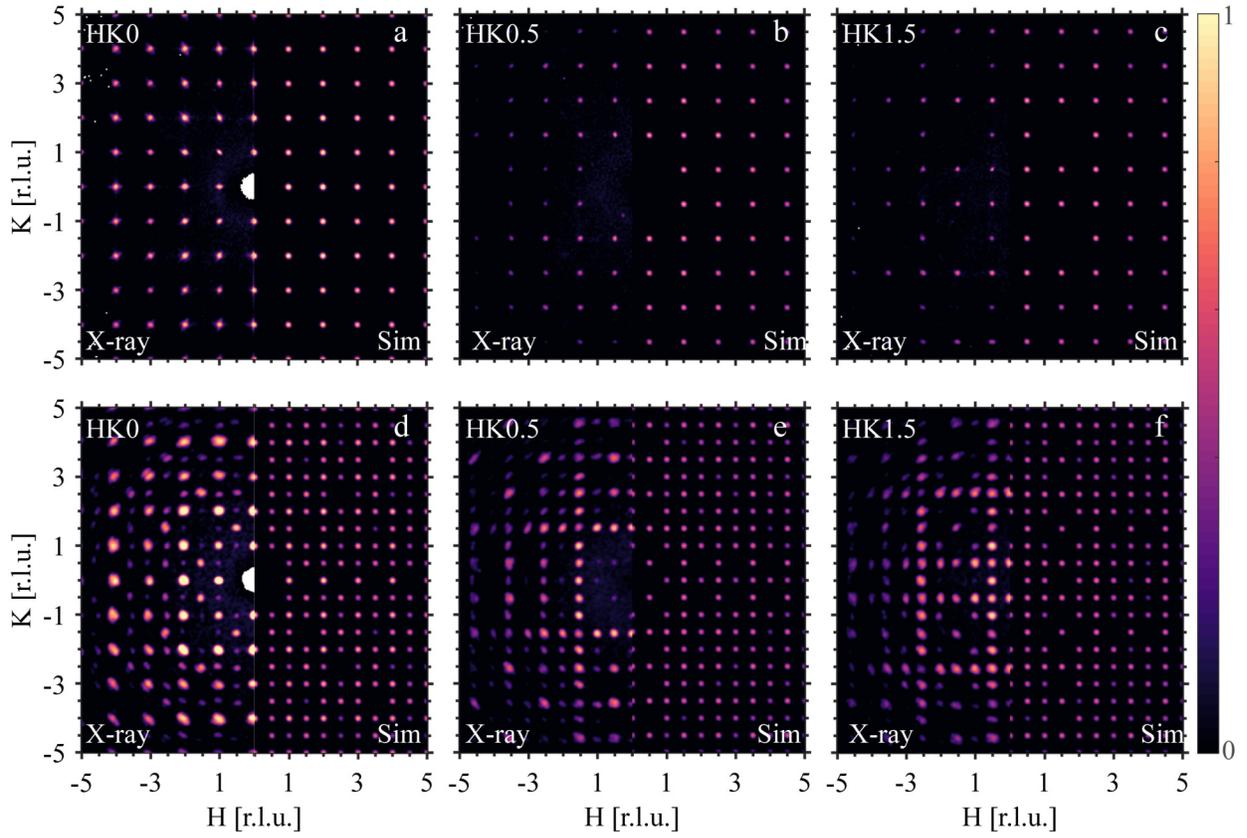

FIG. S3. **Non-merohedral twinning in MAPbBr₃ observed in reciprocal space.** The left panels in (**a**), (**b**), and (**c**) display the experimental scattering function $S(\mathbf{q})$ at $T = 200$ K, while the corresponding right panels illustrate the simulated twinned average $I4/mcm$ phase scattering function in the HK0, HK0.5, and $HK1.5$ planes, respectively. Similarly, the left panels in (**d**), (**e**), and (**f**) showcase the experimental scattering functions at $T = 100$ K, with the corresponding right panels depicting the simulated twinned average $Pnma$ phase scattering function in the HK0, HK0.5, and $HK1.5$ planes, respectively. The twinning operators for $I4/mcm$ twins and $Pnma$ twins are provided in Table I and Table II, respectively. A noticeable discrepancy between the simulated and experimental Bragg peak intensities is apparent. This arises due to the assumption in our simulation that all twin components have an equal likelihood of occurring within the crystal.

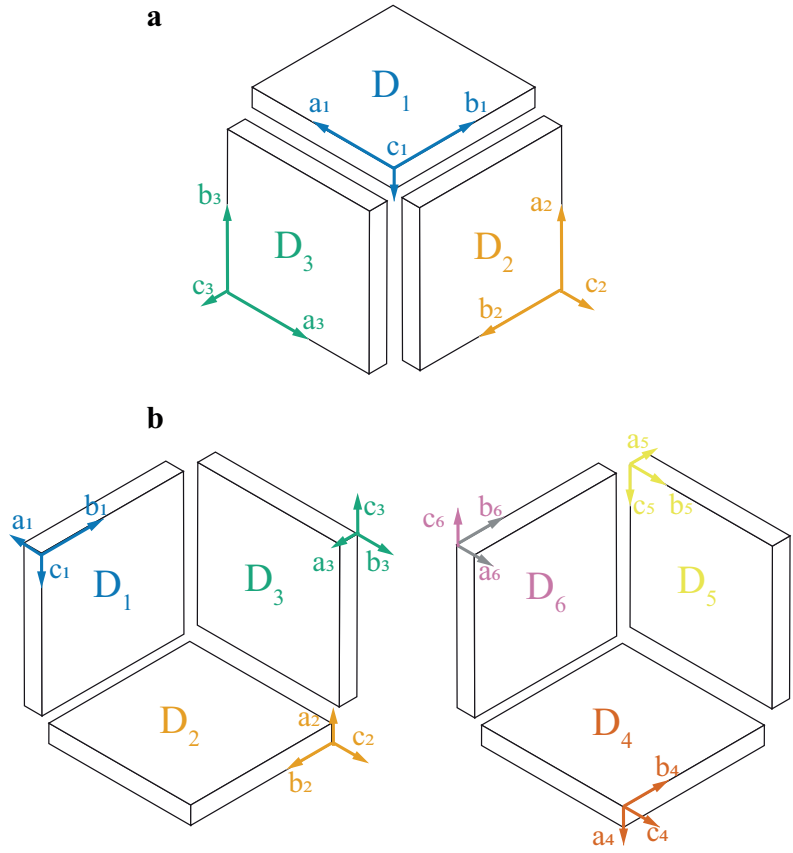

FIG. S4. **Illustration of twinning transformations.** (**a**) The diagram illustrates the relative orientations of the coordinate systems for three twin components that emerge during cubic to tetragonal ferroelastic phase transition in MAPbBr$_3$. The transformation operators can be found in Table I. (**b**) The diagram shows the relative orientations of the coordinate systems for six twin components that occur during the transition from a tetragonal to an orthorhombic phase in MAPbBr$_3$. The corresponding transformation operators are provided in Table II. The same twin laws apply to the corresponding local structure i.e. local tetragonal/orthorhombic planar nanodomains in average cubic and tetragonal phases, respectively. The reciprocal space representation of the six local structure components in the average tetragonal phase is depicted in Fig. S6.

### iii. MODIFIED GLAZER NOTATION FOR LOCAL OCTAHEDRAL TILTING CORRELATIONS

The Glazer notation offers a systematic approach to describe octahedral tilting patterns in perovskite structures [18]. In the Glazer notation, the sequence of symbols directly corresponds to the pseudocubic crystallographic axes: the first symbol for tilts along $a$ [100], the second along $b$ [010], and the third along $c$ [001]. The set of three letters (e.g., $abc$) is also used to symbolise the magnitudes of tilt about these axes, with identical letters indicating equal tilt amplitudes. Each symbol is accompanied by a superscript denoting the tilt's ordering across adjacent octahedral layers: '0' for no tilt, '+' in-phase ordering (where adjacent layers tilt in the same direction), and '-' anti-phase ordering (where adjacent layers tilt in opposite directions). For instance, $a^0a^0c^-$ indicates no tilt along the $a$ and $b$ axes and an infinite anti-phase ordering of c-axis tilts along the $c$ axis. Due to the rigid octahedra rules, if one octahedron is arbitrarily rotated along e.g. the $c$ axis, the adjacent octahedra always anti-phase order along $a$ and $b$ while the ordering along the $c$ axis is not constrained [18].

When the local structure is present, this as a consequence leads naturally to the occurrence of pancake local nanodomains in perovskite structures. If one octahedron is arbitrarily rotated along the $c$ axis, instead of infinite ordering along $a$ and $b$ axes we have long-range ordering which we denote with subscript letter L. Tilts are naturally less coupled along $c$ and thus we would have short-range ordering along $c$ denoted with letter S in the subscript. This phenomenon can be described using the notation $a^0a^0c_{LLS}^{---}$ for short-range anti-phase ordering of c-axis tilts along the $c$ axis. The local structure resembles $I4/mcm$ symmetry, observable in the average cubic phase of MAPbBr$_3$. The three subscripts under $c$ indicate whether the range of ordering of c-axis tilts along the [100], [010] and [001] pseudocubic axes is long ('L') or short ('S'). Similarly, $a^0a^0c_{LLS}^{--+}$ corresponds to short-range in-phase ordering of c-axis tilts along the $c$ axis, with a local structure akin to $P4/mbm$ symmetry, as found in the average cubic phase of FAPbBr$_3$."

We find the local disc-shaped nanodomains akin to $Pnma$ symmetry in the average tetragonal phase of MAPbBr$_3$. In the modified Glazer notation we can denote the local structure as $a_{SLL}^{+--} b_{LSL}^{---} b_{\infty\infty\infty}^{---}$, which is graphically represented in Fig. S6.

Given that long-range ordered octahedra are always in plane and always exhibit anti-phase ordering, we can exclude L subscripts and just denote the sign of short range (S) ordering. For infinite correlations we do not use L or S but just denote the ordering sign. Thus, $a^0a^0c_{LLS}^{---}$ can be reduced to $a^0a^0c_S^-$, $a^0a^0c_{LLS}^{--+}$ to $a^0a^0c_S^+$ and $a_{SLL}^{+--} b_{LSL}^{---} b_{\infty\infty\infty}^{---}$ to $a_S^+ b_S^- b^-$.

### iv. THE CLASSICAL MODEL OF LOCAL OCTAHEDRAL TILTING

#### A. Implementation of the model for quasi-elastic diffuse scattering simulations

To accurately model the 3D volumetric diffuse scattering distribution, the following steps were taken: Standard single crystal XRD structural refinements were performed across all measured crystallographic phases to extract structure factors and Miller indices of the Bragg peaks. This method is not sensitive to the local structure. We determined that the symmetry of the local structure within a global crystallographic phase is the same as the symmetry of the next lower temperature global crystallographic phase. As a result, the total scattering intensity, which contains the diffuse scattering, can be modelled by adding the contributions from both the high-symmetry global structure (infinitely-ordered) and the lower-symmetry local structure (short-range ordered), while adjusting the Bragg peaks' FWHM of the lower-symmetry phase to account for the short-range ordering of the local structure.

We will first describe a procedure to generate 3D volumetric reciprocal space intensities using the CIF file of a known space group, assuming no short-range order is present. Given a CIF file with structure factors ($F_{hkl}$) for each Miller index triplet $(h, k, l)$, we can calculate the volumetric 3D reciprocal space intensities by following these steps:

1. We first apply the transformation matrix to convert structure factors $F_{hkl}$ and their corresponding Miller indices $(h, k, l)$ from the CIF file, defined in the original space group, to the equivalent values when the unit cell is expressed in pseudocubic notation.

2. For the sake of simplicity, we assume that each reflection will result in a Gaussian peak in reciprocal space with

instrument resolution-limited variance $\sigma^2$ to calculate the intensity distribution $I_{hkl}(\mathbf{q}^*)$ for each reflection:

$$I_{hkl}(q_H^*, q_K^*, q_L^*) = |F_{hkl}|^2 \frac{1}{(2\pi)^{3/2}\sigma^3} \exp\left(-\frac{((q_H^* - h)^2 + (q_K^* - k)^2 + (q_L^* - l)^2)}{2\sigma^2}\right), \tag{S1}$$

where $\mathbf{q}^* = (q_H^*, q_K^*, q_L^*)$ are the components of the scattering vector $\mathbf{q}$ reduced to reciprocal space units, given by $\mathbf{q}^* = \frac{a\mathbf{q}}{2\pi}$, where $a$ is the pseudocubic lattice constant.

3. We sum the intensity distributions for all reflections $I_{hkl}(\mathbf{q}^*)$ to obtain the volumetric intensity distribution $I(\mathbf{q}^*)$:

$$I(\mathbf{q}*) = \sum_{h,k,l} I_{hkl}(\mathbf{q}*). \tag{S2}$$

To generate the local structure contribution to the volumetric intensity distribution, we again, for the sake of simplicity, assume that peak intensities maintain the Gaussian distribution even in the case of local order. Thus, we can define the 3D elliptical Gaussian function by extending the instrument resolution-limited standard deviation $\sigma$ (values for each instrument used are given in Table III) by the amount $\Delta\sigma_i$, which is inversely proportional to the correlation length $\xi_i$ along a particular direction $i = H, K, L$ of the reciprocal space. Thus, the Eq. (S1) is modified into:

$$I_{hkl}^{\mathrm{loc}}(\mathbf{q}) = |F_{hkl}^{\mathrm{loc}}|^2 \frac{1}{(2\pi)^{3/2}(\sigma + \Delta\sigma_H)(\sigma + \Delta\sigma_K)(\sigma + \Delta\sigma_L)} \exp\left(-\frac{(q_H^* - h)^2}{2(\sigma + \Delta\sigma_H)^2} - \frac{(q_K^* - k)^2}{2(\sigma + \Delta\sigma_K)^2} - \frac{(q_L^* - l)r}{2(\sigma + \Delta\sigma_L)^2}\right). \tag{S3}$$

To obtain the total intensity distribution $I_{\mathrm{tot}}$, we need to account for the modifications in the structure factors for specific $h$, $k$, and $l$. We can do this by introducing two intensity components, $I_{hkl}$ and $I_{hkl}^{\mathrm{loc}}$, each of which satisfies different conditions on the Miller indices $(h, k, l)$. $I_{hkl}$ represents the parent average symmetry phase, thus all Bragg peaks will be resolution limited $I_{hkl}(\mathbf{q}^*, \sigma)$. $I_{hkl}^{\mathrm{loc}}$ represents the local structure and as a result, each Bragg peak width will be expanded by $\Delta\sigma_i$. The total intensity distribution can then be expressed as a combination of these two components:

$$I^{\mathrm{tot}}(\mathbf{q}*) = \sum_{Rot}[AI_{hkl}(\mathbf{q}*, \sigma) + BI_{hkl}^{\mathrm{loc}}(\mathbf{q}*, \Delta\sigma_H, \Delta\sigma_K, \Delta\sigma_L)] + CI_{000}(\mathbf{q}*, \sigma_{bgr}). \tag{S4}$$

In this expression, the term $\sum_{\mathrm{Rot}}$ represents a summation of all possible rotations of the 3D volumetric intensities originating from the local structure. The rotation matrices are determined by the symmetry relationships between the possible local structure relative orientations. A scaling constant, $A$, is incorporated to adjust for any non-perfect relationships between the measured scattered intensities and structure factors. The term $BI_{000}(\mathbf{q}^*, \sigma_{\mathrm{bgr}})$ represents a Gaussian function centred at the reciprocal space origin. This Gaussian accounts for any imperfections in background subtraction encountered during the X-ray measurement.

To simulate the local $I4/mcm$ structure within the global $Pm\bar{3}m$ structure, we can split the total intensity into two components: $I_{hkl}$ and $I_{hkl}^{\mathrm{loc}}$. $I_{hkl}$ corresponds to the Miller indices $(h, k, l)$ that satisfy the following condition:

$$(h, k, l) \in \mathbb{Z}^3, \tag{S5}$$

i.e. when Miller indices $(h, k, l)$ are simultaneously integer values. All of the above Miller indices would be precisely contained in the global $Pm\bar{3}m$ structure. The $I_{hkl}^{\mathrm{loc}}$ component represents the intensities of the superstructure reflections that would arise when the $Pm\bar{3}m$ structure is transitioning to the $I4/mcm$ structure. The Miller indices of $I_{hkl}^{\mathrm{loc}}$ are all indices that do not satisfy the first condition, i.e. all Miller indices that are not simultaneously integer values. It was necessary to divide the total intensity into $I_{hkl}$ and $I_{hkl}^{\mathrm{loc}}$, as the structure factors arising from the local octahedra tilting $I_{hkl}^{\mathrm{loc}}$ will differ from the structure factors that originate from the non-tilted cubic matrix encompassed in $I_{hkl}$.

To simulate the local $Pnma$ structure within the global $I4/mcm$ structure, we can again split the total intensity into two components: $I_{hkl}$ and $I_{hkl}^{\mathrm{loc}}$. $I_{hkl}$ corresponds to the Miller indices $(h, k, l)$ that satisfy the following condition:

$$\big((h, k, l) \notin \mathbb{Z}^3\big) \ \lor \big((h, k, l) \in \mathbb{Z}^3\big). \tag{S6}$$

i.e. when Miller indices $(h, k, l)$ are simultaneously integer or non-integer values. The $I_{hkl}^{\mathrm{loc}}$ component corresponds to all other $(h, k, l)$ that do not satisfy the condition mentioned above. These correspond to the superstructure reflections that would emerge when the $I4/mcm$ structure is transitioning to the $Pnma$ structure.

We have previously established that the local structure symmetry follows the twinning non-merohedral symmetry. Thus, there are three components of the local $I4/mcm$ structure in global $Pm\bar{3}m$ and 6 components of the local $Pnma$ in global $I4/mcm$. The following table summarises the symmetry operations that were taken to calculate the total intensity in the case of two different local structures.

TABLE I. Components of the $I4/mcm$ and $P4/mbm$ local structure

| Local symmetry | Glazer notation | $D_1$ | | $D_2$ | $D_3$ |
| | | $I_{hkl}$ | $I_{hkl}^{\text{loc}}$ | | |
|---|---|---|---|---|---|
| $I4/mcm$, $P4/mbm$ | $a^0a^0c_S^-$, $a^0a^0c_S^+$ | $(h,k,l)\in\mathbb{Z}^3$ | other | $\mathcal{R}(\pi,[10\bar{1}])$ | $\mathcal{R}(\pi,[01\bar{1}])$ |

In this table $\mathcal{R}(\pi,[10\bar{1}])$ denotes that the $D_2$ component was obtained by the rotation of the $D_1$ intensity component for $180°$ around $[10\bar{1}]$ direction in the reciprocal space. The relationship between correlation length $\xi_i$ and $\sigma+\Delta\sigma_i$ can be derived by assuming the FWHM of a Gaussian function is approximately equal to the FWHM of a Lorentzian function, which accurately represents the function arising from short-range order. This approximation is valid in our case (when instrument response broadening is on the order of Lorentzian linewidth) according to Kielkopf's approximate formula [19] for the FWHM of a Vogit profile. The relationship between $\xi_i$ and FWHM (when FHWM is measured in reciprocal lattice units) is:

$$\xi_i = \frac{2a}{\pi FWHM_i}. \tag{S7}$$

TABLE II. Components of the $Pnma$ local structure

| Local symmetry | Glazer notation | $D_1$ | | $D_2$ | $D_3$ | $D_4$ | $D_5$ | $D_6$ |
| | | $I_{hkl}$ | $I_{hkl}^{\text{loc}}$ | | | | | |
|---|---|---|---|---|---|---|---|---|
| $Pnma$ | $a_S^+b_S^-b^-$ | $((h,k,l)\notin\mathbb{Z}^3)\ \vee\ ((h,k,l)\in\mathbb{Z}^3)$ | other | $\mathcal{R}(\pi,[10\bar{1}])$ | $\mathcal{R}(\pi,[1\bar{1}0])$ | $\mathcal{R}(\pi,[\bar{1}0\bar{1}])$ | $\mathcal{R}(\pi,[\bar{1}\bar{1}0])$ | $\mathcal{R}(\pi,[010])$ |

Note that this expression represents the correlation length diameter and is twice the correlation length radius which is obtained as:

$$\xi_i^r = \frac{a}{2\pi HWHM_i}, \tag{S8}$$

or in real space as:

$$R(x) = \exp\left(-\frac{x}{\xi_i^r}\right), \tag{S9}$$

where $x$ is a spatial coordinate and $R(x)$ is the spatial tilting correlation function [20]. Now, in Eq. (S7) we express the FWHM in terms of total standard deviation $\sigma+\Delta\sigma_i$:

$$FWHM_i = 2\sqrt{2\ln 2}(\sigma+\Delta\sigma_i). \tag{S10}$$

To find the relationship between $\xi_i$ and total standard deviation:

$$\xi_i = \frac{a}{\pi\sqrt{2\ln 2}(\sigma+\Delta\sigma_i)}. \tag{S11}$$

This is the relationship between the correlation length $\xi_i$ and the total standard deviation of the Gaussian function in reciprocal space, where for e.g. $a^0a^0c_{LLS}^{---}$ ($a^0a^0c_S^-$): $\Delta\sigma_\perp=\Delta\sigma_L$ and $\Delta\sigma_\parallel=\Delta\sigma_H=\Delta\sigma_K$ and for $a_{SLL}^{+--}b_{LSL}^{---}b_{\infty\infty\infty}^{---}$ ($a_S^+b_S^-b^-$): $\Delta\sigma_\perp=\Delta\sigma_H$, $\Delta\sigma_\parallel=\Delta\sigma_K$ and $\Delta\sigma_L=0$.

We assume that all components of the local structure are equally probable. We inherit the CIF file of the lower symmetry average phase and use those atomic positions to generate X-ray scattering factors of the local structure. It should be noted that this approximation could introduce discrepancies, as the actual degree of local octahedra tilt might not perfectly correspond to the inherited tilt from the low-temperature average phase. This deviation could then manifest as a minor disparity between the simulated structure factor intensities and the experimentally measured values. Additionally, it should be noted that this model does not differentiate between statically and dynamically tilted octahedra, but rather simulates experimentally observed X-ray scattering function $S(\mathbf{q})$ which is a sum of both elastic (static) and inelastic (dynamic) scattering. To finally estimate the correlation lengths of the local structure based on the experimental data, we fit our 3D volumetric intensity equation to the 3D experimental X-ray scattering function. The obtained fitting parameters are shown in Table III and the 95 % confidence intervals in Table IV.

TABLE III. Fitted parameters

| Sample | T [K] | Glazer Notation | $\sigma$ [r.l.u.] | $\Delta\sigma_\perp$ [r.l.u.] | $\Delta\sigma_\parallel$ [r.l.u.] | $A$ | $B$ | $C$ | $\sigma_{\mathrm{bgr}}$ | $a$ [Å] | $\xi_\perp$ [Å] | $\xi_\parallel$ [Å] |
|---|---|---|---|---|---|---|---|---|---|---|---|---|
| MAPbBr$_3$ | 300 | $a^0a^0c_S^-$ | 3e-3 | 0.2571 | 0.0734 | 0.151 | 0.151 | 13 | 1.117 | 5.941 | 6.176 | 21.024 |
| MAPbBr$_3$ | 200 | $a_S^+b_S^-b^-$ | 8.8e-3 | 0.2876 | 0.0926 | 2.05e-3 | 3.04e-2 | 28.1 | 1.9785 | 5.921 | 5.401 | 15.786 |
| FAPbBr$_3$ | 300 | $a^0a^0c_S^+$ | 20e-3 | 0.0625 | 0.0358 | 1.3e-3 | 1.3e-3 | 125.52 | 0.7698 | 6.007 | 19.686 | 29.106 |
| FAPbBr$_3$ | 200 | $a^+a^+a^+$ | 6.4e-3 | 0 | 0 | 1 | 1 | 0 | / | 5.958 | $\infty$ | $\infty$ |

TABLE IV. Confidence intervals

| Sample | T[K] | $\Delta\sigma_\perp$ [r.l.u.] | $\Delta\sigma_\parallel$ [r.l.u.] | $A$ | $B$ | $C$ | $\sigma_{\mathrm{bgr}}$ | $\xi_\perp$ [Å] | $\xi_\parallel$ [Å] |
|---|---|---|---|---|---|---|---|---|---|
| MAPbBr$_3$ | 300 | 1.2e-3 | 3.6e-4 | 1.2e-3 | 1.2e-3 | 1.2914 | 3.8e-2 | 0.028 | 0.1 |
| MAPbBr$_3$ | 200 | 16.8e-3 | 1.7e-3 | 1.8e-3 | 6.93e-4 | 9.1e-1 | 1.1e-2 | 0.307 | 0.266 |
| FAPbBr$_3$ | 300 | 2.43e-3 | 16.46e-3 | 7e-5 | 7e-5 | 9.55 | 12e-3 | 0.58 | 8.585 |
| FAPbBr$_3$ | 200 | / | / | / | / | / | / | / | / |

We extend the same model applied in the average cubic phase to shed light on the local structure within the average tetragonal phases. To accurately account for the diffuse scattering within these phases, our model introduces the formation of six twin components. In essence, each of the three local tetragonal twin components forks into two additional ones as visualised in Fig. S4. Similar to the case of the average $Pm3m$ cubic structure, which comprises local $I4/mcm$ tetragonal twins, we demonstrate that the average $I4/mcm$ tetragonal structure is composed of local $Pnma$ orthorhombic twins (Fig. S6). These relative orientations of these six components in real space are presented in Fig. S4 **b** and in reciprocal space in Fig. S6. Twins can be understood as local clusters of octahedra tilted such that $a_S^+b_S^-b^-$, like visualised in Fig. S6**b**.

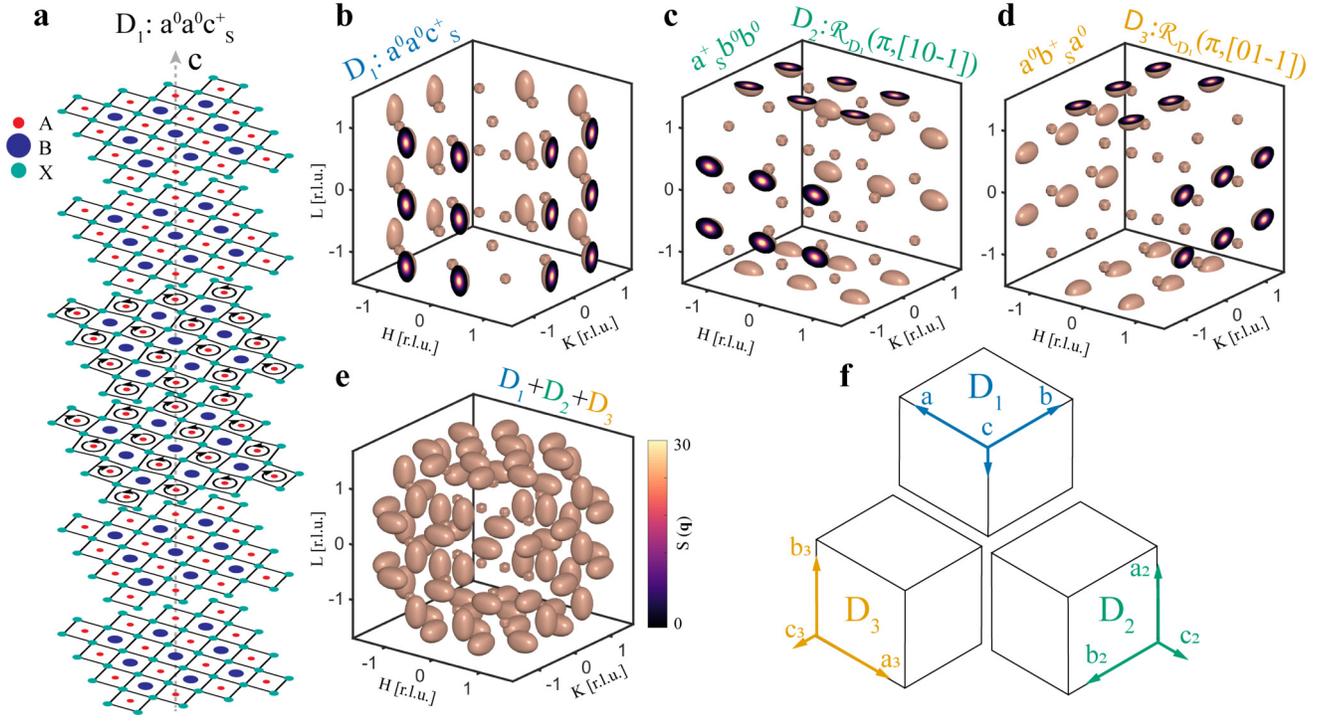

FIG. S5. **Components of local in-phase octahedral tilting with** $P4/mbm$ **symmetry in real and reciprocal space.** (**a**) Real-space depiction of the $D_1$ component of in-phase perovskite $ABX_3$ octahedral local structure, designated as $a^0a^0c_S^+$ in modified Glazer notation. The octahedra display non-zero tilt along the $c$ axis, with short-range in-phase ordering (one $P4/mbm$ $c$ axis unit) along $c$ and long-range ordering along $a$ and $b$. (**b**) Computed 3D volumetric X-ray scattering intensities in the reciprocal space of the structure in (**a**) represent the $D_1$ diffuse scattering component. (**c**) The reciprocal space of the second local structure component ($D_2$) is derived by rotating ($D_1$) in (**b**) 180° around the $[10\bar{1}]$ vector of the $D_1$ coordinate system. (**d**) The reciprocal space of the third local structure component ($D_3$) is achieved by rotating ($D_1$) in (**b**) 180° around the $[01\bar{1}]$ vector of the $D_1$ coordinate system. (**e**) The complete 3D diffuse scattering signal is the cumulative sum of all three components. (**f**) In real space, the local structure is characterized as the combination of three distinct local $P4/mbm$ planar components. The thickness of these components is determined by the degree of short-range anti-phase local octahedral tilting along the $c$ axes of the local coordinate system for each component. The relationship between the axes is demonstrated in (**f**).

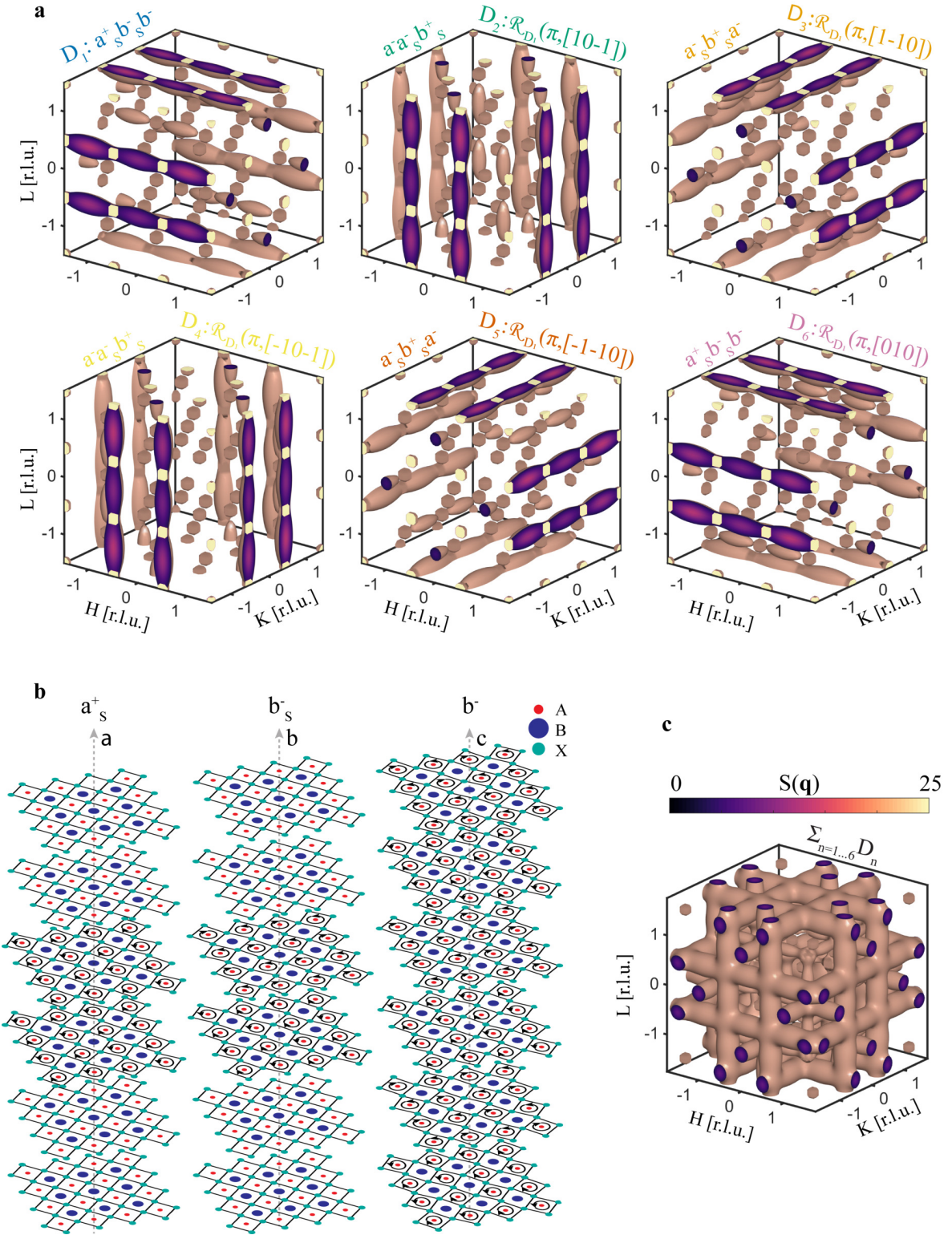

FIG. S6. **Components of local** $Pnma$ **symmetry in real and reciprocal space.** (**a**) Six 3D volumetric scattering factor intensities in reciprocal space that correspond to 6 twin components. These are obtained via the coordinate system transformation denoted above the images. (**b**) Real-space depiction of the $D_1$ component of perovskite $ABX_3$ octahedral local structure, designated as $a_S^+ b_S^- b^-$ in modified Glazer notation. Octahedra exhibit non-zero short-range in-phase tilt along the $a$ axis, complemented by short-range anti-phase ordering along $b$, and infinite ordering along $c$. (**c**) The cumulative sum of all six components yields the complete 3D volumetric structure factors that account for the diffuse scattering signal. Isovalue used for all plots is 3.

## B. Comparison of experimental and classically modelled diffuse scattering patterns

The presence of additional diffracted intensities within the half-integer 2D reciprocal space planes (Fig. S7) signifies a unit cell doubling, a characteristic indicative of a reduction in symmetry. In that context, the diffuse scattering rods and ellipsoids form due to broadened superstructure peaks within reciprocal space, with the degree of broadening serving as a metric for local order.

To accurately determine the symmetry of the local structure, it is crucial to fully understand the symmetry inherent to the diffuse scattering pattern. We establish that the symmetry of the QEDS within the average cubic phases of these materials mirrors the symmetry of the Bragg peaks (Laue symmetry) found in their average tetragonal phases. This Laue symmetry of the average tetragonal structure is however higher than the inherent symmetry of the crystal, which is common in the case of non-merohedral twinning (for details see Section ii).

To test our classical model we compare simulated Quasi-Elastic Diffuse Scattering (QEDS) $S(\mathbf{q})$ across $HK1.5$, $HK0.5$, $HK0$ reciprocal space planes as shown in Fig. S7, Fig. S8 and Fig. S9.

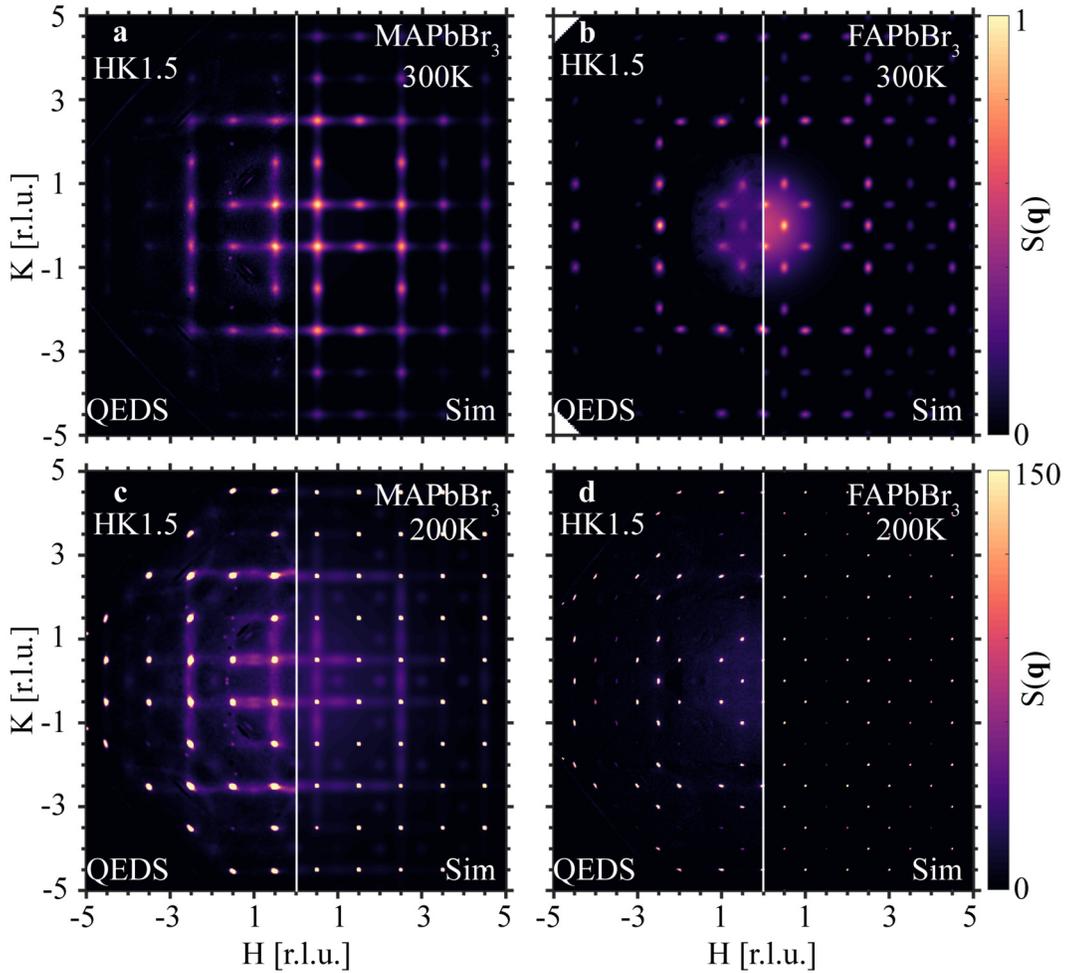

FIG. S7. **Relationship between phase transitions and diffuse scattering in HK1.5 planes.** Experimental QEDS (left panels) and simulated (right panels) signals across HK1.5 reciprocal space planes show great agreement. (**a**) and (**b**) MAPbBr$_3$ and FAPbBr$_3$ at $T = 300$ K in average cubic $Pm\bar{3}m$ phase but with local planar $I4/mcm$ and spherical $P4/mbm$ nanodomains, respectively. (**c**) and (**d**) MAPbBr$_3$ and FAPbBr$_3$ at $T = 200$ K in average tetragonal $I4/mcm$ (with local $Pnma$ nanodomains) and average $Im\bar{3}$ phase, respectively. The broad QEDS peaks in (**a**) and (**b**) narrow down to Bragg peaks in (**c**) and (**d**) upon the phase transition.

In Fig. S9 **a** and **b** we present experimental total diffuse scattering (left panels) and simulated QEDS (right panels) signals across HK0 reciprocal space planes in MAPbBr$_3$ and FAPbBr$_3$, respectively. It is evident that the QEDS simulation does not reproduce the rods of diffuse scattering linking the Bragg peaks in Fig. S9 **a**. This outcome is anticipated since the simulation excludes the TDS, which is responsible for these rods. Nonetheless, the QEDS simulation successfully identifies the emergence of additional broad reflections at the M points in reciprocal space, highlighted by white circles in Fig. S9 **a** and **b**. These QEDS features in the HK0 plane arise from the cross sections of R-M diffuse rods within the e.g. H0.5L and H1.5L planes, that are perpendicular to HK0 (note that all set of all perpendicular planes such as e.g. HK$n$, H$n$L 0K$n$, where $n$ are integer and half-integer values, are always identical in the average cubic phase even in the presence of diffuse scattering due to effective twinning of the dynamic local nanodomains as explained in detail in the main text). Conversely, the MD simulation depicted in Fig. S13 **e** and **f** encompasses both TDS and QEDS, thereby accurately mirroring the complete experimental diffuse scattering observed in the HK0 planes.

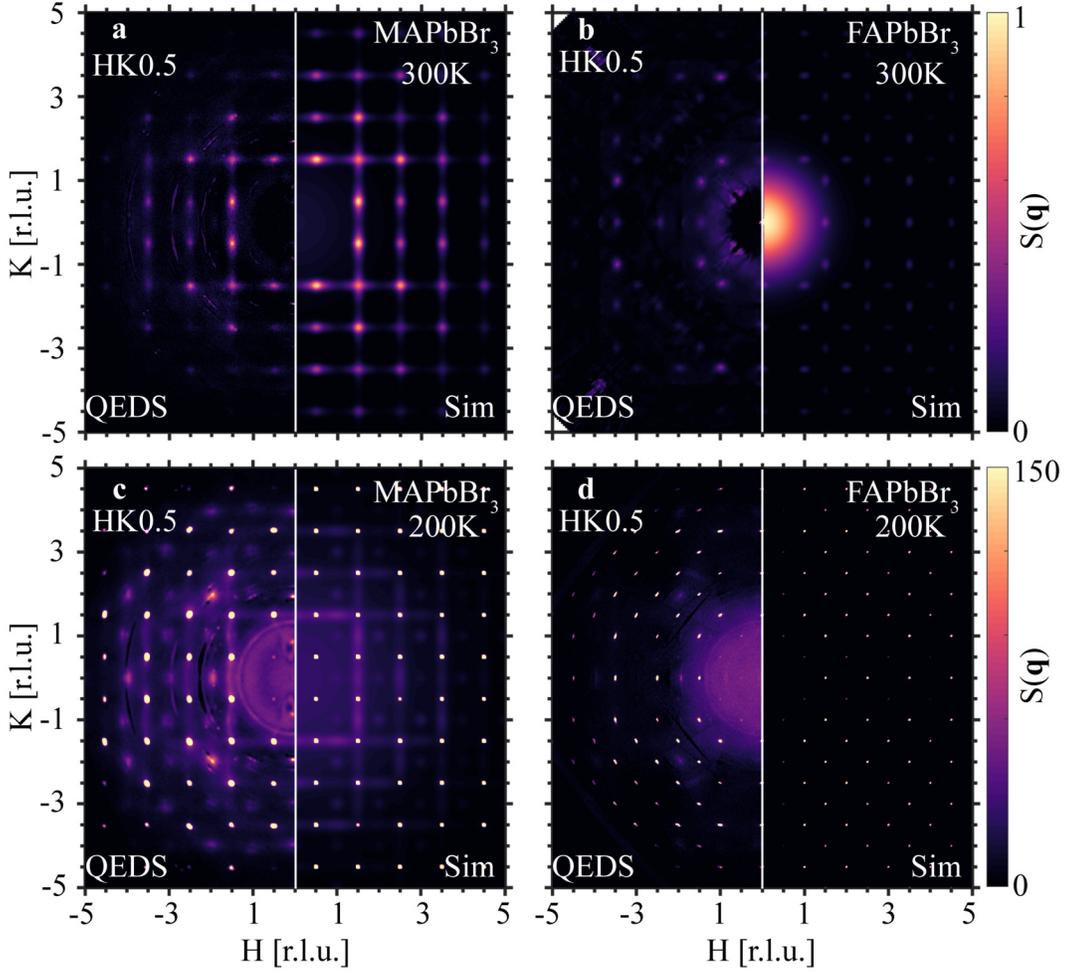

FIG. S8. **Relationship between phase transitions and diffuse scattering in HK0.5 planes.** Experimental QEDS (left panels) and simulated (right panels) signals across HK0.5 reciprocal space planes. (**a**) and (**b**) MAPbBr$_3$ and FAPbBr$_3$ at $T = 300$ K in average cubic $Pm\bar{3}m$ phase but with local planar $I4/mcm$ and spherical $P4/mbm$ nanodomains, respectively. (**c**) and (**d**) MAPbBr$_3$ and FAPbBr$_3$ at $T = 200$ K in average tetragonal $I4/mcm$ (with local $Pnma$ nanodomains) and average $Im\bar{3}$ phase, respectively. The broad QEDS peaks in (**a**) and (**b**) narrow down to Bragg peaks in (**c**) and (**d**) upon the phase transition.

We quantitatively test our classical model on several selected reciprocal space planes in MAPbBr$_3$ and FAPbBr$_3$ as shown in Fig. S10 and Fig. S11, respectively. The information about $\xi_\perp$ is predominantly captured in the horizontal cross-section of the diffuse rods across the R-M direction in the Brillouin zone. Conversely, $\xi_\parallel$ is discerned from the vertical cross-section of these rods. Thus, in MAPbBr$_3$ we present a 1D cross-section of the 2D $S(\mathbf{q})$ HK1.5 at K = 2.5 (along the rod) and K = 1.5 (vertical to the rod), in Fig. S10 **e**. We observe great agreement between the experiment and simulation, which implies that the correlation length values we derive are accurate. The same is true for HK0.5 plane. Similar consideration can be found for FAPbBr$_3$ in e.g. Fig. S11 **e**, which shows 1D cross sections at certain

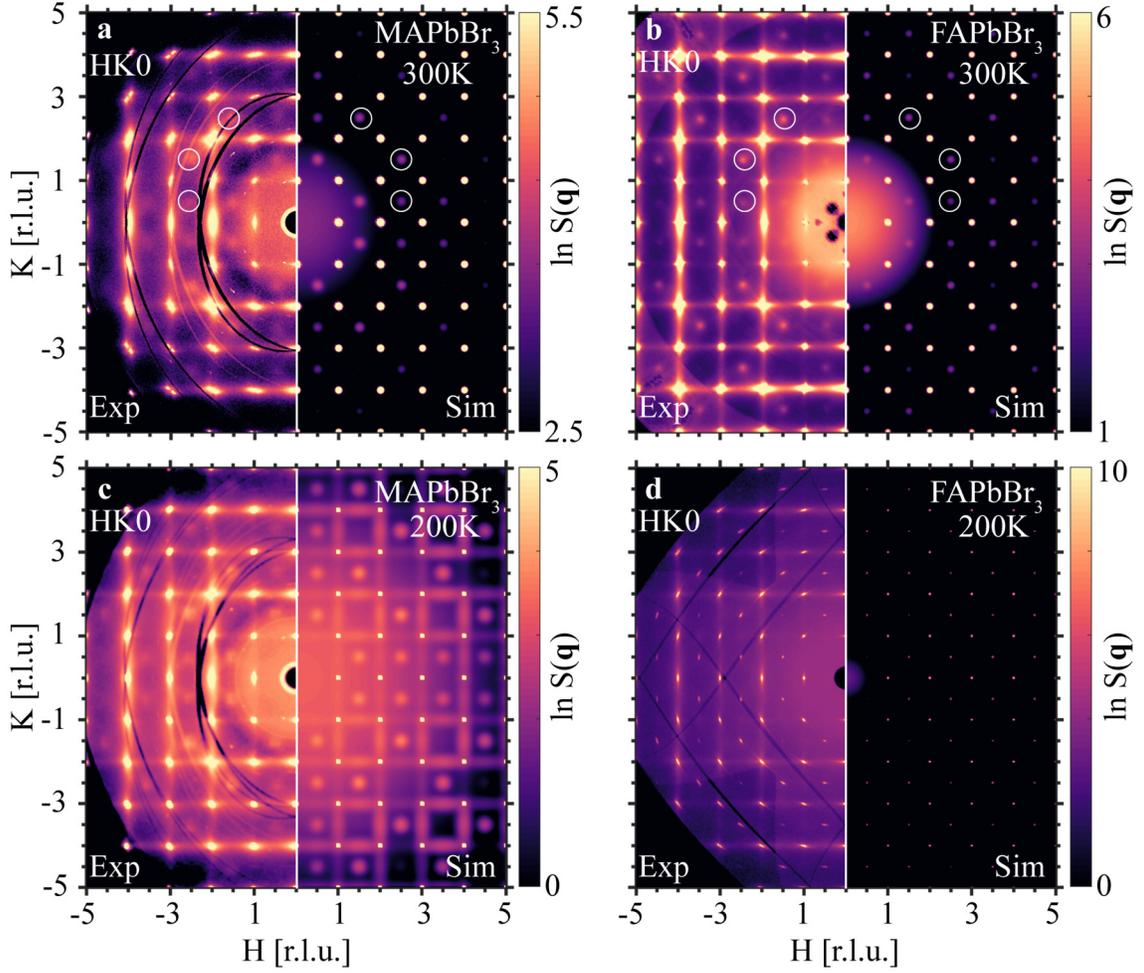

FIG. S9. **Relationship between phase transitions and diffuse scattering in HK0 planes.** Experimental total diffuse scattering (left panels) and simulated QEDS (right panels) signals across HK0 reciprocal space planes in MAPbBr$_3$ and FAPbBr$_3$ are shown in (**a**)-(**d**). Our simulations do not capture TDS signals that are present in HK0, however, the broadened QEDS superstructure reflections are captured in (**a**)-(**d**). Diffuse superstructure reflections at 300 K at M points in the Brillouin zone are circled in (**a**) and (**b**).

K values of the HK1.5 plane. While the peaks in FAPbBr$_3$ are more isotropic, we can still compare cross sections along the rod $K = 3.5$ and vertical to the rod $K = 1.5$. The agreement between the simulation and experiment gives confidence in the accuracy of our estimated correlation lengths. We conducted two independent experiments to characterise the local structure of FAPbBr$_3$ at 300 K. We used both laboratory X-ray source (CuKα radiation to avoid fluorescence from bromine) and synchrotron radiation to verify that in both cases we observe the same diffuse scattering patterns. The results are presented in Fig. S12. We further used both datasets to extract the correlation lengths. Using the X-ray source dataset we obtain $\xi_\perp = 19.686$ Å and $\xi_\parallel = 29.106$ Å and with the synchrotron source we obtain $\xi_\perp = 19.969$ Å and $\xi_\parallel = 27.191$ Å, showing remarkable agreement.

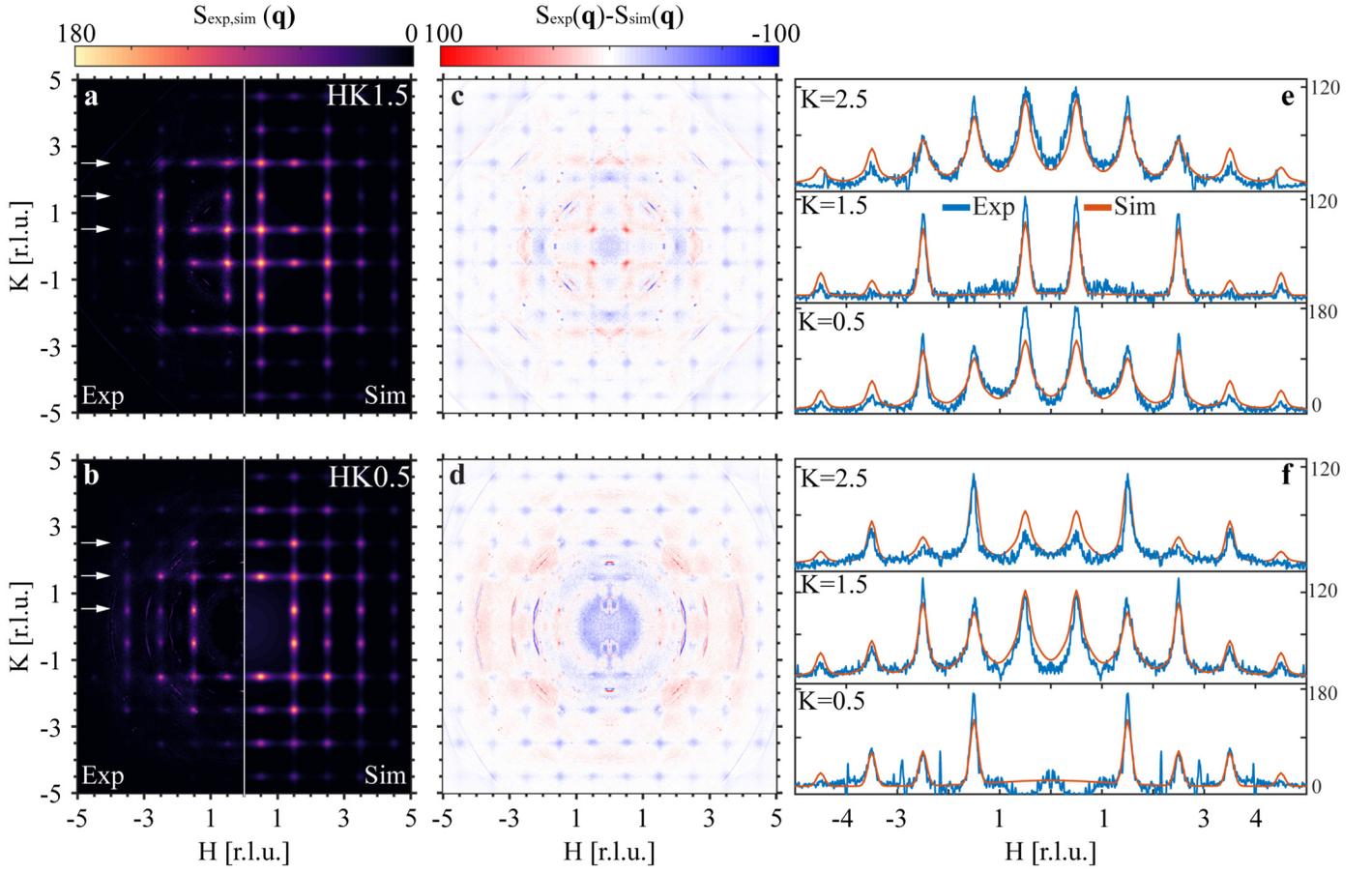

FIG. S10. **Comparison of experimental and modelled quasi-elastic diffuse scattering signals in MAPbBr$_3$ at T = 300 K.** The panels on the left represent experimental scattering functions $S_{exp}(\mathbf{q})$ and those on the right illustrate the simulated scattering function $S_{sim}(\mathbf{q})$ for the $HK1.5$ (**a**) and HK0.5 (**b**) reciprocal space planes of MAPbBr$_3$. The difference between the experimental and simulated data for HK1.5, HK0.5 reciprocal space planes is displayed in (**c**) and (**d**), respectively. Cross sections along $H$ at constant $K = 2.5$, $K = 1.5$ and $K = 0$ in $HK1.5$ (**e**), HK0.5 (**f**) reciprocal space planes. The arrows in (**a**) and (**b**) indicate the directions of the displayed cross sections.

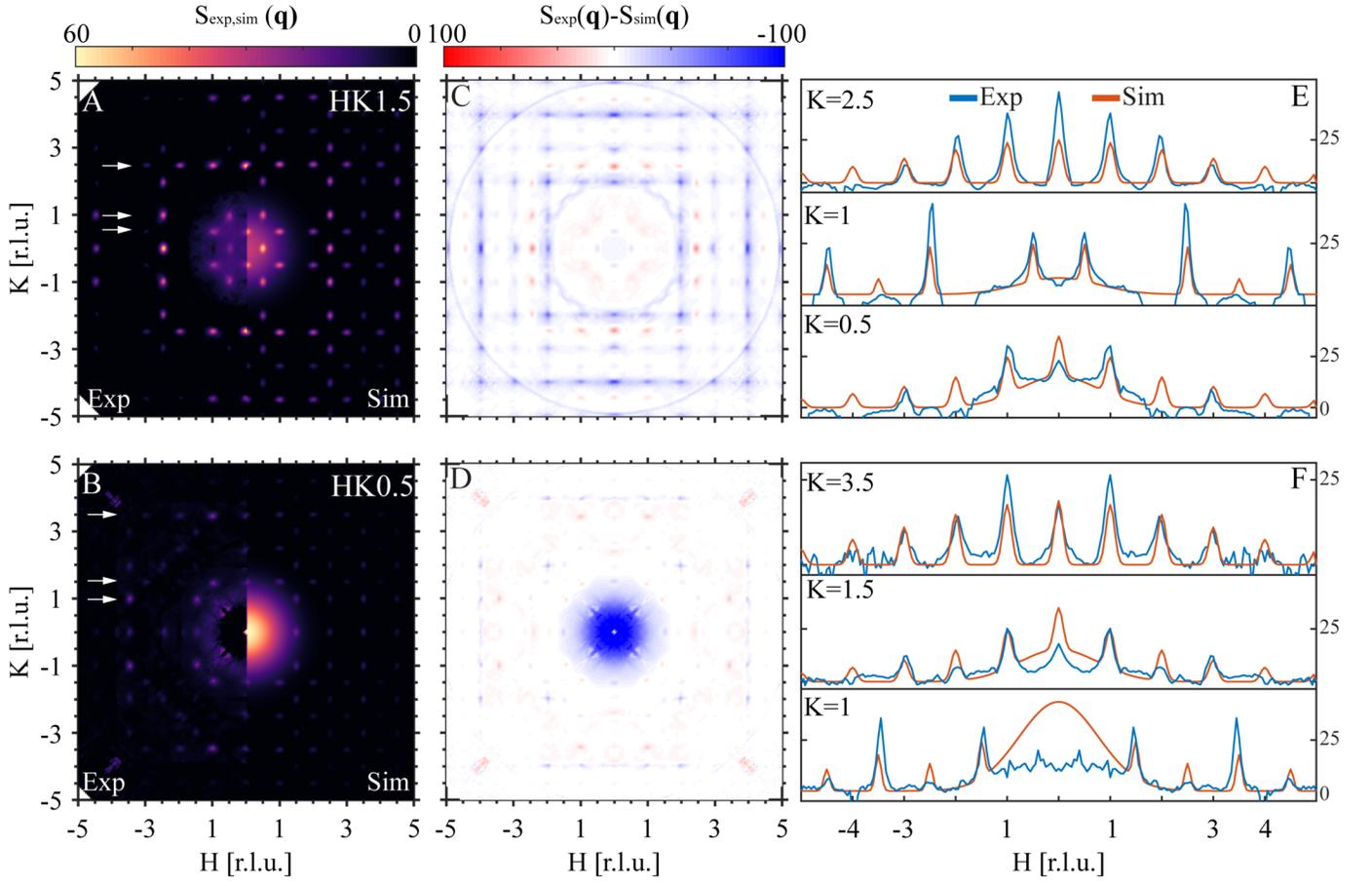

FIG. S11. **Comparison of experimental and modelled quasi-elastic diffuse scattering signals in FAPbBr$_3$ at** $T = 300$ K. The panels on the left represent experimental scattering function $S_{exp}(\mathbf{q})$ and those on the right illustrate the simulated scattering function $S_{sim}(\mathbf{q})$ for the $HK1.5$ (**a**) and HK0.5 (**b**) reciprocal space planes of FAPbBr$_3$. The difference between the experimental and simulated data for HK1.5, HK0.5 reciprocal space planes is displayed in (**c**) and (**d**), respectively. Cross sections along $H$ at constant $K$ values in $HK1.5$ (**e**), HK0.5 (**f**) reciprocal space planes. The arrows in (**a**) and (**b**) indicate the directions of the displayed cross sections.

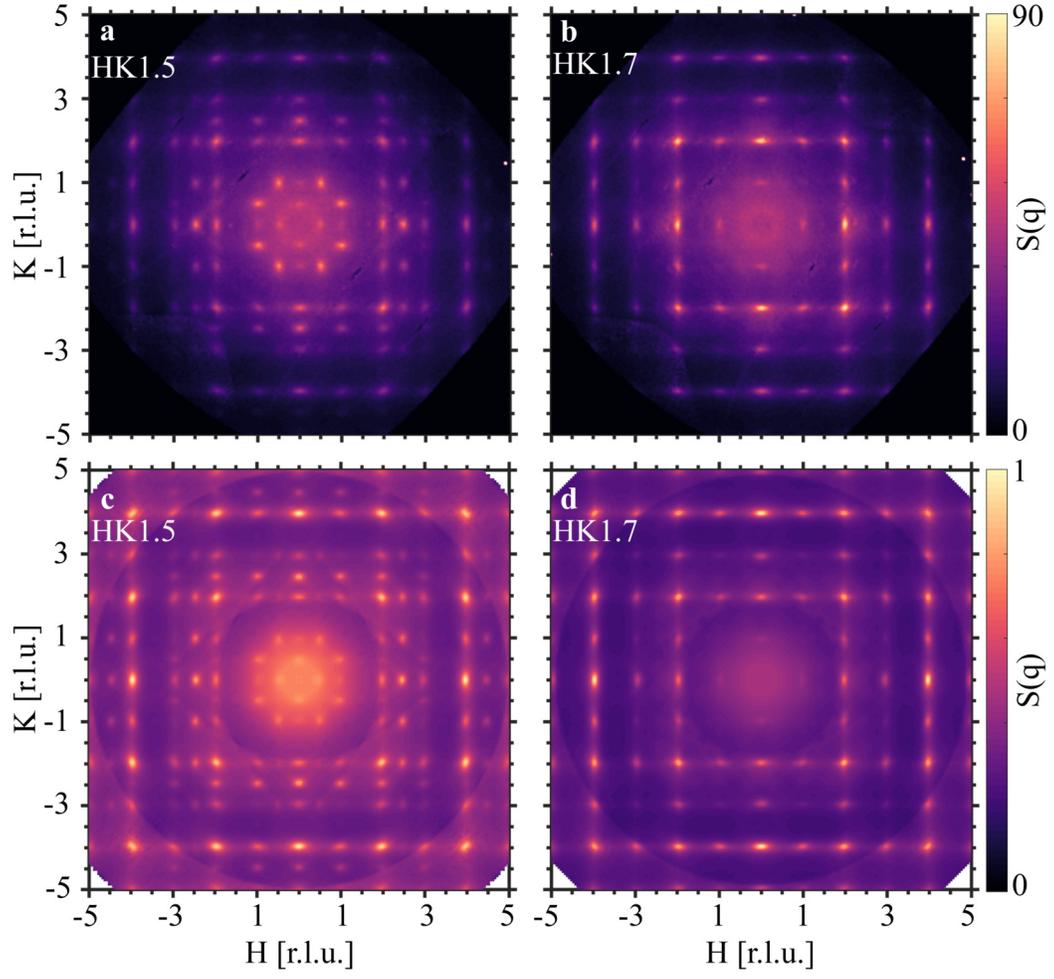

FIG. S12. **Two independent measurements confirm the presence of the same diffuse scattering patterns in FAPbBr₃ at** 300 K. (**a**) and (**b**), show $S(\mathbf{q})$ of HK1.5 and HK1.7 reciprocal space planes, respectively, obtained using synchrotron radiation at MX1 at Australian Synchrotron. (**c**) and (**d**), show $S(\mathbf{q})$ of HK1.5 and HK1.7 reciprocal space planes, respectively, obtained using a laboratory X-ray source (CuKα radiation).

## v. MOLECULAR DYNAMIC SIMULATIONS

### A. Comparison of experimental and MD computed diffuse scattering patterns

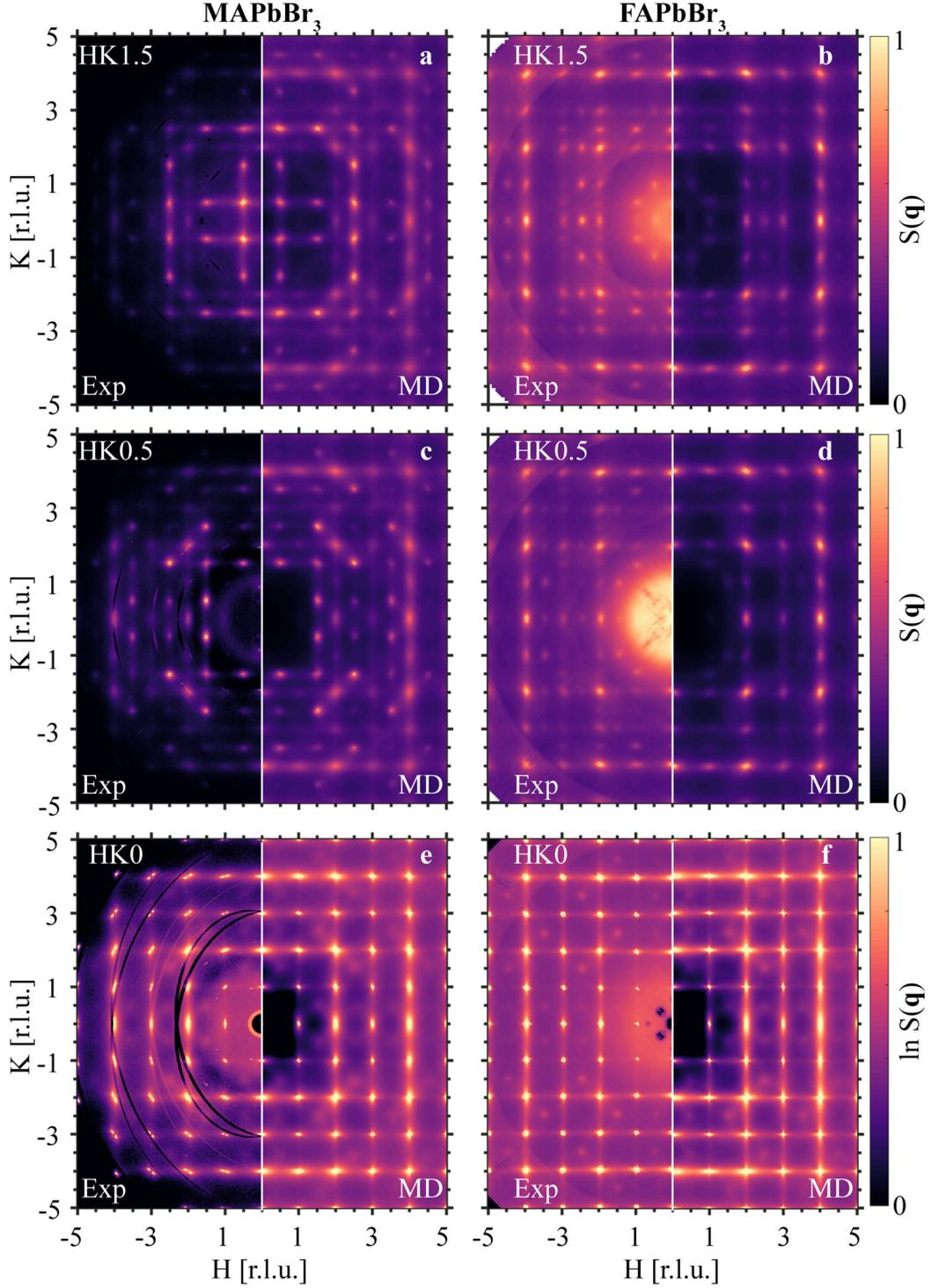

FIG. S13. **Comparison of experimental and MD computed diffuse scattering in HK1.5 planes.** Experimental $S(\mathbf{q})$ and MD simulated (right panels) signals across HK1.5 reciprocal space planes show great agreement, in both MAPbBr$_3$ and FAPbBr$_3$. **(a)** and **(b)** comparison of experimental and MD $S(\mathbf{q})$ in HK1.5 in MAPbBr$_3$ and FAPbBr$_3$ at $T = 300$ K, respectively. **(c)** and **(d)** comparison of experimental and MD $S(\mathbf{q})$ in HK0.5 in MAPbBr$_3$ and FAPbBr$_3$ at $T = 300$ K, respectively. **(e)** and **(f)** comparison of experimental and MD $S(\mathbf{q})$ in HK0 in MAPbBr$_3$ and FAPbBr$_3$ at $T = 300$ K, respectively. To make a more fair comparison with the 300 K experimental diffuse scattering patterns we utilised MD simulation at 340 K with rationale behind it given in Section vii.

## B. The origin of quasi-elastic diffuse scattering at X points

In Fig. S14 **a-d**, we present MD-simulated $S(\mathbf{q})$ across two planes, namely HK1 and HK1.5, for MAPbBr$_3$ and FAPbBr$_3$ at 300 K, highlighting high-symmetry points across the Brillouin zones. To encompass all the high-symmetry points, it is essential to consider two parallel 2D planes separated by 0.5 reciprocal lattice units (r.l.u.), which is depicted in Fig. S14 **e**, showcasing the first Brillouin zone of a pseudocubic real unit cell. We determine that the diffuse scattering in MAPbBr$_3$ emanates from the broadened superstructure diffuse peaks at R points, whereas in FAPbBr$_3$, these peaks are located at M points, as demonstrated in Fig. S14 **a-d**. Zone edge acoustic phonons at R points are characterised by pure out-of-phase octahedral tilt modes, while at M points, they are indicative of pure in-phase octahedral tilts, as illustrated in Fig. S14 **e**. This aligns with our concluding findings that the local structure in MAPbBr$_3$ and FAPbBr$_3$, in the average cubic phase, consists of local $I4/mcm$ and local $P4/mbm$ phases, respectively.

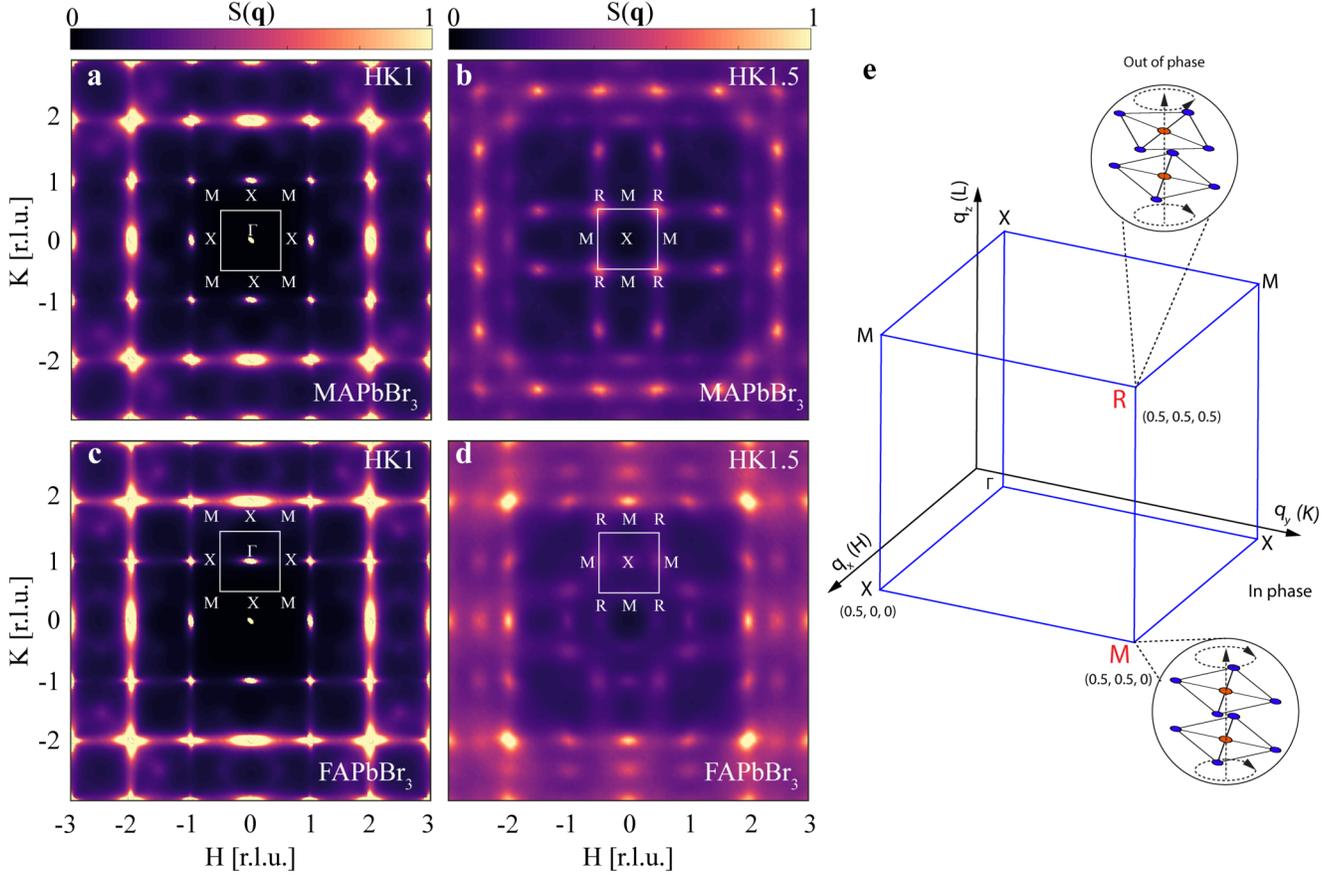

FIG. S14. **Visualisation of high symmetry points in the Brillouin zone and their relationship with diffuse scattering in halide perovskites.** (**a**) and (**b**) MD simulated $S(\mathbf{q})$ across HK1 and HK1.5 planes in MAPbBr$_3$ at 300 K. High symmetry points across four Brillouin zones centered at (001) are depicted. This approach facilitates the visualisation of the relationship between diffuse scattering and high symmetry points. (**c**) and (**d**) MD simulated $S(\mathbf{q})$ across HK1 and HK1.5 planes in FAPbBr$_3$ at 300 K along with high symmetry points. (**e**) 3D representation of the first Brillouin zone of a pseudocubic real unit cell. In halide perovskites, zone edge acoustic phonon modes at R points are pure out-of-phase tilts along the c-axis, whereas at M points, they are pure in-phase tilts along the c-direction.

In the Main Text Fig.1 MD simulations reveal that QEDS in HK1.5 planes of MAPbBr$_3$ and FAPbBr$_3$, energy-integrated from $-1$ to 1 meV shows pronounced R and M-point peaks. Crucially, it is also observed that the patterns includes X-point peaks, which are significantly more pronounced in FAPbBr$_3$ than in MAPbBr$_3$. As we established that dynamic local nanodomains of lower symmetry fluctuating in higher symmetry phase scatter quasi-elastically with X-rays and give rise to R and M QEDS peaks, this would imply that there is also additional local order that gives rise to X-point QEDS scattering. To elucidate the origins of this X-point QEDS signal, we have performed an integration of the MD-computed $S(\mathbf{q}, E)$ over a narrow energy range from 0 to 0.05 meV, as illustrated in Fig. S15, to primarily capture the elastic scattering component. In Fig. S15 **b** and **d** we observe the same diffuse patterns as in Main Text Fig.1 **a** and **c**, bottom right quadrants, confirming that the X scattering has a pure elastic component.

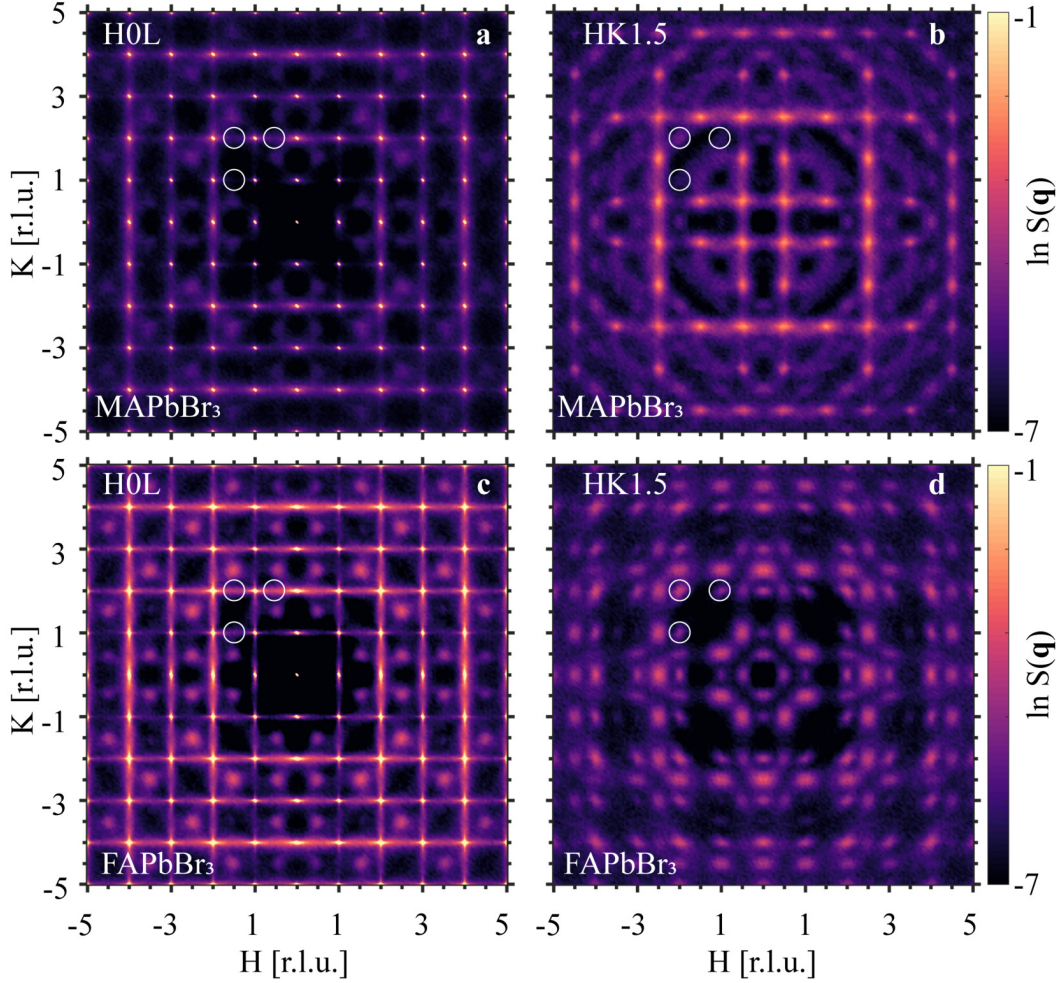

FIG. S15. **MD simulated $S(\mathbf{q})$ obtained by integrating $S(\mathbf{q}, E)$ over $0 < E < 0.05$ meV.** (**a**) and (**b**) $S(\mathbf{q}, E)$ across H0L and HK1.5, respectively, integrated over $0 < E < 0.05$ meV, for MAPbBr$_3$ at 300 K. (**c**) and (**d**) $S(\mathbf{q}, E)$ across H0L and HK1.5, respectively, integrated over $0 < E < 0.05$ meV for FAPbBr$_3$ at 300 K. X points in across the Brillouin zones are denoted with white circles.

In the H0L planes, as illustrated in Fig. S15 panels **a** and **c**, we observe quasi-elastic diffuse scattering (QEDS) at notably low energy transfers. This scattering manifests as rods originating from $\Gamma$ points within the Brillouin zone (at Bragg peaks) extending along the $\Gamma - X - \Gamma$ direction. Given that H0L planes intersect perpendicularly with HK1.5 planes, these $\Gamma - X - \Gamma$ rods intersect the latter at X points, thereby creating observable peaks at these intersections, which are effectively cross-sections of the rods.

By examining two q-cross-sections in H0L centered at the (020) reflection, one vertical in a direction $[0 + q, 2, 0]$ and the other horizontal in a direction $[0, 2 + q, 0]$ (the directions are marked in Fig. S16 **b** and plotting the corresponding $S(\mathbf{q}, E)$ for each, we observe distinct phonon dispersions presented in Fig. S16 **a** and Fig. S16 **e**, respectively. The vertical cross-section reveals the dispersion of longitudinal acoustic (LA) phonons, as shown in Fig. S16 **a**, while the horizontal cross-section exposes the dispersion of transverse acoustic (TA) phonons, as illustrated in Fig. S16 **b**. This distinction arises from the polarization selection rules in $S(\mathbf{q}, E)$, where the vertical cross-section, being longitudinal to the (020) Bragg peak, contains information solely about LA phonons. Conversely, the horizontal direction is associated with TA phonons.

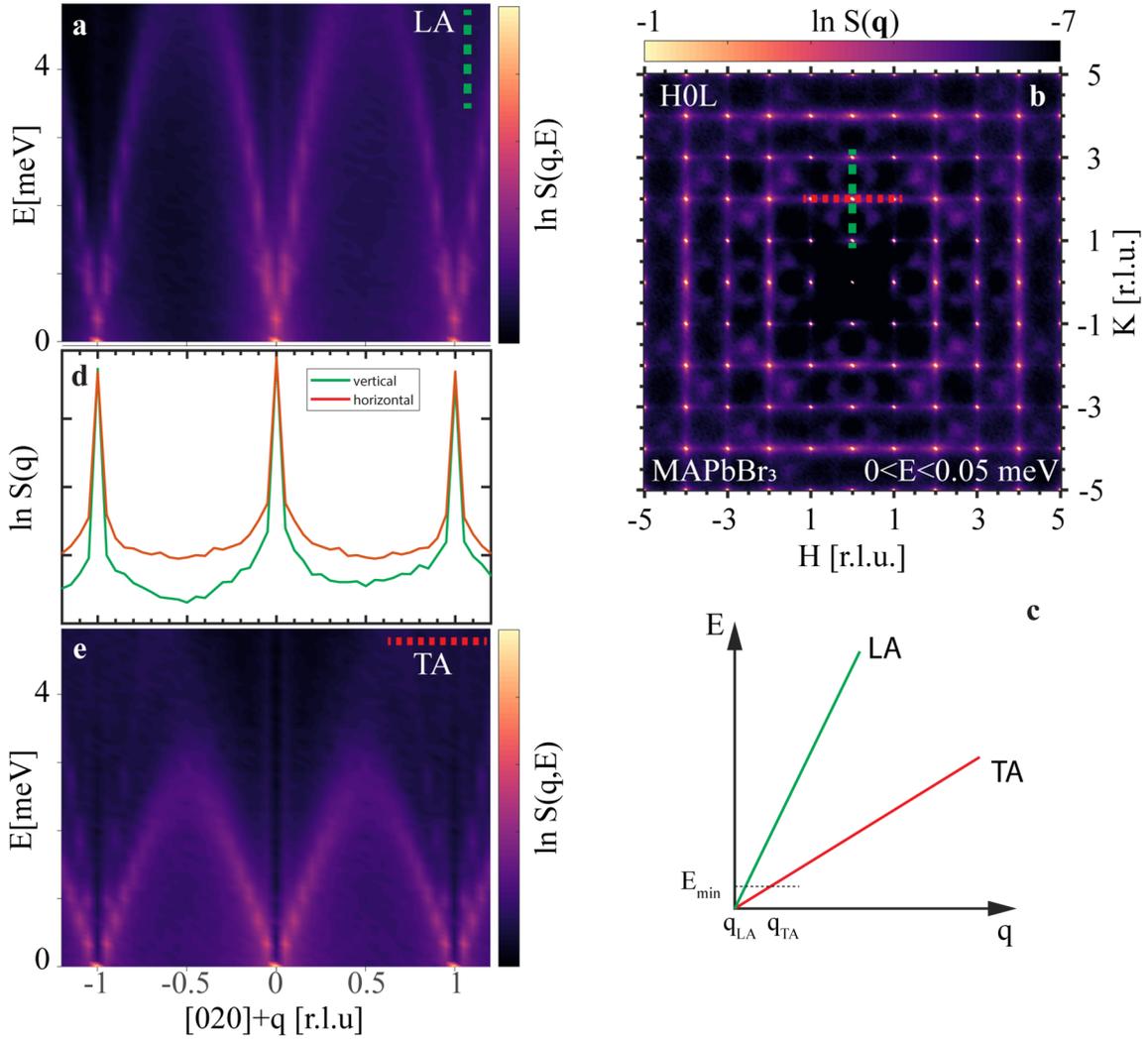

FIG. S16. **The origin of the QEDS scattering in H0L plane demonstrated on the example of MD MAPbBr₃**
$S(\mathbf{q}, E)$ **data.** (**a**) $S(q, E)$ where $q$ is along $[0 + q, 2, 0]$ direction. This map reveals the dispersion of LA phonons. (**e**) $S(q, E)$ where $q$ is along $[0, 2 + q, 0]$ direction. This map reveals the dispersion of TA phonons. The corresponding directions are denoted in **b** in H0L reciprocal space plane where $S(\mathbf{q})$ was obtained by integrating $S(\mathbf{q}, E)$ in energy in a range from 0 to 0.05 meV. (**d**) $S(q)$ along $[0, 2 + q, 0]$ (horizontal) and $[0 + q, 2, 0]$ (vertical) direction with E integrated from 0 to 0.05 meV. Intensities are shown in log scale. (**c**) Demonstration of zone centre LA and TA phonon dispersion. $q_{TA}$ and $q_{LA}$ are wavevectors of TA and LA phonons at $E_{min}$. $q_{TA}$ is always higher than $q_{LA}$.

As observed in Fig. S16 **b**, $\Gamma - \mathrm{X} - \Gamma$ rods emerge along the transverse directions. By integrating the $S(\mathbf{q}, E)$ over an energy range from 0 to 0.05 meV for both vertical and horizontal directions, we observe that while Bragg peaks at integer $q$ values are sharp in the vertical direction, they exhibit broadening in the horizontal (transverse) direction, as shown in Fig. S16 **d**. The broadened Bragg peaks in the transverse direction create a 'valley' between them which we observe as rods of diffuse scattering along $\Gamma - \mathrm{X} - \Gamma$ trajectory. Let us now understand what causes the broadening of these Bragg peaks in $q$ along the transverse direction. By definition, TA phonons are lower energy excitations compared to LA phonons. Thus if we select a very low energy near the zone centre like $E_{min}$ in Fig. S16 **c**, the corresponding wavevector $q$ of TA phonons is always going to be larger than $q$ of LA phonons for the same $E_{min}$. As seen in Fig. S16 **d**, at Brillouin zone centre we expect a very sharp Bragg peak (as it corresponds to infinite correlations of atoms) and a contribution from low energy zone centre acoustic phonons. For TA phonons, the phonon peak will be displaced further away from the adjacent Bragg peaks compared to LA phonons, thus the Bragg peaks will appear as more broadened in the transverse direction. Consequently, zone centre TA phonons invariably broaden Bragg peaks more than LA phonons, leading to the formation of rods in the transverse direction. So, to conclude, zone centre acoustic phonons are responsible for the QEDS $\Gamma - \mathrm{X} - \Gamma$ rods and as a result, also are responsible for the emergence of QEDS X-point peaks in HK1.5 planes. While our analysis has elucidated the emergence of X point peaks using MAPbBr$_3$ as a example, this phenomenon is similarly observed in FAPbBr$_3$. It is important to highlight that FAPbBr$_3$ exhibits softer phonons compared to MAPbBr$_3$ (i.e. LA and TA phonons in FAPbBr$_3$ are always lower in energy than in MAPbBr$_3$), leading to a more pronounced broadening of Bragg peaks along the transverse direction. This can be clearly observed by comparing Fig. S15 a and b, where $\Gamma - X - \Gamma$ rods are more pronounced in FAPbBr$_3$ than in MAPbBr$_3$. Consequently, this results in higher intensity X point peaks in FAPbBr$_3$, at HK1.5 planes.

## vi. PROCEDURE TO SEPARATE TDS AND QEDS IN THE EXPERIMENTAL DATA AND THE PHYSICAL INTERPRETATION OF THEIR RELATIVE INTENSITIES

In our analysis, we distinguish the quasi-elastic diffuse scattering (QEDS) component within the HKn plane of reciprocal space by subtracting the total diffuse scattering observed at HKn from the values measured at HKn + 0.2, where $n$ is a half-integer. This approach assumes that the HKn + 0.2 plane is predominantly composed of Thermal Diffuse Scattering (TDS) signals. Utilising the $S(\mathbf{q}, E)$ derived from molecular dynamics (MD) trajectories, we validate this assumption for the materials under investigation. Fig. S17(a) and (c) present the MD-derived $S(\mathbf{q})$ for FAPbBr$_3$ at 300 K at HK1.5 and HK0.5, respectively, obtained by integrating $S(\mathbf{q}, E)$ over the energy range of 1 to 10 meV. This integration captures the TDS signal, attributed primarily to zone edge acoustic and optical phonons, while effectively isolating the QEDS components. The resemblance of these MD-derived diffuse scattering patterns to the experimentally collected patterns at HK1.7 and HK0.7, as shown in Fig. S17(b) and (d), confirms that the experimental HKn + 0.2 planes predominantly contain TDS. Thus, our subtraction method for isolating QEDS components is substantiated.

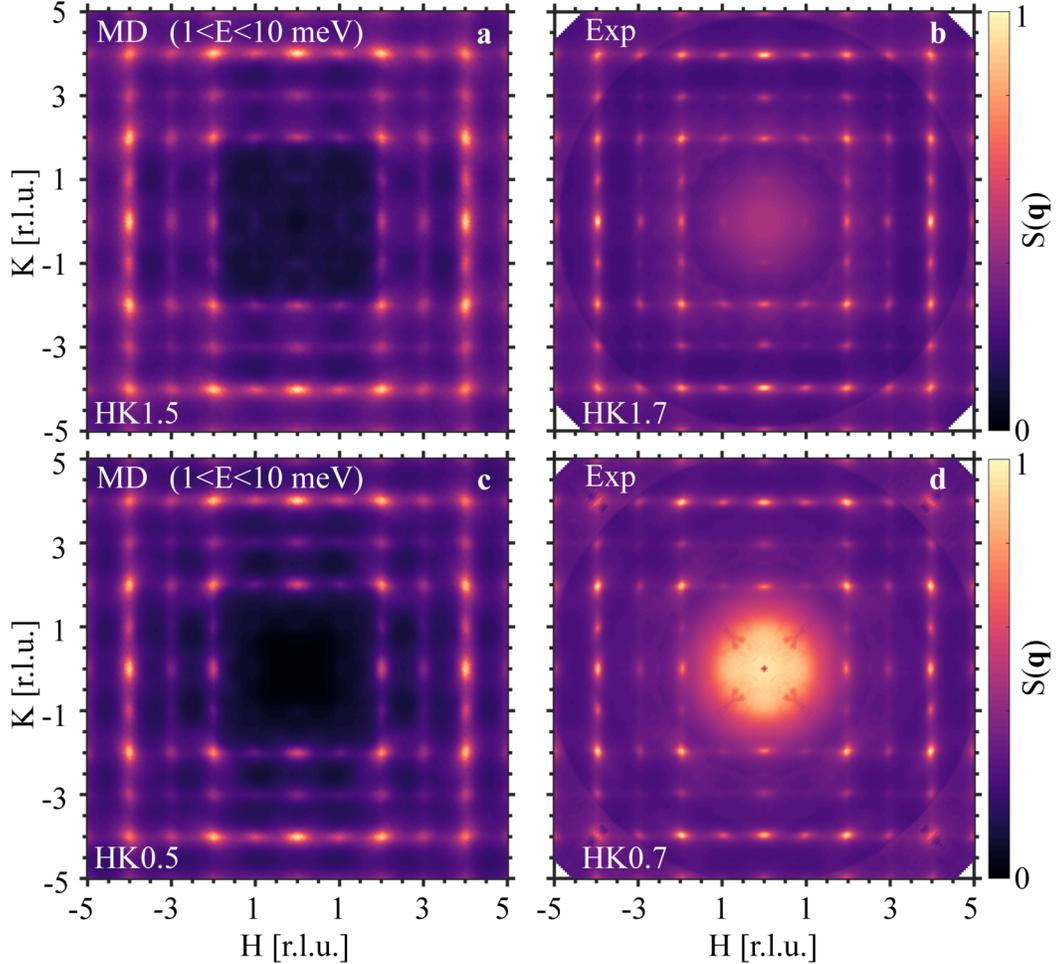

FIG. S17. **Comparison of MD simulated and experimental TDS in FAPbBr$_3$ at** 300 K. (**a**) and (**c**), show MD derived HK1.5 and HK0.5 $S(\mathbf{q})$ in FAPbBr$_3$ at 300 K obtained by integrating $S(\mathbf{q}, E)$ in energy range from 1 to 10 meV. (**b**) and (**d**), show experimental HK1.7 and HK0.7 $S(\mathbf{q})$ in FAPbBr$_3$ at 300 K.

TDS manifests across every two-dimensional cross-section of reciprocal space. In contrast, the local structural phenomena contributing to QEDS manifest predominantly in half-integer planes. Consequently, while a plane such as HK1.5 contains both QEDS and TDS, a plane like HK1.7, being sufficiently distant from HK1.5, is expected to encompass only TDS. We provide a further rationale for confidently attributing the scattering intensity at HKX + 0.2 specifically to TDS. In Fig. S18, we present comparative data of total diffuse scattering in the HK1.5 and HK1.7 planes at 300 K and 200 K. The QEDS signal observed at 300 K (Fig. S18 **a**) converges to Bragg peak scattering at 200 K (Fig. S18 **b**), while the TDS patterns remain consistent through the phase transition as shown in (Fig. S18 **c** and **d**). The same TDS pattern is also evident in the HK1.5 plane at 200 K.

At lower temperatures, a reduction in TDS intensity is anticipated due to the diminishing phonon population, as evident in Fig. S18 **c** and **d**. Furthermore, the 1D cross-section at K = 2 for these two temperatures, shown in Fig. S18 **e**, corroborates this observation. Additionally, the quadratic dependence of $S(\mathbf{q})$ on $\mathbf{q}$ for harmonic phonons, characteristic of TDS, is evident in Fig. S18 **f**. Here, the cross-section at K = 2 in different 2D planes, separated by 0.2 in the L direction of reciprocal space, shows a variance in intensity, with a higher intensity at the elevated $\mathbf{q}$ value plane, aligning with the expected behaviour of TDS.

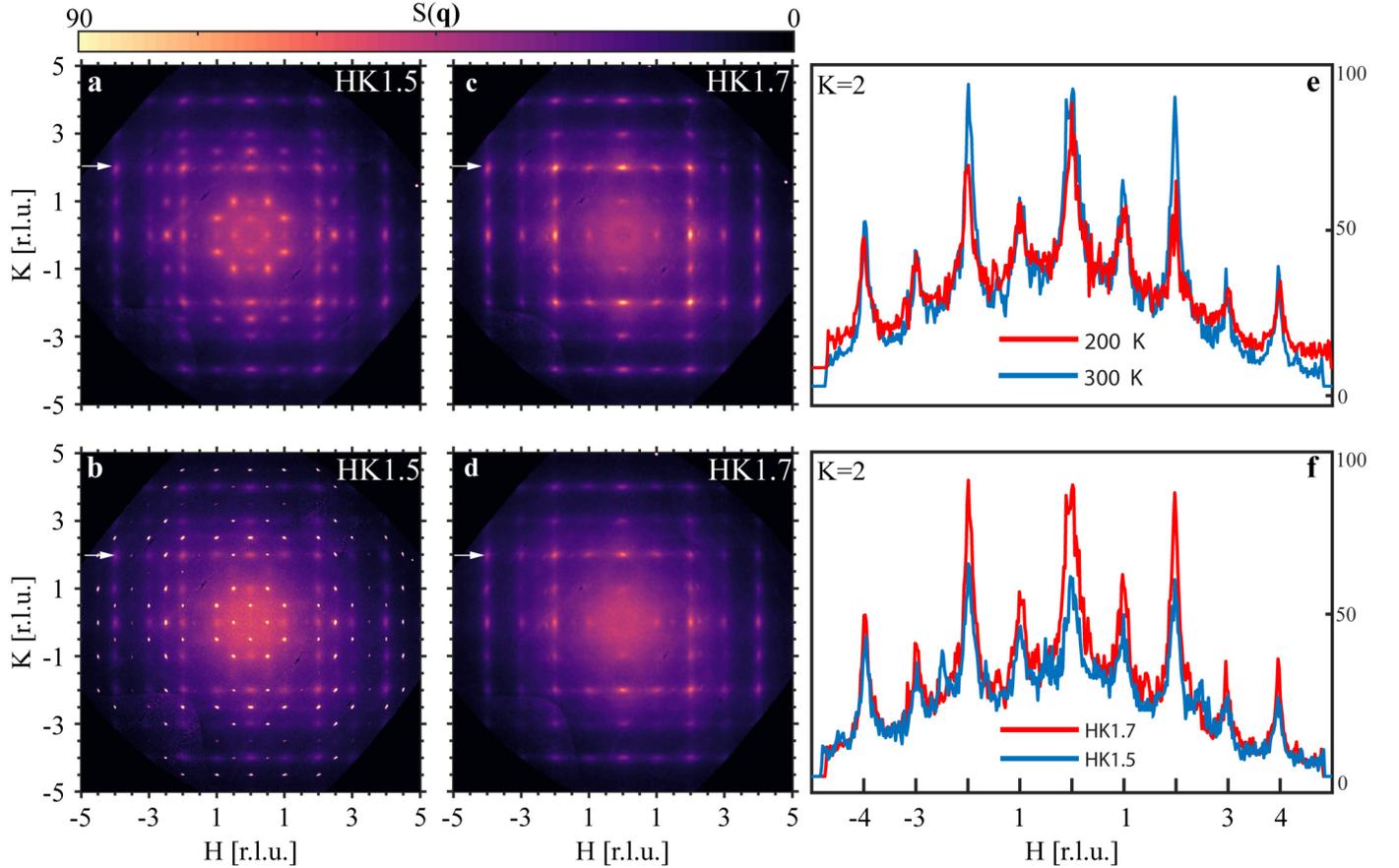

FIG. S18. **The observation of TDS in FAPbBr$_3$.** (**a**) and (**b**) show $S(\mathbf{q})$ of HK1.5 reciprocal space planes at 300 and 200 K, respectively. (**c**) and (**d**), show $S(\mathbf{q})$ of HK1.7 reciprocal space planes at 300 and 200 K, respectively. (**e**) 1D cross sections at $K = 2$ of $HK1.7$ planes at 300 and 200 K. (**d**) 1D cross sections at $K = 2$ of $HK1.5$ $HK1.7$ planes at 300 K. The data is obtained using synchrotron radiation at MX1 at Australian Synchrotron.

In the investigation of diffuse scattering contributions, Inelastic Neutron Scattering (INS) plays a crucial role in classifying TDS and QEDS. INS enables the measurement of excitations' energies associated with diffuse scattering. Particularly, QEDS is attributed to near-static excitations, detectable in the elastic regime centered at zero energy transfers. In Fig. S19 **a** and **b** we mark the directions in the reciprocal space indicated by white rectangles, in which we performed line scans using triple axis spectrometer, in both elastic and inelastic regime. With single crystal X-ray diffraction employed to generate these $S(\mathbf{q}, E)$ maps we are unable to differentiate elastic from inelastic scattering and as a result, both TDS and QEDS signals will be captured. However, INS selectively filters QEDS, as evidenced in Fig. S19 **c** and **d** where the observed peaks at zero energy transfers correspond to QEDS identified in the X-ray 2D $S(\mathbf{q})$ maps. Fig. S19 **e** and **f** demonstrate that X-ray scattering captures both TDS and QEDS peaks, corroborating this classification approach.

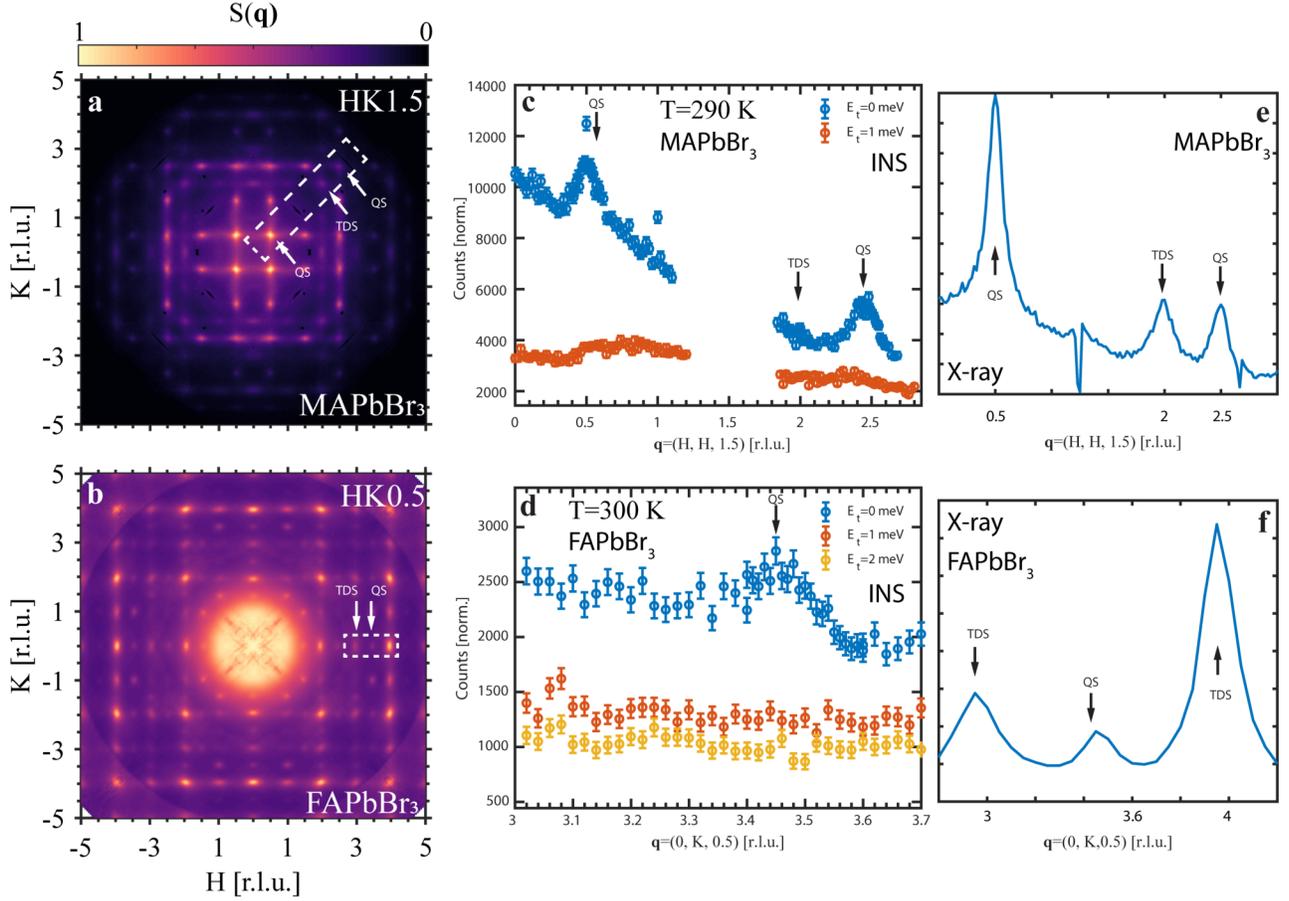

FIG. S19. **Comparison of TDS and QEDS from INS and X-ray scattering.** (**a**) shows $S(\mathbf{q})$ of HK1.5 MAPbBr$_3$ reciprocal space plane at 300. (**b**) shows $S(\mathbf{q})$ of HK0.5 FAPbBr$_3$ reciprocal space plane at 300. The regions which were scanned with INS are denoted with a white rectangle. Arrows show TDS and QEDS peaks. (**c**) Elastic ($E_t = 0$ meV) and inelastic ($E_t = 1$ meV) q scans across the $[-0.5, -0.5, L]$ direction of MAPbBr$_3$ at ($T = 290$ K). (**d**) Elastic ($E_t = 0$ meV) and inelastic ($E_t = 1$ meV, $E_t = 2$ meV) q scans across the $[0, K, 0.5]$ direction of FAPbBr$_3$ at ($T = 300$ K). (**e**) The same 1D scan as in (**c**) but extracted from 2D X-ray $S(\mathbf{q})$ in (**a**). (**f**) The same 1D scan as in (**e**) but extracted from 2D X-ray $S(\mathbf{q})$ in (**b**).

In the main text, we claim that the probability of finding local dynamic nanodomains in FAPbBr$_3$ is lower than in MAPbBr$_3$. This claim is grounded in the observation that QEDS signals are significantly weaker in FAPbBr$_3$ than in MAPbBr$_3$. The contrasting values in TDS and QEDS intensity are evident when comparing the $S(\mathbf{q})$ between MAPbBr$_3$ and FAPbBr$_3$ in the top quadrants of Main Text Fig.1 **a** and **c**, where in both materials X point scattering approximates the TDS signal and R and M point scattering represents the QEDS signals in MAPbBr$_3$ and FAPbBr$_3$, respectively. The R diffuse peaks are significantly stronger than the X peaks in MAPbBr$_3$, whereas in FAPbBr$_3$, their intensities are comparable. Thus, the intensity of the QEDS signal can serve as a measure of disorder and dynamic local symmetry breaking in halide perovskite materials.

To facilitate a direct comparison of TDS and QEDS intensities between these two compounds, we have integrated $S(\mathbf{q})$ in a narrow $\mathbf{q}$ range originating in three distinct regions in reciprocal space each of which is expected to be dominated by either TDS, QEDS or Bragg scattering. We denote the integrated intensity with $S$. This integration has been performed for several representative peaks, and the results are detailed in Table V and Table VI. In our analysis, we introduced a column named "Ratio" in which we divide the obtained scattering function $S(\mathbf{q})$ by the average integrated intensity of representative Bragg peaks. This approach allows for a more equitable comparison between MAPbBr$_3$ and FAPbBr$_3$, considering potential variations in their structure factors and the total volumes of the crystals we measured. From the data presented in the two tables, we observe distinct trends in the relative strengths of QEDS and TDS for each compound. Specifically, in MAPbBr₃, QEDS appears to be approximately twice as strong as TDS. In contrast, for FAPbBr₃, TDS is about 1.35 times stronger than QEDS. QEDS intensity in MAPbBr₃ is around 4 times stronger than in FAPbBr₃, aligning with our hypothesis that the density of dynamic nanodomains in MAPbBr₃ is higher than FAPbBr₃. The same analysis has been employed for a number of different peaks where the same trend was observed.

TABLE V. Quantitative comparison between TDS and QEDS intensisites in MAPbBr₃ at 300 K.

| Plane | K and L | Type | $S$ | Ratio |
|-------|---------|------|-----|-------|
| 1.5KL | 1.5 1.5 | QEDS | 200e3 | 8e-3 |
| 1.5KL | 2 1 | TDS | 100e3 | 4e-3 |
| 1.7KL | 2 1 | TDS | 200e3 | 4e-3 |
| 0KL | 2 1 | Bragg | 2.5e7 | |

TABLE VI. Quantitative comparison between TDS and QEDS intensisites in FAPbBr₃ at 300 K.

| Plane | K and L | Type | $S(\mathbf{q})$ | Ratio |
|-------|---------|------|------|-------|
| HK1.5 | 0.5 0.5 | QEDS | 50e3 | 1.7e-3 |
| HK1.5 | 2 2 | TDS | 70e3 | 2.3e-3 |
| HK1.7 | 2 2 | TDS | 100e3 | 3.3e-3 |
| HK0 | 2 2 | Bragg | 3e7 | |

## vii. THE QUANTITATIVE ANALYSIS OF SPATIAL CORRELATIONS FROM MD DATA

Through a quantitative analysis of real-space MD trajectories, we have estimated the volumetric density of dynamic nanodomains within the simulated supercell. As depicted in Fig. S20 **a**, the distribution of tilt angles for MAPbBr₃ and FAPbBr₃ is collected over a specified period of 1 ns. Our empirical evaluation identifies tilt angles below 5 degrees as part of the octahedra's random tilt background, while angles above this threshold are indicative of dynamic local nanodomains exhibiting lower symmetry. By integrating the data presented in the histogram of Fig. S20 **a**, we calculate the volumetric density of octahedra up to a specific tilt angle threshold, represented on the x-axis in Fig. S20 **b**. This analysis reveals a consistently higher density of tilted octahedra in MAPbBr₃ compared to FAPbBr₃, across any chosen threshold. Employing a 5-degree threshold, we find that 19.1% of octahedra are tilted beyond this angle in FAPbBr₃ and 32.9% in MAPbBr₃. Consequently, the volumetric density of dynamic nanodomains is approximately 1.7 times greater in MAPbBr₃ than in FAPbBr₃.

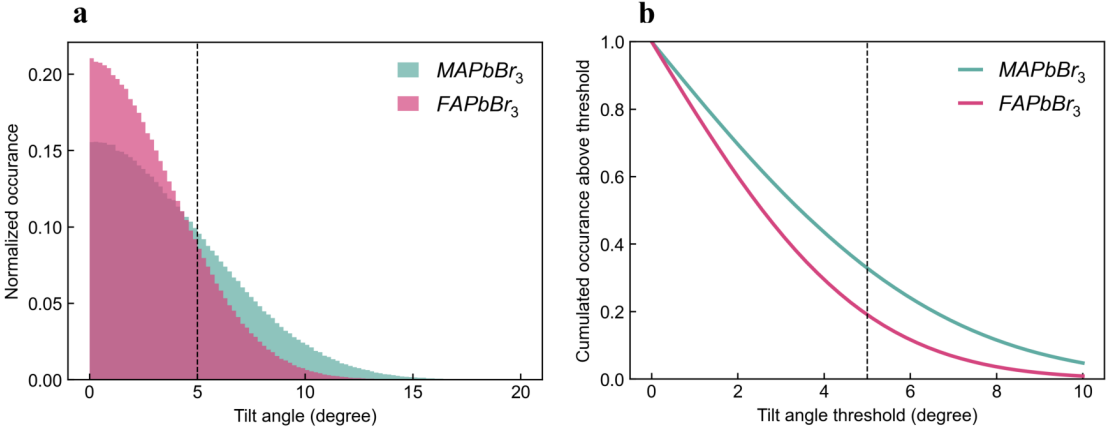

FIG. S20. **The estimation of volumetric density of dynamic nanodomains from MD real space trajectories. a** Histogram of octahedral tilt distributions in MAPbBr₃ and FAPbBr₃ at 300 K with vertical line denoting the threshold angle. **b** Fraction of octahedra that exhibit tilting as a function of threshold tilting angle in MAPbBr₃ and FAPbBr₃ at 300 K.

We further apply quantitative analysis to determine the shape and symmetry of these dynamic nanodomains. From MD trajectories we extract the spatial correlation functions and present them in Fig. S21.

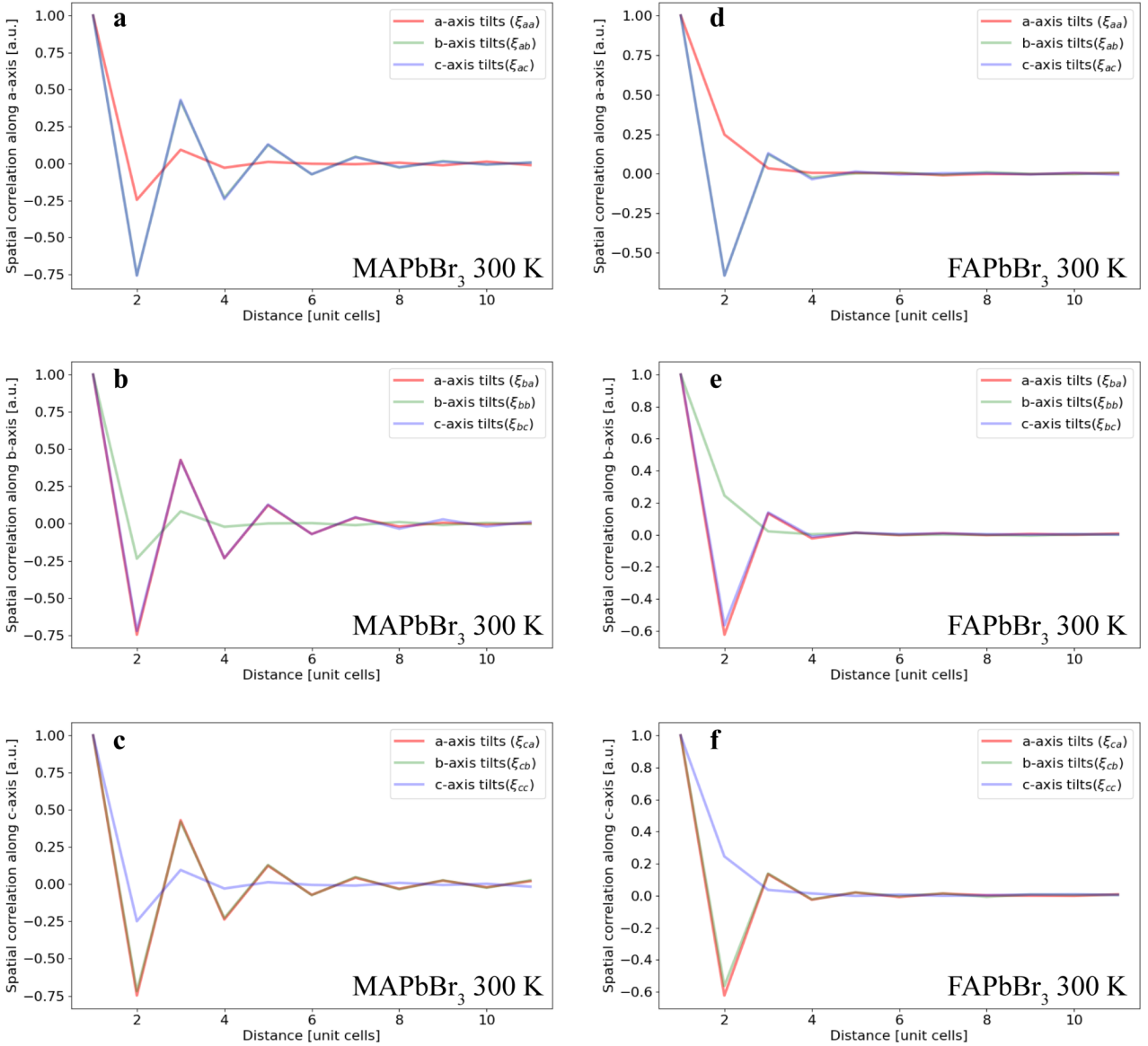

FIG. S21. **Spatial correlation functions determine symmetry and shape of dynamic nanodomains at** $300$ **K.** (**a**)-(**c**) Spatial correlation functions of three types of octahedral tilts along three crystallographic axes for MAPbBr$_3$ at 300 K derived from MD trajectories. For example, in (**a**) we plot three spatial correlation functions. Each function measures spatial correlations of a certain type of octahedral tilt (either $a$, $b$ or $c$-axis tilts) along $a$ crystallographic axis. The corresponding correlation lengths derived from these curves by fitting them to exponentially decaying functions are thus $\xi_{aa}$ $\xi_{ab}$ $\xi_{ac}$. (**d**)-(**f**) Spatial correlation functions of three types of octahedral tilts along three crystallographic axes for FAPbBr$_3$ at 300 K derived from MD trajectories.

The negative correlation value in the plots corresponds to the case of out-of-phase tilting correlation while positive to the in-phase tilting correlation. From the plots, we observe that in MAPbBr$_3$ spatial correlation along axis $n$ is always the shortest for the axis n tilts, while the other two axis tilts exhibit longer correlation. The correlation values alternate in a sign for each case in MAPbBr$_3$ and thus we can conclude that dynamic nanodomains possess local $I4/mcm$ ($a^0a^0c_S^-$) symmetry. However, in the case of FAPbBr3, n axis tilt always exhibits a positive correlation while the two-axis tilts exhibit a negative correlation. This corresponds to local $P4/mbm$ symmetry or ($a^0a^0c_S^+$) in modified Glazer notation. This analysis helps us confirm that the symmetry of the local structure aligns with our experimental results, derived by utilising a classical model.

We fit each spatial correlation function in Fig. S21 to exponential decay function [20] to derive the spatial correlation length tensor $\Xi$. This is because three types of octahedral tilts are possible (rotation along $a$, $b$ or $c$ axis) and we need to track spatial correlations of three tilt types along all three crystallographic axes. The spatial correlation length tensor $\Xi$ derived from spatial correlation functions of MAPbBr$_3$ at 300 K is given below. Each correaltion length value is given as a number of unit cells.

$$\Xi = \begin{bmatrix} \xi_{aa} & \xi_{ab} & \xi_{ac} \\ \xi_{ba} & \xi_{bb} & \xi_{bc} \\ \xi_{ca} & \xi_{cb} & \xi_{cc} \end{bmatrix} = \begin{bmatrix} 0.682 & 2.169 & 2.166 \\ 2.116 & 0.683 & 2.123 \\ 2.094 & 2.148 & 0.696 \end{bmatrix} \tag{S12}$$

where: $\xi_{aa}$, $\xi_{bb}$, and $\xi_{cc}$ are the diagonal components representing normal correlation length $\xi_\perp$ or the half the thickness of dynamic nanodomain disc, and the off-diagonal elements ($\xi_{ab}$, $\xi_{ac}$, etc.) correspond to $\xi_\parallel$, or the radius of the discs. The fact that all diagonal elements have nearly the same value and that also all off-diagonal elements have the same value confirms that these discs are oriented orthogonally relative to each other i.e. form dynamic local twins which we also show in the main text using the classical simulation. Thus to derive average $\xi_\parallel$ and $\xi_\perp$ we average all the off-diagonal and diagonal elements, respectively. Finally, we present a table where we compare correlation length diameters (i.e. two times the MD correlation lengths above) derived from the real space analysis of MD trajectories, single crystal X-ray diffuse scattering. We note that the cubic-tetragonal phase transition temperature we obtain from MD simulations in MAPbBr$_3$ is approximately 270 K (as shown in Fig. S22 **a** when values from heating and cooling cycles are averaged) which is around 35 K higher than the experimentally observed phase transition. Thus, to make a fair comparison we compare the correlation lengths from MD trajectories to the 300 K X-ray values both at the same absolute temperature (300 K) and at approximately the same offset from the phase transition temperature (340 K) MAPbBr$_3$, as presented in Table VII.

| MAPbBr$_3$ | MD 300K | MD 340K | Exp X-ray 300K |
|---|---|---|---|
| $\xi_\perp$ [Å] | 8.5892 | 5.808 | 6.17±0.03 |
| $\xi_\parallel$ [Å] | 23.86 | 20.898 | 21.02±0.1 |
| Ratio | 2.78 | 3.59 | 3.4 |

TABLE VII. Comparison of MD and experimentally derived spatial correlation lengths for MAPbBr$_3$. $\xi_\perp$ and $\xi_\parallel$ are the thickness and diameter, respectively of the disc-shaped dynamic nanodomains. Ratio corresponds to $\frac{\xi_\parallel}{\xi_\perp}$. Note that in the MD case we compare two times the correlation lengths, to represent the diameter, rather than the radius, of the dynamic nanodomains.

We also observe that the diffuse scattering profiles we simulate from MD trajectories at 340 K in MAPbBr$_3$ match better with experimental observations. This is obvious from Fig. S23 where 340 K MD simulated $S(\mathbf{q})$ exhibit more pronounced rods along the R-M direction in the reciprocal space compared to 300 K MD simulated $S(\mathbf{q})$, which we utilise in Main Text Fig.1. This is in agreement with data presented in Table VII where correlation lengths diameters at 340 K MD match better the experimentally derived diameters. Additionally, in Fig. S24 we show that dynamic nanodomain lifetimes are shorter at 300 K compared to lifetimes at 340 K, indicating that the dynamic nanodomains slow down as the tetragonal phase is approached by cooling.

For FAPbBr$_3$, we observe no clear indication of the phase transition. At 200 K, we do see an asymmetry in the tilt correlation towards positive values for all 3 axes, this may indicate that the system is starting to transform towards the $a^+a^+a^+$. As the experimentally observed transition occurs at 265 K, it is thus possible that MLFF underestimates the phase transition temperature, which could explain the discrepancy between experimentally observed and MD derived correlation length diameters as shown in Table VIII. However, observing phase transitions directly from the inherently limited simulation time accessible to MD is often difficult if there is a significant first-order character to the transition. Indeed, in this case, the system can remain in the phase it was initiated in ($a^0a^0a^0$ in this case) for a time that is significantly longer compared to the length of the MD simulation.

Nevertheless, regardless of the MD transition temperature, the ratio of the two correlation length diameters is comparable with experimentally derived, which is also significantly lower than the ratio in MAPbBr$_3$. Thus, we conclude that MAPbBr$_3$ dynamic nanodomains form discs of pancakes as the ratio of normal correlation to parallel is significant, while in FAPbBr$_3$ nanodomains are more isotropic and exhibit lower ratio values. Another factor contributing to the discrepancy between MD-derived and experimental correlation length diameters stems from the methodology employed to obtain correlation lengths. In experiments, correlation lengths are determined by fitting the $S(\mathbf{q})$ at R or M points in the Brillouin zone to a Lorentzian function, thereby exclusively probing the correlation lengths of purely in-phase and out-of-phase tilted local nanodomains. Conversely, when analysing octahedral correlations in real space from MD data, we inherently include contributions from all possible tilt correlations, not just the in- and out-of-phase ones. As a result, the correlation lengths derived using those two methodologies might differ.

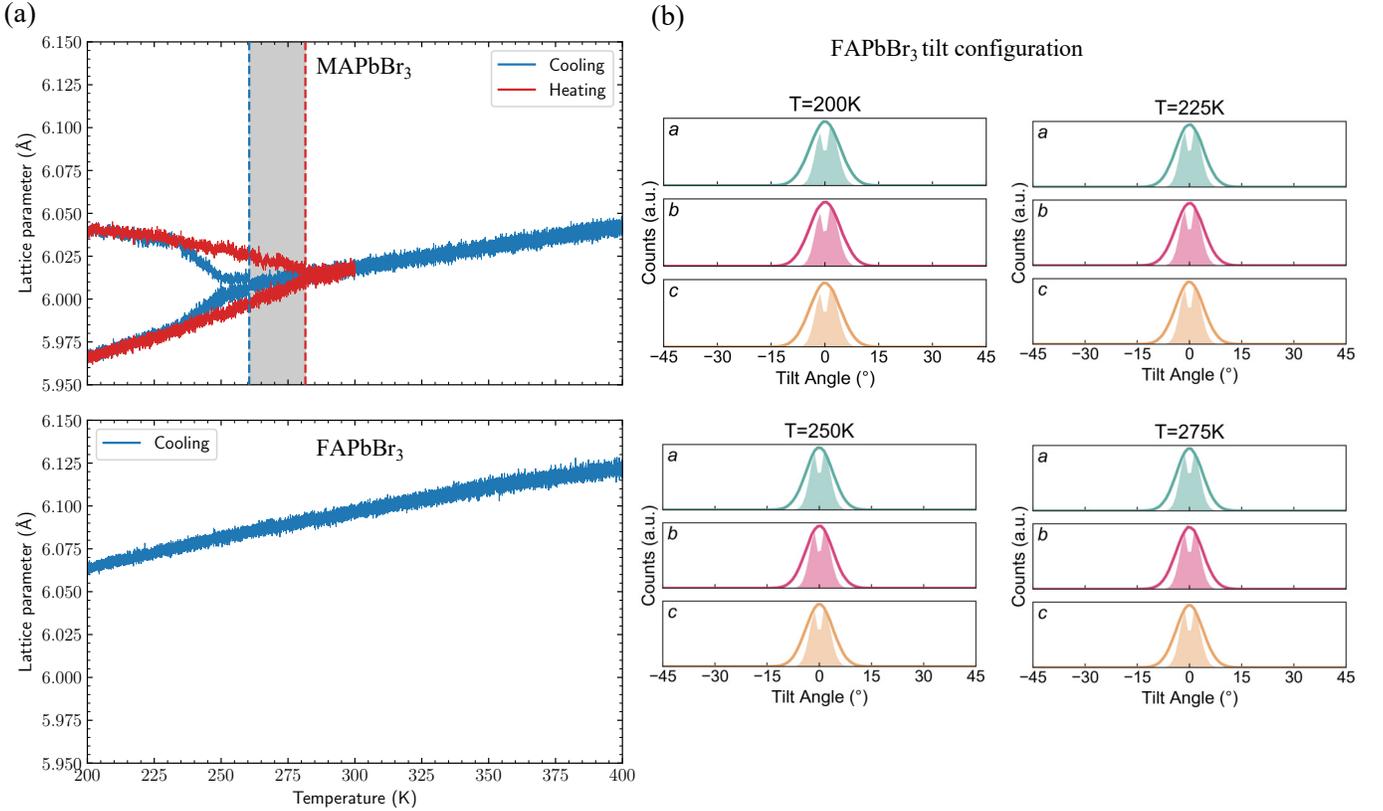

FIG. S22. **Phase transition temperatures from MD simulations.** **(a)** Lattice parameters in MAPbBr$_3$ and FAPbBr$_3$ during cooling (blue lines) and heating (red lines) NPT MD runs. The cooling/heating rates were 400 K/ns. As a result of the rapid cooling/heating rates, there is substantial hysteresis in the phase transition temperature of MAPBBr$_3$ (grey shaded area). The average of the cooling and heating simulations gives a phase transformation temperature of $\sim$ 270 K. No direct signature of phase transformation is visible from the lattice parameters in the FAPbBr$_3$ case. **(b)** To probe whether the phase transition occurs, we analysed NPT MD trajectories and quantified octahedral tilting in FAPbBr$_3$ at 200 K, 225 K, 250 K and 275 K. These simulations started from configurations at the corresponding temperature from the cooling run in (a), and were then ran for 100 ps. Each panel corresponds to one axis. The solid lines denote the dynamic distribution of tilting. The shaded area below the solid lines is the distribution of the tilt angle correlation with the next nearest neighbour along the same direction (see [20] for details). The corresponding global Glazer tilting pattern is $a^0a^0a^0$ at all temperatures. At 200 K, we observe a clear asymmetry in the tilt correlation towards positive values for all 3 axes, which is potentially a precursor of a transition to $a^+a^+a^+$.

| FAPbBr$_3$ | MD 300K | Exp X-ray 300K |
|---|---|---|
| $\xi_\perp$ [Å] | 8.23 | 19.67±0.58 |
| $\xi_\parallel$ [Å] | 15.78 | 29.06±8.6 |
| Ratio | 1.91 | 1.47 |

TABLE VIII. Comparison of MD and experimentally derived spatial correlation lengths for FAPbBr$_3$. $\xi_\perp$ and $\xi_\parallel$ are the thickness and diameter, respectively of the disc-shaped dynamic nanodomains. Ratio corresponds to $\frac{\xi_\parallel}{\xi_\perp}$. Note that in the MD case we compare two times the correlation lengths, to represent the diameter, rather than the radius, of the dynamic nanodomains.

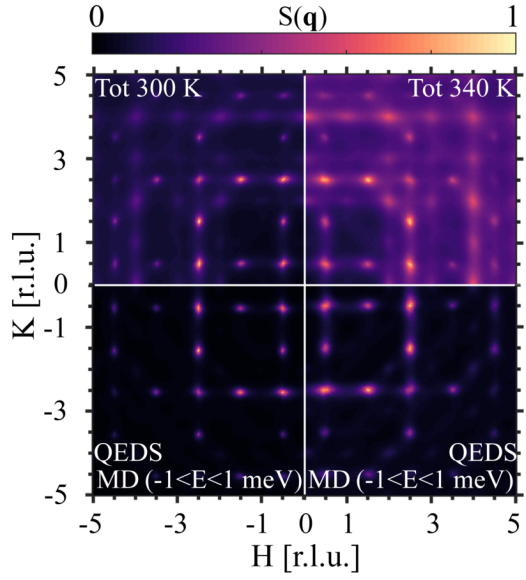

FIG. S23. **Comparison of 340 K and 300 K MD simulated diffuse scattering in MAPbBr₃.** Left quadrants correspond to total $S(\mathbf{q})$ and QEDS component of $S(\mathbf{q})$ at 300 K, while right quadrants correspond to data simulated at 340 K.

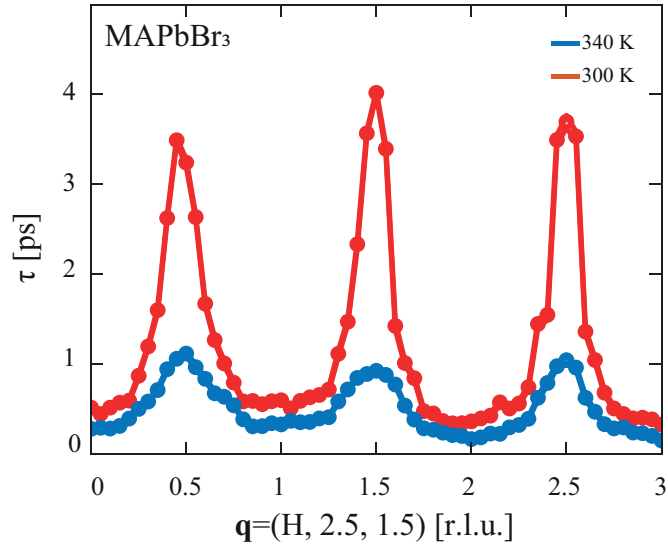

FIG. S24. **Comparison of dynamic nanodomain lifetimes along R-M diffuse rods at 340 K and 300 K from MD simulated $S(\mathbf{q}, E)$ in MAPbBr₃.**

## viii.  THE ASSIGNMENT OF LOWER TEMPERATURE PHASE OF FAPbBr₃

Although the Bragg pattern produced by assuming three tetragonal $P4/mbm$ macroscopic twin components is almost identical to the Bragg pattern produced with one $Im\bar{3}$ component, the appearance of additional superstructure peaks that only exist in $Im\bar{3}$ and not in twinned $P4/mbm$ confirms the absence of non-merohedral twinning in this material (e.g., the (1.5 1.5 0) reflection referenced to the cubic cell). This also rationalizes the origin of a metrically cubic unit cell measured in high-resolution powder synchrotron X-ray and neutron diffraction experiments at the same temperature [13].

An alternative hypothesis for the origin of these superstructure peaks could be that of multiple scattering, especially given the heavy elements in FAPbBr₃. Multiple elastic scattering can occur when a reflected wave matches any other Bragg conditions (e.g., a sphere of radius $|\mathbf{k}_i|$ centered at $\mathbf{k}_i$ passes through reciprocal lattice points other than the origin). We calculated the possibility of matching such a condition for $(hkl)$ reflections in the range, $-8 \leq h, k, l \leq 8$ with respect to the enlarged cubic supercell ($a = 11.9171$ Å). With the tolerances given by either the best (0.0094 Å$^{-1}$) or worst (0.0367 Å$^{-1}$) measured FWHM of the Bragg peaks and the experimental wavelength (0.961121 Å), no multiple scattering conditions were found for the observed (110) (as shown in Fig. S25) or (330) reflections of the $Im\bar{3}$ supercell. This analysis is equivalent to searching for half-integer reflections of the primitive, parent $Pm\bar{3}m$ unit cell.

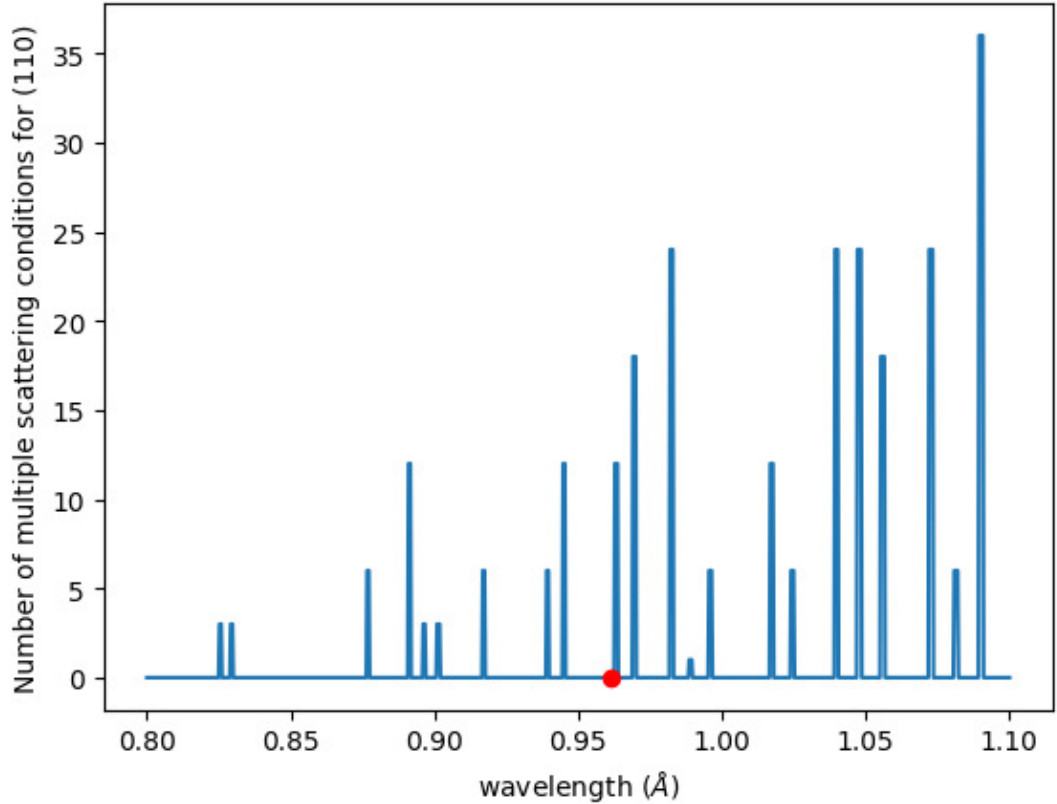

FIG. S25. **Number of multiple scattering events for (110) reflection of the $Im\bar{3}$ FAPbBr$_3$ supercell as a function of incident X-ray wavelength.** The red dot marks the position of 0.961121 Å.

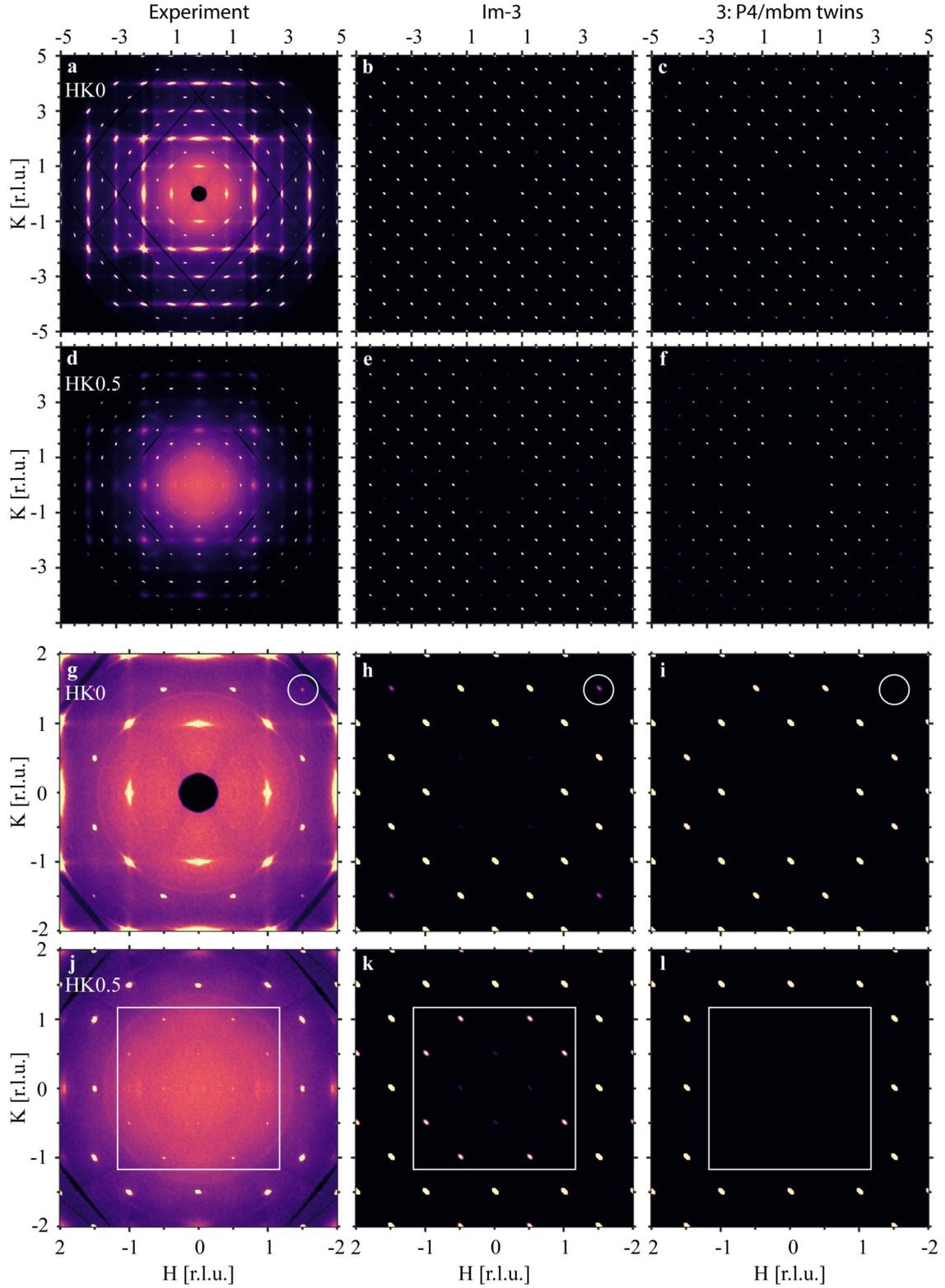

FIG. S26. **Accurate assignment of low symmetry phase in FAPbBr$_3$ at $T = 200K$.** In the first column 2D experimental diffraction patterns are shown. The second column shows the simulated patterns utilising $Im\bar{3}$ structure we obtained by refining the experimental data, while the third column shows the simulated patterns utilising the structure factors (and incorporating three twins) from $P4/mbm$ cif file which we obtained by refining data. $HK0$ reciprocal space planes are shown in (**a**)-(**c**) and (**g**)-(**i**) and HK0.5 reciprocal space planes are shown in (**d**)-(**f**) and (**j**)-(**l**). The positions of superstructure peaks expected in $Im\bar{3}$ space group are marked with white circles and rectangles.

## ix. THE INFLUENCE OF BEAM DAMAGE ON LOCAL STRUCTURE

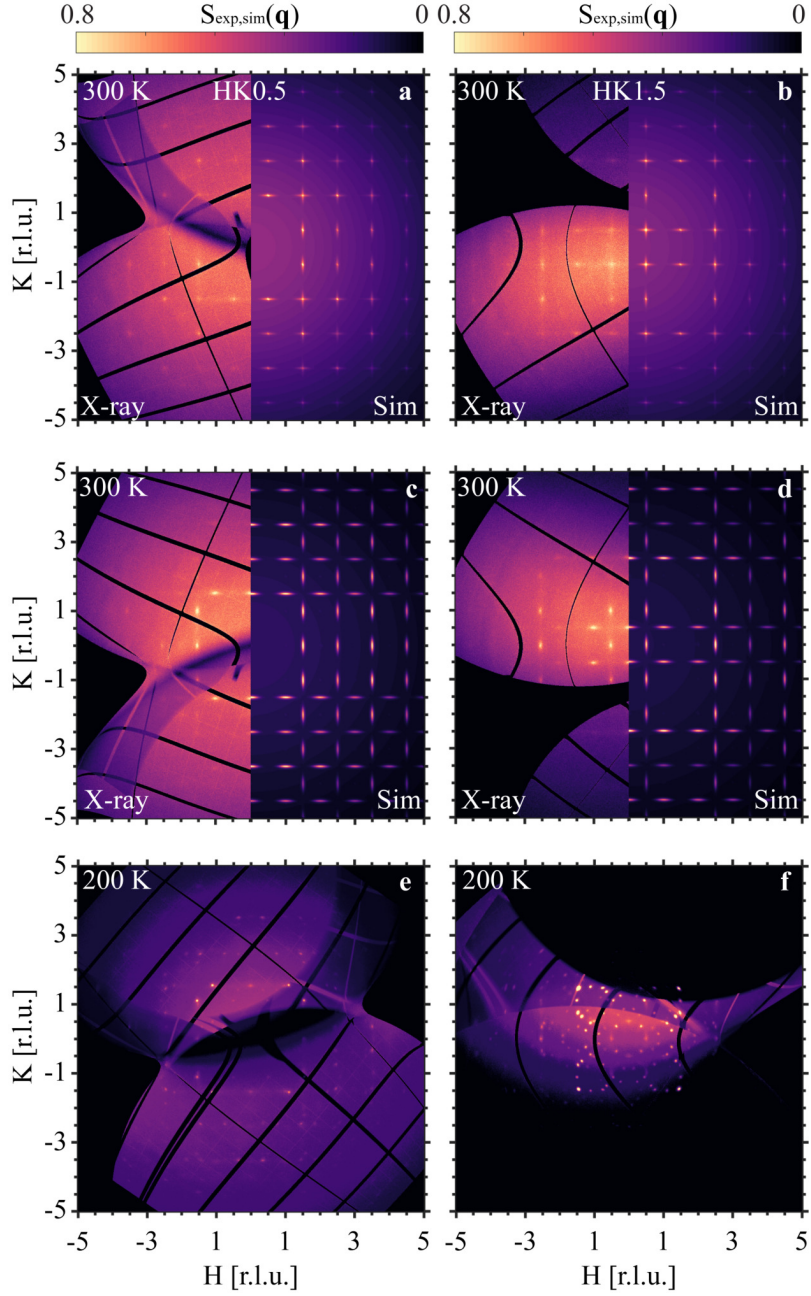

FIG. S27. **The effect of beam damage on experimental scattering function $S(\mathbf{q})$ in MAPbBr$_3$.** (**a**) and (**b**) represent 2D X-ray experimental scattering function (left panels) and simulated for local $I4/mcm$ structure (right panels). This data was collected at $18\,\mathrm{keV}$ within $35\,\mathrm{min}$ at X-ray flux of $1.542 \times 10^{13}\,\mathrm{s^{-1}}$ incident on an area of $100\,\mu\mathrm{m}$ x $100\,\mu\mathrm{m}$. Repeating the experiment after the first exposure under the same conditions, we observe the change in diffuse scattering signals (**c** and **d**, left panels) which we assign to the formation of local $P4/mbm$ structure (as simulated in right panels). We repeated the experiment again but this time at $T = 200\,\mathrm{K}$, and we observed the appearance of additional intensities in the X-ray scattering function (**e** and **f**), which we assign to another beam damage mechanism. We confirm that the sample did not transition to average $I4/mcm$ phase as expected at $T = 200\,\mathrm{K}$.

We observed that prolonged exposure to X-ray synchrotron radiation of MAPbBr$_3$ in the average cubic phase induces a transformation of the local out-of-phase octahedral tilting to in-phase, thus effectively changing the symmetry of the local structure from $I4/mcm$ to $P4/mbm$, whilst maintaining the same twin operators, as demonstrated in Fig. S27.

It is important to note that this occurs without the significant loss of intensity of Bragg peaks, which is a common observation under beam damage. With further exposure, additional reflections appear in the majority of reciprocal space planes, accompanied by a flower-like diffuse scattering in the $HK1.5$ planes in both MAPbBr$_3$ (Fig. S27 **e** and **f**) and FAPbBr$_3$ (Fig. S28). Our observation is consistent with previous reports which suggested both electron and X-ray beam damage could lead to the appearance of superstructure reflections [21, 22].

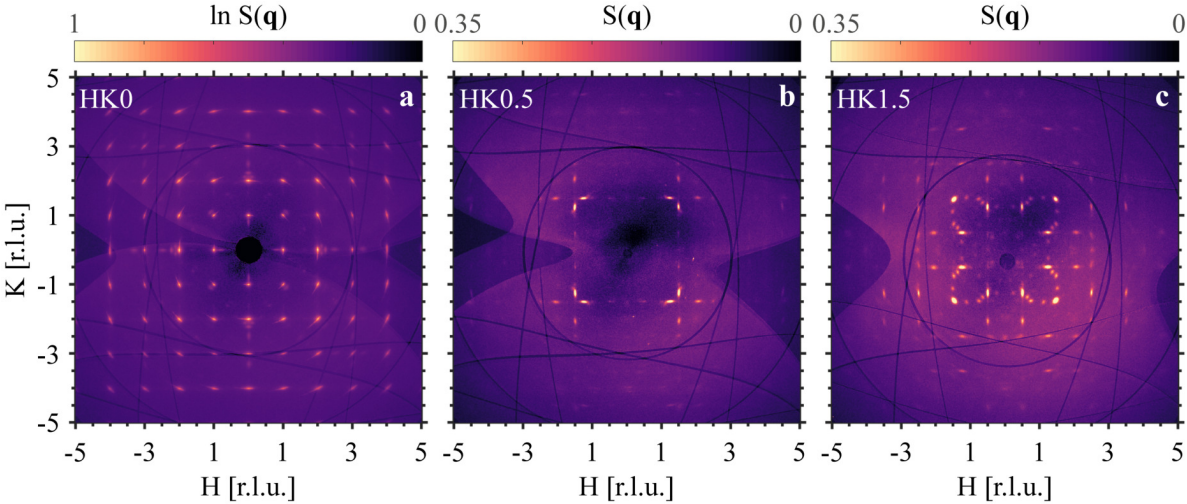

FIG. S28. **The effect of beam damage on experimental scattering function in FAPbBr$_3$.** Due to high symmetry, to improve signal-to-noise ratio, each image was obtained by averaging the intensities across $HKn$, $nKL$ and $HnL$ planes.

## x. INELASTIC NEUTRON SCATTERING

The lifetime of the local structure was calculated using the relationship $\tau = \hbar/FWHM$, where $FWHM$ is the full-width half-maximum of the corresponding Lorentzian peak [23] in energy domain. To estimate the energy windows we are probing with different instruments we utilise the energy resolution as the measure of FWHM. Table IX below shows the dynamic windows we are probing for various types of scans for various instruments. Using Sika we performed elastic $q$ resolved scans which revealed the presence of diffuse scattering peaks at R and M points in MAPbBr$_3$ and FAPbBr$_3$, as shown in Fig. S30 **a** and **b**, respectively. To determine in what inelastic energy range the scattering occurs from the local dynamic nanodomains we conducted additional scans with Taipan at fixed energy transfers of 1 meV and 2 meV, as seen in Fig. S31. Neither scan showed the diffuse scattering peaks in the inelastic domain, confirming the local dynamic nanodomain scattering quasi-elastically in energy windows below 1 meV. For further details on the dynamics windows and their relationship to the instrument resolution in energy see Table IX. To determine the lifetimes of dynamic nanodomains, we performed quasi-elastic scans using a cold triple-axis spectrometer, Sika. As we have outlined in the main text, the R point scattering in MAPbBr$_3$ in the average cubic phase carries information about dynamic tetragonal nanodomains which exhibit out-of-phase correlations of c-axis octahedral tilts along the c crystallographic axis. We fix our q point to the R point in MAPbBr$_3$ but perform a variable energy transfer scan in the quasi-elastic energy window. We repeat the scan for multiple temperatures as we cool down the sample to approach the average tetragonal phase as shown in Fig. S32 **a**. We perform the Lorentzian fit to the data to extract FWHM and thus determine the lifetime of the local nanodomains. However, we find that at each temperature point, the best fit is achieved with FWHM equal to the resolution function. Thus we determine that the quasi-elastic scattering profile is resolution-limited implying that the lifetimes of these excitations are at least approximately 7 ps or higher. In Fig. S32 **b** we repeat the same procedure in FAPbBr$_3$ crystals at M points to capture in-phase dynamic octahedral tilts that are present in this material. However, the results are again resolution-limited leading to the same conclusion about the dynamics of these nanodomains. One should be cautious when interpreting these results as a large contribution to the quasi-elastic line is expected from the stochastic motions of the hydrogen-rich A-site cations which would lead to incoherent quasi-elastic scattering. This signal will be convoluted with the quasi-elastic Lorentzian that would originate from the local structure. As such our estimation of the lower bound of the lifetimes of the dynamic nanodomains might not be completely accurate. The correlation length (diameter) of the local dynamic nanodomains was calculated using the relationship $\xi = \frac{a}{\pi \cdot \text{HWHM}}$, where $HWHM$ is the half-width half-maximum of the corresponding Lorentzian peak measured in reciprocal lattice units and $a$ is the pseudocubic unit cell [24]. We measured resolution in momentum space for various $q$ points at Taipan and converted it to a maximum detectable correlation length diameter. The results are presented in Table X.

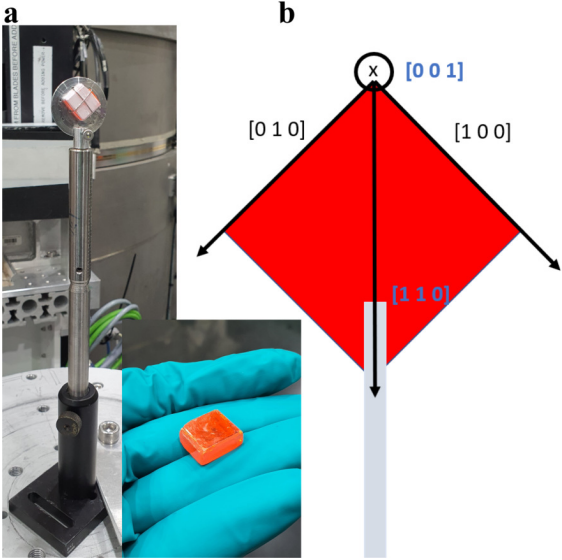

FIG. S29. **MAPbBr$_3$ single crystal mounting for Inelastic Neutron Spectroscopy experiment.** (**a**) To access the HHL plane crystal was mounted diagonally. (**b**) The schematic of crystal orientation where crystallographic directions denoted in blue form the scattering plane which will be accessible by triple axis spectrometer.

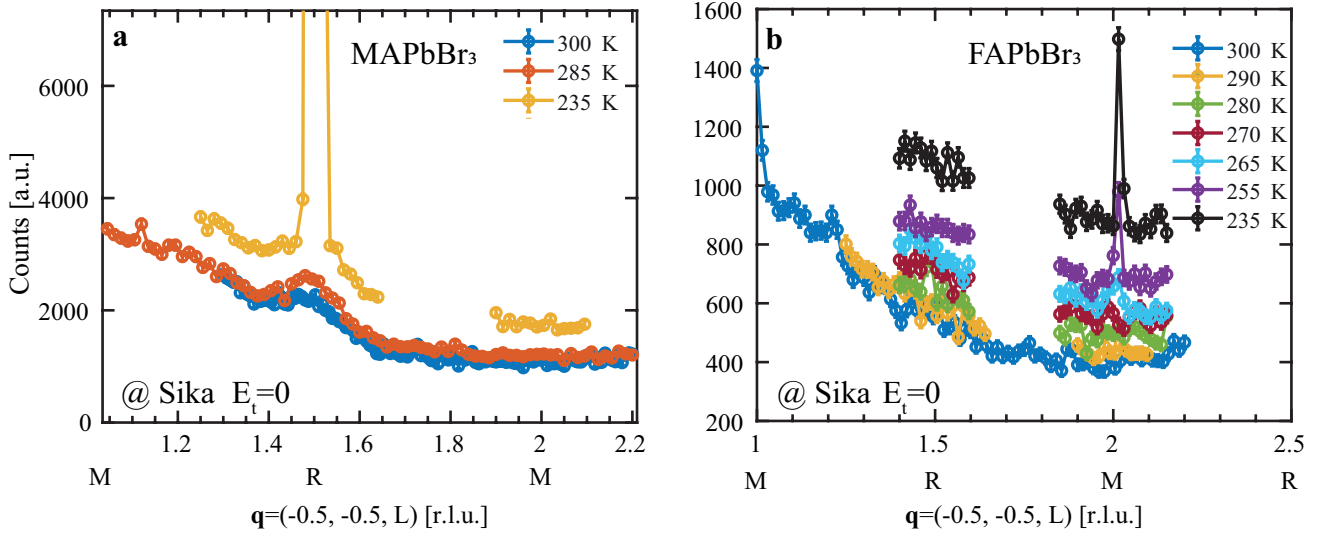

FIG. S30. **Elastic neutron scattering scans with Sika.** (**a**) Elastic scans (zero neutron energy transfer, $E_t = 0$) across the $[-0.5, -0.5, L]$ direction in the reciprocal space of MAPbBr$_3$ for various temperatures in the average cubic phase. (**b**) Elastic scans (zero neutron energy transfer, $E_t = 0$) across the $[-0.5, -0.5, L]$ direction in the reciprocal space of FAPbBr$_3$ for various temperatures in the average cubic phase.

TABLE IX. Dynamic windows at Sika and Taipan.

| Instrument | Resolution HWHM | Dynamic window |
|---|---|---|
| Sika | 0.045 meV | < 7.31 ps |
| Taipan | 0.45 meV | < 0.731 ps |

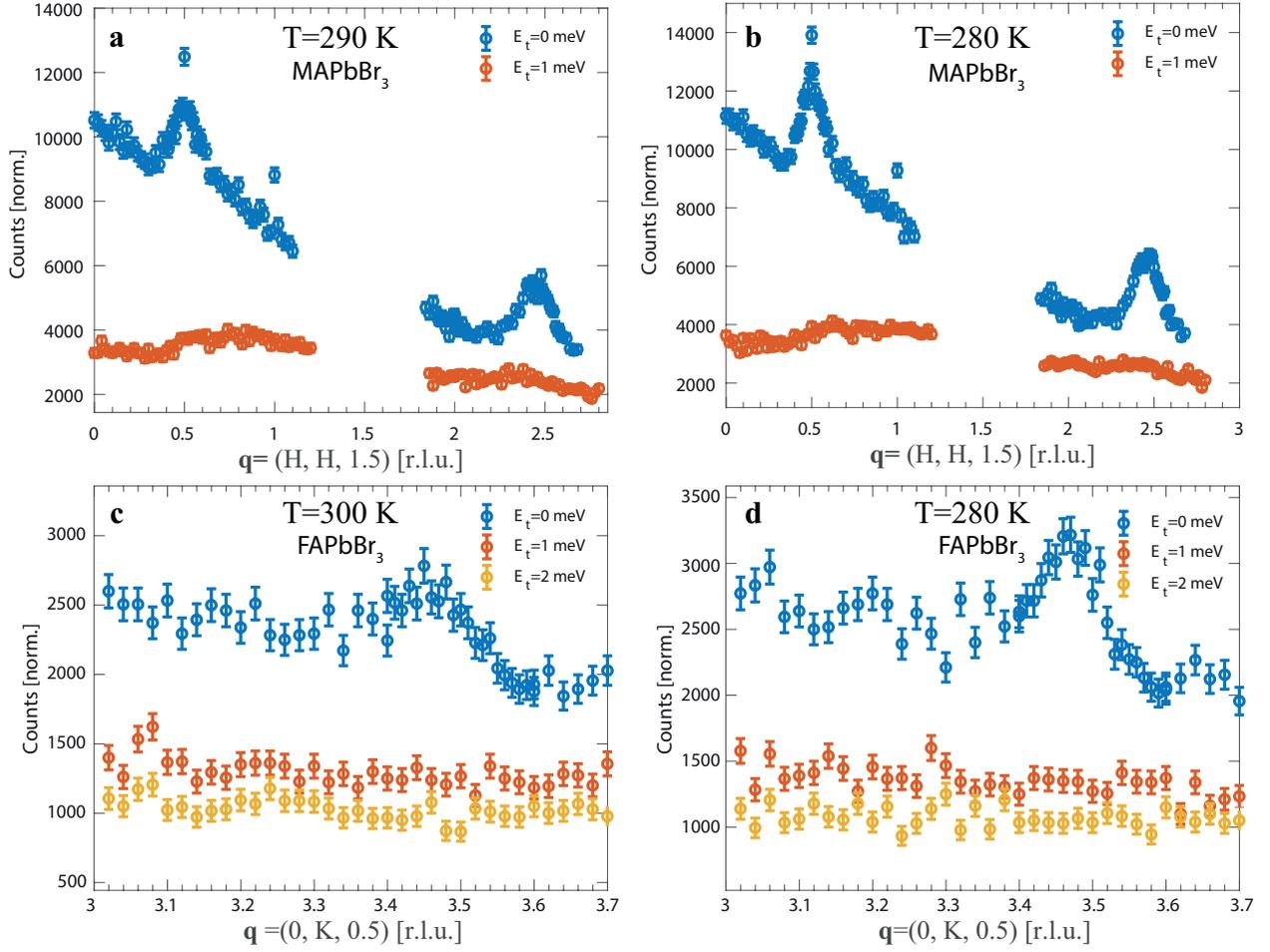

FIG. S31. **Elastic and inelastic constant energy, variable $q$ scans with Taipan.** (**a**) Elastic ($E_t = 0\,\mathrm{meV}$) and inelastic ($E_t = 1\,\mathrm{meV}$) Q scans across the $[H, H, 1.5]$ direction of MAPbBr$_3$ at ($T = 290\,\mathrm{K}$). (**b**) Elastic ($E_t = 0\,\mathrm{meV}$) and inelastic ($E_t = 1\,\mathrm{meV}$) $q$ scans across the $[H, H, 1.5]$ direction of MAPbBr$_3$ at ($T = 280\,\mathrm{K}$). (**c**) Elastic ($E_t = 0\,\mathrm{meV}$) and inelastic ($E_t = 1\,\mathrm{meV}$, $E_t = 2\,\mathrm{meV}$) $q$ scans across the $[0, K, 0.5]$ direction of FAPbBr$_3$ at ($T = 300\,\mathrm{K}$). (**d**) Elastic ($E_t = 0\,\mathrm{meV}$) and inelastic ($E_t = 1\,\mathrm{meV}$, $E_t = 2\,\mathrm{meV}$) $q$ scans across the $[0, K, 0.5]$ direction of FAPbBr$_3$ at ($T = 280\,\mathrm{K}$). It is noteworthy that the diffuse scattering peaks in both MAPbBr$_3$ and FAPbBr$_3$ comprise solely elastic scattering, with no inelastic components observed.

TABLE X. Real space resolution at various $q$ points at Taipan

| q [r.l.u.] | [0, 0.5, 0.5] | [0, 2.5, 0.5] | [0, 3.5, 0.5] | [0.5, 0.5, 1.5] | [2.5, 2.5, 1.5] |
|---|---|---|---|---|---|
| FWHM [r.l.u.] | 0.0192 | 0.0277 | 0.0345 | 0.0135 | 0.0264 |
| $\xi$ [u.c.] | 33.157 | 22.983 | 18.4530 | 47.1570 | 24.114 |

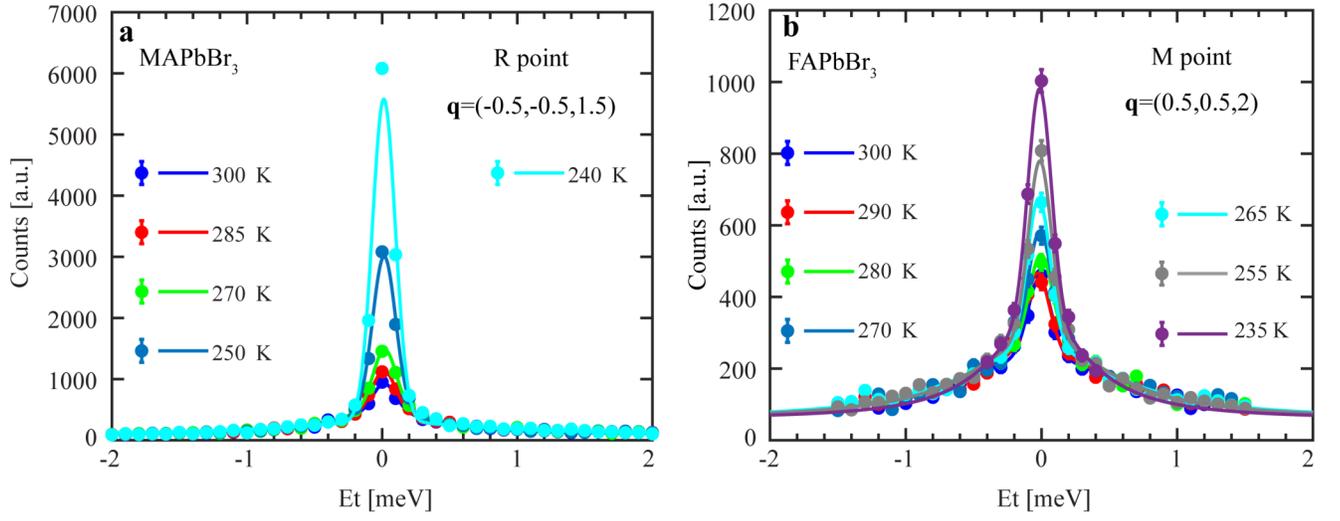

FIG. S32. **Constant** $q$**, variable energy scans of the quasielastic spectrum at the zone edge obtained using Sika.** (**a**) Temperature dependence of MAPbBr$_3$ quasi-elastic spectrum at R point. (**b**) Temperature dependence of FAPbBr$_3$ quasielastic spectrum at M point. Solid lines are best fits to the experimental data.

### xi. DIRECT OBSERVATION OF PRESENCE OF FERROELASTIC TWIN DOMAINS IN MAPBBR₃ AND THEIR ABSENCE IN FAPBBR₃ SINGLE CRYSTALS

Recently, it has been demonstrated that ferroelastic domain walls in halide perovskites can be detected using non-polarized light [25], which differs from the standard technique of using cross-polarized light [26, 27] . The authors propose that in halide perovskites, domain walls form atomically coherent interfaces between twin domains, which share the same composition but exhibit different crystalographic orientations, defined by the mirror twin plane. In materials with low symmetry and optical anisotropy, reflections and refractions may occur due to anisotropy of refractive indices. The optical contrast observed under nonpolarized illumination results from optical reflections and refractions at the domain wall interface, as the refractive indices normal to the interface will differ. We employ the same technique and detect the presence of twin domains in MAPbBr₃ and their absence in FAPbBr₃. We note that for twins to be easily observable with this method, it is crucial to cleave the crystals to surfaces with very low roughness. Figure S34 illustrates the formation of twins when entering the tetragonal phase below 236 K and their subsequent evolution as the temperature decreases. At $T = 200$ K, real-space images reveal twin boundaries at a 45-degree angle relative to the principal pseudocubic axes. These boundaries correspond to the (110) twins, which are also identified in reciprocal space through SC-XRD (Section ii). The intermediate phase between the cubic and tetragonal phases, previously identified as incommensurate, also results in a distinctive twin domain pattern at $T = 150$ K, as shown in Fig. S34. Figure S34 **a** further highlights the significant increase in possible twin orientations upon entering the orthorhombic phase at 145 K. This observation aligns with our twinning model for the orthorhombic structure deduced from reciprocal space (Section ii), which also predicts the increased degrees of freedom for local structure orientations when transitioning to the orthorhombic phase. Figure S34 **a** demonstrates that the formation of twin domains is a macroscopic effect, as it is observable across fields of view (FOVs) spanning several hundreds of microns. On the other hand, we report the absence of ferroelastic twins in Figure S34 **b**. We have scanned several fields of view on this and other FAPbBr₃ crystals and failed to detect ferroelastic twins in a broad temperature range spanning from 100 K to 300. This observation is consistent with the existing literature, where ferroelastic twin domain formation has not been reported.

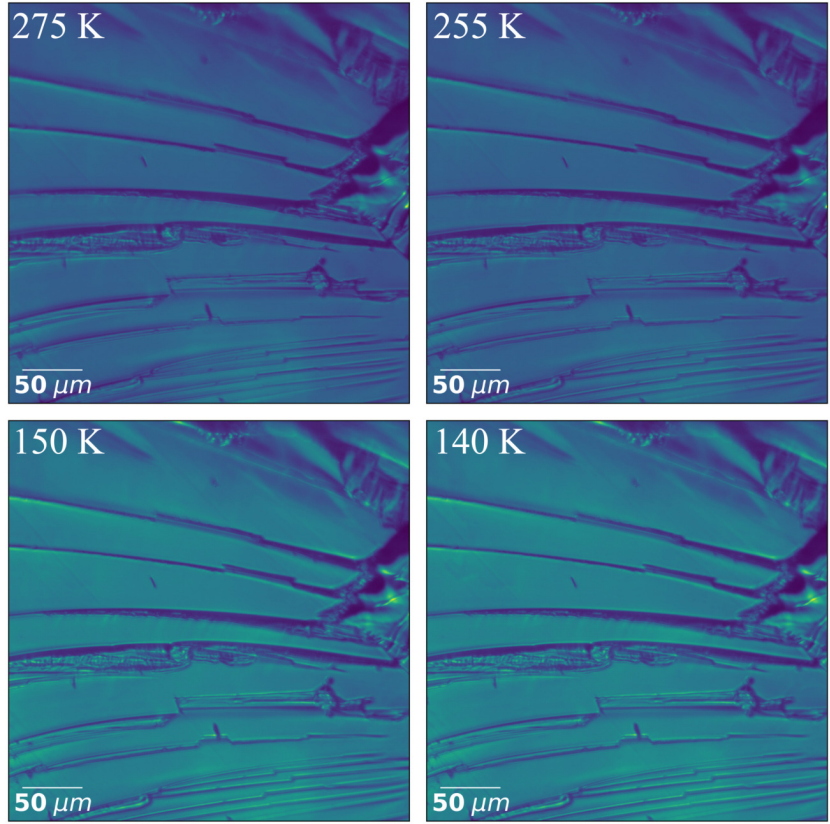

FIG. S33. **The absence of ferroelastic twin domains in FAPbBr₃ was observed using an optical microscope.** Although the system undergoes phase transitions covering the temperature range presented in these images, the formation of ferroelastic twin domains was not observed.

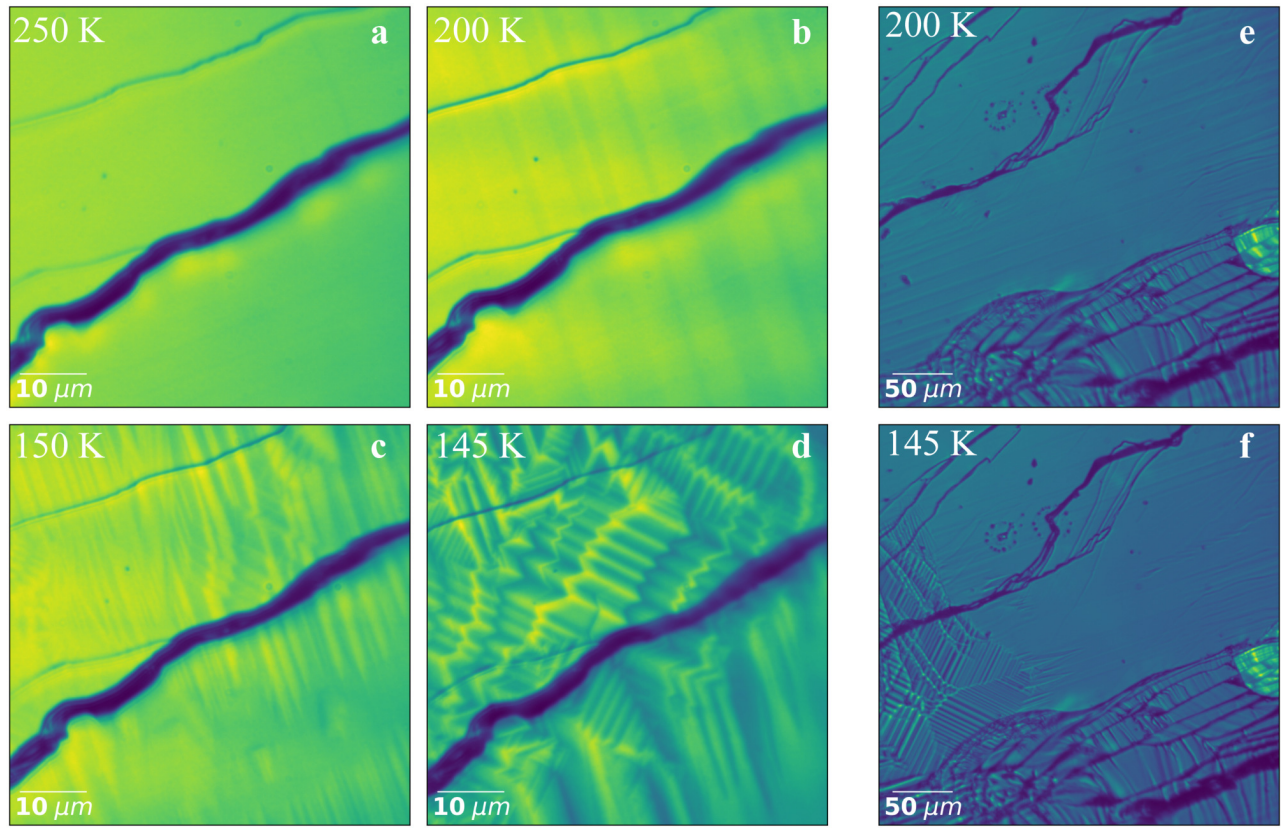

FIG. S34. Formation of ferroelastic twin domains in MAPbBr$_3$ was observed using an optical microscope.(**a**)-(**d**) Diffraction-limited optical images of the same region as a function of temperature. Various twin structures can be observed upon cooling. It is evident that twins are first present at $T = 200$ K (**b**) upon cubic-tetragonal transition and then two types of twins are formed in the orthorhombic phase in (**c**) and (**d**). (**e**) and (**f**) Lower magnification images reveal twin domains in the tetragonal phase at $T = 200$ K and the orthorhombic phase at 145 K.

## xii. VERIFICATION OF THE ALLEGRO MACHINE LEARNED FORCE FIELDS

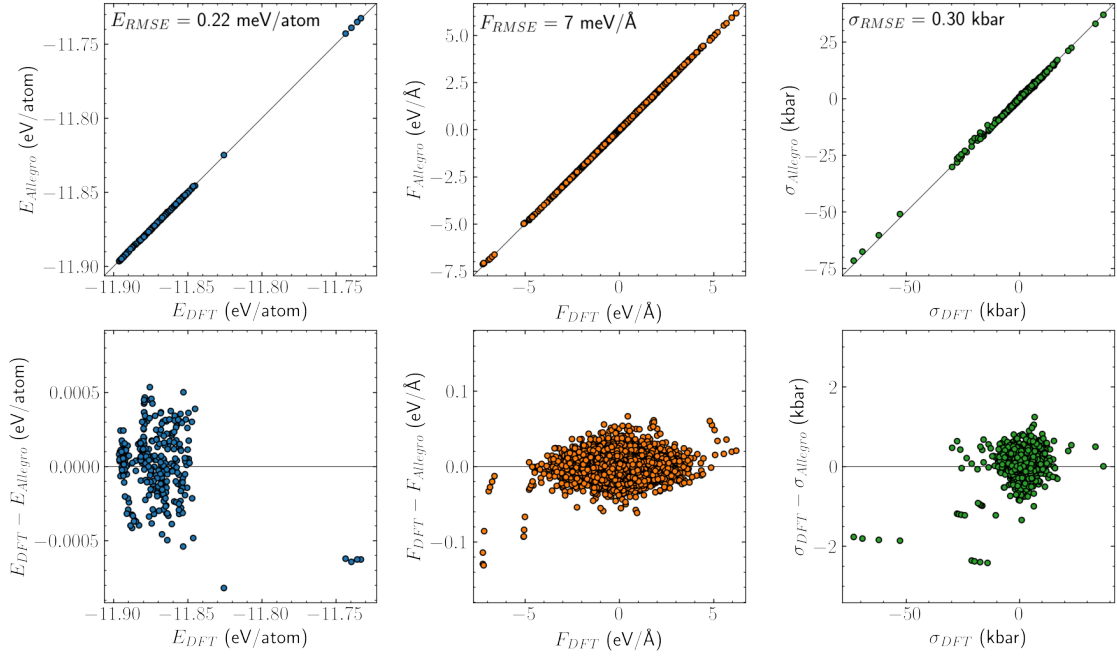

FIG. S35. Parity plot (top row) and error distribution (bottom row) for the MAPbBr$_3$ Allegro MLFF. Root mean squared errors (RMSEs) of energies, force components and stress components are given in the top row.

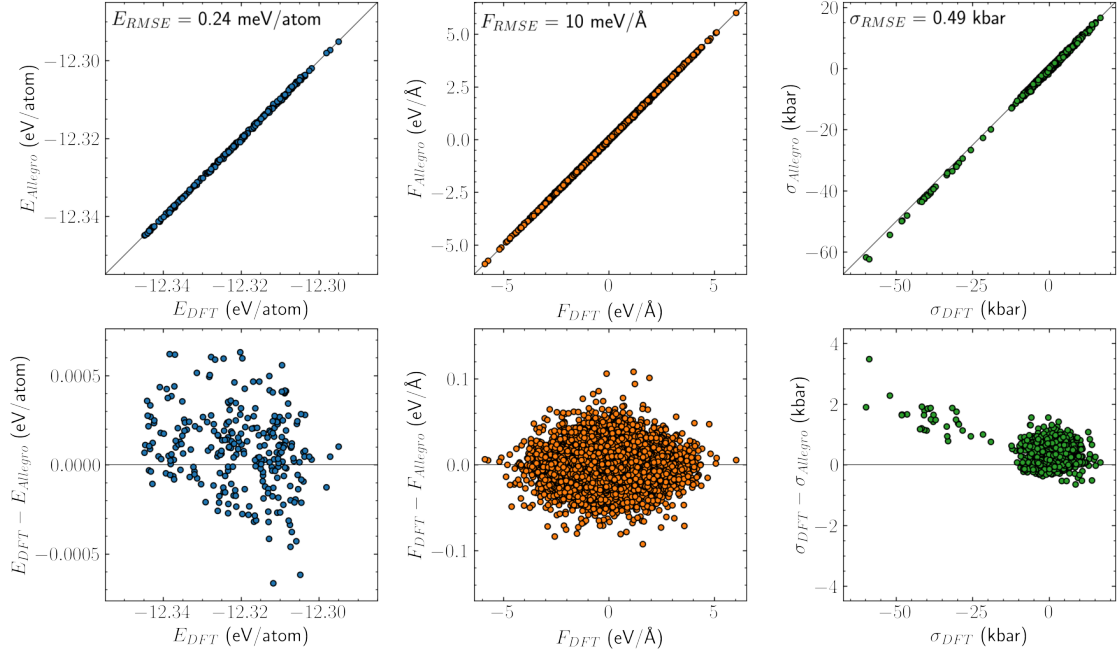

FIG. S36. Parity plot (top row) and error distribution (bottom row) for the FAPbBr$_3$ Allegro MLFF. Root mean squared errors (RMSEs) of energies, force components and stress components are given in the top row.